\documentclass[12pt,b4paper]{article}
\usepackage[utf8]{inputenc}
\usepackage{amsmath,setspace,geometry}
\usepackage{amsfonts}
\usepackage[shortlabels]{enumitem}
\usepackage{mathptmx}
\usepackage[T1]{fontenc}
\usepackage{graphicx}
\usepackage{bbm}
\usepackage[colorlinks,citecolor=purple,urlcolor=blue,bookmarks=false,hypertexnames=true]{hyperref}
\usepackage[]{natbib} 
\usepackage[sectionbib]{chapterbib}
\bibpunct[:]{(}{)}{,}{a}{}{,}
\usepackage[english]{babel}
\usepackage{float}
\usepackage{booktabs}
\usepackage{pdfpages}
\usepackage{threeparttable}
\usepackage{lscape}
\newtheorem{assumption}{Assumption}
\newtheorem{definition}{Definition}

\title{Estimating Endogenous Coalitional Mergers: Merger Costs and Assortativeness of Size and Specialization\thanks{I thank my advisor Jeremy Fox for his valuable advice. I also thank Sen Lu for sharing an early version of code for computing matching equilibria with externality and Fumio Nagasawa for allowing me to refer to the firm-level data he has collected. I benefited from the extensive comments of Robin Sickles, Isabelle Perrigne, Kei Kawai, Takeshi Murooka, Naoki Wakamori, Takuma Matsuda, and Yuri Matsumura as well as seminar participants at Japan Empirical Industrial Organization Workshop, 16th Applied Econometrics Conference, Japan Society of Logistics and Shipping Economics Kanto region, and 24th Moriguchi prize conference. I also thank Koichiro Hayashi and Nobufuku Takahashi for sharing industry knowledge and expertise at Syosen Mitsui and Nippon Yusen groups.
}}
\author{Suguru Otani\thanks{Department of Economics, Rice University. Email: so19@rice.edu}}

\begin{document}

\maketitle
\begin{abstract}
I present a structural empirical model of a one-sided one-to-many matching with complementarities to quantify the effect of subsidy design on endogenous merger matching. I investigate shipping mergers and consolidations in Japan in 1964. At the time, 95 firms formed six large groups. I find that the existence of unmatched firms enables us to recover merger costs, and the importance of technological diversification varies across carrier and firm types. The counterfactual simulations show that 20 \% of government subsidy expenditures could have been cut. The government could have possibly changed the equilibrium number of groups to between one and six.
 
 \vspace{0.1in}
\vspace{0in}

\bigskip
\end{abstract} 

\newpage
\section{Introduction} \label{sec:introduction}

The relationship between the matching mechanism and merger policy is a complex and evolving area of research, and the precise nature of their interaction is still not fully understood.
One case that illustrates the success of special merger encouragement as a mechanism is the shipping mergers and consolidations that occurred in Japan in 1964. Japan relied on domestic shipping firms to transport nearly 90 percent of its trade.\footnote{I am referring to page 10 of \cite{book4}.} Simultaneously, Japan was beset by a chronically inefficient industrial structure caused by postwar entry into the shipping industry. To resolve this inefficiency problem, the Japanese government implemented ``the Two Reconstruction Laws" in 1964, which entitled merged firms to a subsidy if, after mergers and groupings, their total tonnage exceeded a certain threshold.\footnote{After the Peace Treaty came into effect in 1952, Japanese shipping and shipbuilding regained complete freedom in international markets, but this involvement also meant vulnerability to the vagaries of world politics. The Japanese shipping industry faced the closure of the Suez Canal in 1956 and a depression after the Korean War from 1950 to 1953. During the Korean War, Japanese shipping tonnage had expanded at a particularly rapid rate, and then had a large degree of overcapacity when the shipbuilding demand came to an abrupt end. 

At that time, the industry needed to overcome a shortfall in earnings within the industry. The shortfall can be attributed to two principal factors. First, there was a sharp fall in freight rates as the wartime boom turned to a postwar slump in demand. Second, Japanese interest rates were higher than those in other countries, which imposed a heavy burden on firms that had acquired tonnage with the aid of borrowed capital.
For reference, see the detailed history of the Japanese shipping industry from 1952 to 1962 in \cite{chida1990japanese}, Section 4.} The objective was to decrease the number of firms. After the shipping mergers and consolidations, 27 major firms and 68 small firms formed six mutually exclusive groups through mergers and hierarchical relationship contracts, and 33 firms chose to remain unmatched. These six groups satisfied the subsidy requirements of total group tonnage. Thus, the government successfully implemented a simple subsidy mechanism to effectively encourage firms to endogenously initiate and implement horizontal coalitional mergers and decreased the number of firms. After these mergers and consolidations, Japan succeeded in achieving the top share of total tonnage in the world's shipping industry until the 1973 oil shock.\footnote{See global trends of tonnage in the 1960s in Figure \ref{fg:shippingtonnage} in Appendix \ref{sec:institutional_background}.}

Given the implementation of laws as an ideal experiment, I estimate the primitive parameters in the one-sided one-to-many coalitional matching model with complementarities by \cite{fox2018qe} using novel detailed data about Japanese shipping mergers and consolidations and the processes from 1960 to 1964. The data contain the observed characteristics of almost all firms and ships in the industry before and after the mergers and consolidations.

Constructing a coalitional matching model poses some challenges. The first challenge is a tractable computation. My paper contributes to the literature on coalition formation by applying a tractable matching model. To the best of my knowledge, only \cite{fox2019externality} constructed a one-sided one-to-many matching model to explain mergers in the United States mobile carrier industry. The empirical matching model allows us to quantitatively explain why one firm is more likely to merge with multiple other firms endogenously, based on the observed matching patterns and characteristics. This cannot be investigated by alternative non-cooperative game-theoretic models, such as quantity competition models.

Another challenge is estimating the merger costs for each main firm. The one-to-many matching model with unmatched firms can recover the costs that the main firm implicitly incurs by inviting an additional firm into the group. The balance of the subsidy amount and merger cost is crucial for evaluating merger outcomes. To the best of my knowledge, exploiting unmatched firms to estimate matching costs in a single, large market setting is new to the literature.\footnote{\cite{fox2018jpe} investigated the identification with unmatched agents across many matching markets and found that the existence of unmatched agents helped the identification of unobservable complementarities. \cite{bajari2008evaluating} also investigated the estimation of fixed cost parameters as fixed effects in a many-markets setting.} 

My estimation results indicate that assortativeness of both size and specialization contributes to merger incentives. However, the assortativeness of specialization has a significantly different role across carrier and firm types. This implies that the importance of technological diversification and specialization for merger incentives varies. I also find the existence of high merger costs for adding a target firm to a group in a single, one-sided market. 

Finally, I exercise two counterfactual simulations given the estimated parameters. First, I investigate the effect of changes in the subsidy design on the equilibrium number of groups and unmatched firms. The counterfactual simulations reveal the efficiency of the government subsidy in terms of expenditure minimization. Second, I investigate how the subsidy threshold and amount could have changed merger configurations. The simulations open a black box regarding how market participants are determined via coalitional mergers. \footnote{The numerical implementation of a general matching model with complementarities based on \cite{azevedo2018existence} is new to the literature. In Appendix \ref{sec:monte_carlo}, I present identification and estimation results in a rich variety of real-world market situations. In Appendix \ref{sec:comparative_statics}, I demonstrate the numerical results of comparative statics based on simulated data.} 

My simulation results illustrate that, given a subsidy amount, by shifting the subsidy threshold, the government could have changed the number of equilibrium matching groups to between one and six. Conversely, given a subsidy threshold, increasing the subsidy amount would not have stimulated additional merger incentives. The simulations of several policy mix scenarios indicate that the actual subsidy provision in 1964 was too generous, and 20 \% of the subsidy expenditure could have been cut. The cost-savings amount would have been approximately equivalent to the matching production of a hypothetical merger of the two largest groups.

Second, in the counterfactual merger configurations, I find that coalitional mergers would have been driven by large firms that could have overcome the high merger costs given the small subsidy amount. As the subsidy amount had increased, additional mergers of unmatched firms would have occurred, and some groups would have split into multiple subgroups to enjoy the larger subsidy amount shared in the subgroups. For example, for individual firms, the subsidy benefit from a group consisting of four firms would be less than the benefit from its subgroup consisting of two firms because the subsidy amount was shared with all members in the group.

\subsection{Related literature}

This paper examines the effect of a subsidy mechanism on endogenous merger matching. This paper is based on three strands of the literature, namely, one-sided one-to-many matching with complementarities, endogenous merger analysis, and optimal subsidy to decrease the number of firms. See \cite{chiappori2016jel} for a general reference on several empirical matching models.

First, this paper contributes to the literature on empirical one-sided one-to-many transferable utility (TU) matching. The most related econometric model is \cite{fox2018qe}, whose model has been applied to other empirical topics such as banking merger \citep{akkus2015ms,chen2013ijio}, faculty room allocation \citep{baccara2012aer}, executive and firm matching \citep{pan2017determinants}, and buyer and seller relationships in the broadcast television industry \citep{stahl2016aer}. These papers have applied the matching maximum score estimator proposed by \cite{fox2010qe} to two-sided many-to-many and one-to-one matching in a TU matching environment. 
However, in general, the one-sided matching model does not guarantee the existence of stable matching.\footnote{Recently, \cite{chiappori2019human} demonstrated the mild conditions for the existence of a stable equilibrium in the one-sided one-to-one matching model. The present paper takes a different approach to convert the general results of \cite{azevedo2018existence} into our computationally feasible setup.} To overcome this obstacle, the present paper extends the approach of \cite{azevedo2018existence} to the one-sided one-to-many matching model in a single large market with unmatched agents.\footnote{As an alternative approach to estimating the one-to-many matching model, \cite{agarwal2015aer} and \cite{diamond2017qe} showed the theoretical results on identification and estimation and demonstrated an empirical application of medical matches. Their results crucially rely on the properties of one-to-many matching data, whereas my model does not. Instead, my model does not incorporate a rich structure of preference heterogeneity and the differentiation of agents.} The proposed tractable method can also be applied to other settings. The computational applicability is demonstrated in Appendix \ref{sec:monte_carlo}.

A few papers investigated the empirical one-sided matching model and the coalitional matching model with a government subsidy mechanism using various approaches. \cite{weese2015qe} developed a method to estimate the bounds of primitive parameters in the coalition formation model by using data regarding the Japanese municipal governments. \cite{gordon2009jpub} investigated the effect of merger subsidies on a geographically adjacent coalition of school districts for economies of scale. These two papers exploited a non-transferable utility (NTU) matching framework. I do not take their coalitional game approach because it requires superadditivity for the payoffs assigned to each coalition, which ignores the additional merger costs, my main subject of interest.

Second, this paper contributes to the literature on endogenous merger analysis. In particular, this paper focuses on endogenous mergers in a single, static, large matching market because of the institutional background explained in Section \ref{sec:data_background}.\footnote{Endogenous merger analyses in the industrial organization literature are divided into dynamic and static matching models. In terms of dynamic matching models, they follow \cite{gowrisankaran1999dynamic}. \cite{stahl2011dynamic} was the first to estimate a merger activity model using a dynamic, strategic framework. \cite{jeziorski2014effects} estimated the sequential merger process to analyze ownership consolidation in the United States radio industry after the enactment of the Telecommunications Act of 1996. \cite{igami2019mergers} applied a stochastic sequential bargaining model to the merger processes of the hard disk industry. As the most recent paper, \cite{hollenbeck2020horizontal} enriched the Gowrisankaran-type dynamic endogenous merger model. With different dynamic approaches, \cite{nishida2015better} compared post-merger and pre-merger beliefs and equilibrium behaviors in a Markov perfect equilibrium in the Japanese retail chain industry. \cite{perez2015building} incorporated mergers as bidding games by incumbents and investigated the effect of the Reagan-Bush administration's merger policy on the reallocation of assets in the United States cement industry. Conversely, using a static matching model, \cite{uetake2019entry} developed an empirical two-sided NTU matching model with externalities using moment inequalities and investigated the effect of entry deregulation on the ``with whom"-decisions of bank mergers  by the Riegle-Neal Act. \cite{akkus2015ms} tackled the same Act with a different approach. They added transfer data and constructed a one-to-one matching model with transfer utility and found that merger value increased from cost efficiencies in overlapping markets, relaxing regulations, and the network effects exhibited by acquirer-target matching.} This paper constructs a coalitional matching model by converting the abstract model of \cite{azevedo2018existence} in which the equilibrium solution is guaranteed to exist uniquely into a numerically tractable model as a classical assignment problem \citep{shapley1971assignment}. In addition, the present paper quantifies the relative importance of post-merger size effects and technological specialization, which are the main economic forces driving firms to pursue mergers to gain cost efficiency in the shipping industry \citep{notteboom2004container}. The cost-side incentives such as decreasing costs are indirectly incorporated in the merger incentive. 

Third, this paper empirically evaluates the effect of using a subsidy mechanism to resolve the excess entry problem. \cite{mankiw1986rand} proposed the so-called ``Excess Entry Theorem". This theorem states that the business-stealing effects and
fixed costs in oligopoly markets result in an excessive number of entrants at a free-entry Cournot and Bertrand equilibrium, and they also induce social inefficiency. In the same spirit as \cite{mankiw1986rand}, \cite{suzumura1987res} proved the inefficiency of excessive entrants under a free-entry quasi-Cournot equilibrium and stated that the paper was motivated by the regulation of excessive competition throughout Japan's postwar period. Although the institutional background and history are shared with this article, I emphasize that my cooperative game-theoretic model does not address the excess entry problem based on non-cooperative game-theoretic models, but rather focuses on merger patterns, which are of interest to the Japanese government.

In many industries, excess entry is reported to be problematic, so many researchers have analyzed it using several structural models and quantitative methods. \cite{waldfogel1999free} studied the radio industry to quantify the inefficiency of excess entry and exercised counterfactual simulations to make comparisons of the number of entrants under free-entry, monopoly, and socially optimal. They proposed two alternatives to resolve excess entry -- the direct regulation of entries and levying an entry tax. By using real-world data about special permissions and government interventions, I investigate the third alternative -- that a regulator encourages the formation of mergers and coalitions.\footnote{\cite{nishiwaki2016horizontal} and \cite{okazaki2022} answered similar policy questions in a related industry in Japan. \cite{nishiwaki2016horizontal} models the divestment dynamics of firms with chronic excess capital in the Japanese cement industry from 1998 to 2009 and examines whether merger-induced divestment improves total welfare. He focuses on endogenous divestment as a key device to resolve the excess entry problem by assuming mergers as exogenous events. Conversely, I utilize the two laws as a natural experiment inducing matchings and focus on endogenous mergers as a device to resolve the excess entry problem. \cite{okazaki2022} investigate the aspect of the capacity coordination policies.} I also evaluate the efficiency of subsidy design in the sense of minimizing government expenditures given targeted matching outcomes. Recently, \cite{miller2019optimal} empirically evaluated the optimality of subsidies in health insurance markets by incorporating market competition and demand changes into the channels of subsidy effects. \cite{clark2020incentivized} evaluate the impact of government-provided incentives by transfer tax reductions on consolidation patterns of local electricity distribution companies in Ontario. Aside from industrial and institutional differences, the focus of my paper -- the channels of a subsidy design's effects on matching outcomes -- complements their insights.

The remainder of this article is organized as follows. I first describe the data, institutional details, and conduct preliminary analysis in Section \ref{sec:data_background}. Next, I detail the one-to-many matching merger model in Section \ref{sec:model}. Section \ref{sec:estimation} introduces the estimation procedures. Section \ref{sec:results} discusses the estimation results. Section \ref{sec:counter_factual} reports the results of the counterfactual experiments. I conclude in Section \ref{sec:conclusion}. The institutional and estimation details, Monte Carlo simulations, comparative statics, supplemental estimations, and robustness checks are shown in Appendices \ref{sec:institutional_background}, \ref{sec:algorithm}, \ref{sec:monte_carlo}, \ref{sec:comparative_statics}, \ref{sec:supplemental_estimation}, and  \ref{sec:robustness_check}, respectively. 
\section{Data and Institutional Background}\label{sec:data_background}

This section provides data descriptions of my dataset and provides a brief overview of the industry background before and after the mergers and consolidations. Next, I present the descriptive statistics. Institutional evidence and preliminary regression results to explore merger patterns are shown in Appendix \ref{sec:institutional_background}.\footnote{Before introducing specific data details, I make some remarks about the broad historical background.
This paper focuses on the period between 1960 and April 1964, which  is before global containerization (1966) and just after the start of the foreign-flag convenience ship boom in Japan (1962). This paper also focuses on the shipping industry, not the shipbuilding industry, although the main firms overlapped both industries indirectly through their stockholders during this period. }

\subsection{Data}\label{subsec:dataset_information}

My dataset consists of two parts. The first part records information about the firms and ships owned by each firm. The dataset is based mainly on ``The list of shipping tonnage of ocean-going ship industry \textit{"(``Gaikousen-Unkougyoubetsu-Senpuku-Kouseihyou," \citeyear{listgaikou}}, in Japanese), published biannually by the Japanese Shipowners' Association (JSA). The data source provides ship-level characteristics, which consist of contract type and carrier types as follows. 

First, a ship's contract type can fall into three different categories -- (i) time charters, (ii) bareboat charters, and (iii) owned ships. A time charter is a contractual agreement between a shipowner and a charterer, where the charterer hires the vessel for a specified period of time and pays a fixed rate for each day the vessel is under charter. A bareboat charter, also known as a demise charter, is a contract in which the charterer takes full control and responsibility for the vessel, including crewing and maintenance. These categorizations are aggregated because contract types are not important from the viewpoint of subsidy requirements.

Second, according to the classifications of the \cite{syuuyaku}, the carrier type of a ship is categorized into four types.\footnote{According to \cite{listgaikou}, the carrier type is categorized into nine detailed types as follows: (1) cargo, (2) ore carrier, (3) oil tanker, (4) bauxite carrier, (5) nickel carrier, (6) lumber carrier, (7) LPG carrier, (8) passenger cargo, and (9) refrigerated transportation. For example, Ship A is classified as a pair (time charter, oil tanker) and Ship B is classified as a pair (bareboat charter, lumber carrier). However, because there are many missing values in the classification information in \cite{listgaikou}, I follow the categories in \cite{syuuyaku}.} The first carrier type is a cargo liner, which is operated regularly to carry cargo according to a schedule and route set in advance. The second type is a special vessel that specializes in transporting a particular type of cargo such as timber or iron ore. This special vessel is in service with an optimal structure and equipment for this purpose and works exclusively on certain routes. The third is a tramper, which is irregularly operated without a specific route, depending on the availability of cargo. The fourth is a tanker, which is equipped with a tank in its belly to carry oil or other liquids.

Carrier types, such as the special vessel and tanker, determine exclusively what types of goods the owner firms can carry. For example, only tankers can carry oil, whereas only some special vessels can carry bauxite. Therefore, the carrier type represents the degree of shipping ability specialization. For each carrier type, I collect data regarding the firm's key characteristics measured in dead-weight (D/W) tonnage tons.\footnote{Although the data about shipyard id, ship class, power (in horsepower), speed (in knots), age, and main shipping route are recorded, I focus on the key independent variables because the present model considers firm-level decisions based on the firm's tonnage. Technically, the ship class records the inspector company's name, which is similar to car inspection but does not imply a vertical quality measure. The main shipping route records specific destinations such as country or ocean, but is not consistent within the data, and some ships are defined as irregular tramper ships, so I omit this information.} I filled in the missing values for some small firms by referring to \cite{nostalgic}.\footnote{I confirmed that the information about the main ships in \cite{nostalgic} is precise and consistent with my raw data, although he collects a consistent dataset privately. The owner of the website kindly allowed me to refer to the dataset.} The present paper focuses on ocean-going ships and related subsidies; therefore, information about non-international, coastal trading vessels operated by small domestic coastal trading firms is omitted.

The second data source is ``Documents related to the consolidation of shipping companies" (\textit{``Kaiun Kigyou Syuyaku Kankei Shiryo," \citeyear{syuuyaku}}, in Japanese), published by the \cite{syuuyaku}, which includes  information about firm type. Firms are of three firm types: (i) a main firm that has the right of capital control over the group, (ii) an affiliate firm that receives capital controlled led through the merged company, and (iii) a wholly controlled firm that has a close relationship with the merged company or its affiliates and is on a long-term charter or that delegates the operation of a vessel to the merged company or its affiliates. The firm types are important for satisfying subsidy requirements, which is explained in Section \ref{sec:model}.\footnote{For preliminary analysis in Appendix \ref{subsec:preliminary_regressions}, we use ``The report of shipowner's management" (\textit{``Keiei no jissou"}, \citeyear{jissou}, in Japanese), published by the JSA, which includes basic financial data. For the analysis of coalitional mergers, the information about the main banks of each firm is potentially important to explain the incentives for coalitional matchings. Finally, to create Figure \ref{fg:shippingtonnage}, I use world shipping data published by the Ministry of Transport Shipping Bureau, Lloyd statistics, and \cite{nostalgic}'s complementary auxiliary dataset, which is partially common with recent papers \citep{kalouptsidi2014aer,kalouptsidi2017res,jeon2017learning,barwick2019china}. I refer to \cite{book3} and \cite{book4} for institutional background knowledge and detailed histories of consolidations of my interest.} 

This paper focuses on firm-level mergers and group-level subsidy criteria. Therefore, I use firm-level total tonnage for each carrier type in the main analysis after aggregating ship-level total tonnage. I manually construct a dataset combining these data sources from the Japan Maritime Center library. Note that this paper focuses on the static matching model in a single large market. Thus, I mainly use the data in 1963 and 1964. I use data from other years only to check the accuracy and consistency of the main pre- and post-merger data.

\subsection{Reconstruction Maintenance Law and Interest Subsidy Law}\label{subsec:subsidy_laws_details}

In this subsection, I explain the institutional background of shipping mergers and consolidations and the details of Japan's policy according to \cite{chida1990japanese}, Sections 4 and 5. See Table \ref{tb:industry_history} in Appendix \ref{sec:institutional_background} for reference of major events related to my sample.

The story of the modern Japanese shipping industry dates back to the postwar period. After World War II, the Japanese government implemented a planned shipbuilding policy (\textit{``Keikaku Zousen"} in Japanese) to compensate for major losses of domestic ships during the war. Many entrants chose to enter the shipping and shipbuilding market after Japan participated in the global ocean-going shipping market. Japanese shipping firms enjoyed the fruit of high demand for shipping and shipbuilding during the Korean War from 1950--1953 and the Suez Crisis in 1956, but they depended heavily on planned shipbuilding and government loans with a low interest rate. As a result, firms did not make an effort to catch up to technological innovation and reduce their operational costs, and their financial condition worsened in the 1950s.\footnote{The detailed history is as follows. The shipping industry recovered to its prewar level as early as 1957 in terms of ship tonnage  by implementing planned shipbuilding, but most funding for construction still depended on loans. Although the government had launched policies such as the interest subsidy system and improved the financing conditions of the fiscal fund for planned shipbuilding, these policies alone were insufficient to achieve satisfactory results. Moreover, since this interest subsidy was suspended due to the temporarily high profits of the Suez boom period, the difference in international competitiveness quickly led to a deterioration in performance. After 1956, when the economy recovered after the war, the shipping capacity required by the Japanese economy continued to increase rapidly even during the shipping recession after the summer of 1957, due to the rapid growth of the heavy chemical industry.
In December 1960, the Cabinet approved a plan to double the shipping industry's national income, which called for the construction of 13.35 million gross tons of ocean-going shipping capacity by the target year of 1970. This tonnage was determined to be the minimum amount necessary to bring Japan's international balance of payments into the black by \$200 million in the target year. }

To change the chronically inefficient industrial structure, the Japanese government planned to implement the so-called ``Two Laws for the Reconstruction of the Shipping Industry" (\textit{``Kaiun Saiken Niho"} in Japanese) in 1960 as one of the industrial policy interventions in the Income-Doubling Plan announced nationally that year.\footnote{\cite{suzumura2012excess} explained the institutional details of the relationship between industrial policy interventions and the excess entry theorem of \cite{suzumura1987res} in major industries in postwar Japan.} The laws, which were proclaimed and enforced on July 1, 1963, are divided into two parts.\footnote{For reference, \cite{chida1990japanese}, Section 5, stated:

\textit{``the main points of the new legislation can be summed up under three headings: 
\begin{enumerate}
    \item There should be mergers between the major operators to form new companies, each owning at least half a million d.w.t. The new firms, called ``nucleus companies" (\textit{Chukaku-tai}), would then be expected to act as a kind of ``parent" company within the new groupings.
    \item Each nucleus company would include other operators and owners within its group so that its total capacity was over one million tons d.w.t.
    \item Government assistance of varying types:
    \begin{itemize}[(a)]
        \item The provision of capital on favorable terms for building vessels under the Programmed Shipbuilding Scheme.
        \item Favorable terms under the Interest Subsidy Scheme for new construction with capital raised from both public and city banks.
        \item The right to postpone over five years interest payments on outstanding past loans received from the Development Bank of Japan. The city banks were expected to adopt similar policies. The cash equivalent of these sums was to be used to reduce their capital debt with respective financial institutions."(p.141)
    \end{itemize}
\end{enumerate}} Note that the above tonnage is calculated based on ocean-going ships.
 }

The first part was the Reconstruction Maintenance Law, which addressed shipping industry's reconstruction (\textit{``Saikenseibi"} in Japanese) and was to take effect for 5 years starting in April 1964. The Reconstruction Maintenance Law set a goal and method for the rationalization of an industry. The objective of the law was to resolve the underdepreciation of domestic firms in 5 years. The government announced that firms were granted a 5-year moratorium period for interest expenses related to shipbuilding if the firms satisfied the following two requirements. The first requirement was that a merged group must own more than 0.5 million tons of shipping tonnage of an ocean-going ship after mergers. The second requirement was that the group whose central firm was the merger firm must own more than 1 million tons of shipping tonnage of an ocean-going ship, including that of affiliate and wholly controlled firms. 

The second part was the Interest Subsidy Law, which provided the interest subsidies. This was a revision of an earlier law that offered terms that were more favorable. The Interest Subsidy Law extended the term of interest subsidy for a development bank loan from 5 to 10 years and subsidized 1.0 \% out of 5.0 \% of its interest expenses. Furthermore, the law extended the term of interest subsidies for a city bank loan to 7 years and subsidized 1.5 \% out of 7.5 \% of its interest expenses. Hereafter, I refer to these two financial supports simply as ``subsidies" for the sake of simplicity.

To receive subsidies under the law, qualified merger firms submitted a merger plan or extension petition by December 31, 1963. As a result, by May 31, 1964, 44 firms had submitted and completed merger plans, and 53 firms participated in grouping without submitting a plan. This paper treats these firms as merging and merged firms because the operation of the latter 53 firms' ships was shared within the group, so that merged firms and group-joining firms become indistinguishable.\footnote{I noticed that the number of firms participating in mergers is slightly different across papers and data sources because some firms retained their names after mergers. For example, \cite{chida1990japanese} showed that 88 ocean-going shipping firms formed coalitional groups while keeping the name of the buyer firm. Strictly speaking, the market consisted of 94 matched firms (six main buyer firms, six main seller firms, 27 affiliated firms, and 55 wholly controlled firms) and 31 unmatched firms after dropping seven missing unmatched firms. This is consistent with my data, as shown in Table \ref{tb:total_tonnage_size_group_table}.}

In this study, I interpret the amount of two types of subsidies with calculated real interest payments for the total cost of shipbuilding within 10 years for each firm as a single constant dummy variable indicating whether the group qualified for the subsidy.\footnote{Alternative specifications of the subsidy are shown in Appendix \ref{sec:monte_carlo}.} Note that the firm's decision about building ships is dynamic \citep{kalouptsidi2014aer, jeon2017learning, barwick2019china} and merger decisions must be affected by the cost-side incentives of the expectation of future cost savings and subsidies. However, based on the institutional background of the law, it is reasonable to construct a static matching model with complete information.

\subsection{Assortativeness of size and technological specialization as proxies for economies of scale and scope.}\label{subsec:assortativeness_of_size_and_specialization}

The anecdotal main drivers of mergers in the shipping industry are economies of scale and scope \citep{notteboom2004container}. Theoretically, economies of scale and scope are defined as cost efficiencies. Economies of scale exist when the average production cost of a single product decreases with the number of units produced, whereas economies of scope are cost-saving externalities between product lines. Analogously, a merger between two firms will generate economies of scale if, post-merger, the merged firm can provide the same quality at a lower cost or higher quality at the same cost. This implies that even if a merger decision is modeled as a single agent problem, the decision is based on the expectation of post-merger cost efficiencies in all possible future paths. In a matching environment like my data, the dynamic merger consideration is more complicated. In addition, measuring post-merger cost efficiencies in itself is challenging.\footnote{As far as I know, \cite{jeziorski2014estimation} is a notable and exceptional study quantifying post-merger cost efficiencies with assumptions on a firm's sequential dynamic acquisition and repositioning decisions. His model assumes that merger execution costs are zero, aside from the merger transfer price based on the institutional background, so that he can estimate a cost-efficiency parameter. My model takes a different approach to explicitly quantify merger execution costs. From the different approach focusing on indivisibility in shipping, \cite{holmes2018indivisibilities} investigated the relationship between consolidation and scale economies in distribution.}

Instead, I focus on how much the observed proxy variables of post-merger size and technological specialization affect merger incentives as actual pre-merger firms would expect before the enactment of the subsidy laws. Specifically, quantifying the assortativeness of the proxy variables in a single matching market is the main interest of this paper.

I specify the proxy variables as follows. The proxy variable for economies of scale is measured by firm-level ship tonnage for each carrier type. Since shipping firms gain profits from transportation and the shipping industry is modeled as a homogeneous good or a competitive market, tonnage size reflects a potential scale effect that buyer firms expect at the time of their merger decision. The proxy variable for economies of scope is measured by the firm-level ship tonnage share of each carrier type. This proxy captures firm-level pre-merger technological specialization. If a firm with a low level of specialization in liners merges with another firm with a low level of specialization in liners, the merged group may gain cost efficiency through technological diversification across carrier types because of technology spillover. Similarly, if a firm with a high level of specialization in tankers merges another firm with a high level of specialization in tankers, the merged group may also achieve cost efficiency through technological specialization. Thus, the technological share information is useful for pre-merger firms to predict potential (dis-)economies of scope. In Section \ref{sec:estimation}, I evaluate which carrier types contribute to merger incentives.

\subsection{Descriptive statistics}\label{subsec:descriptive_statistics}

\begin{figure}[!ht]
\begin{center}
\includegraphics[height = 0.45\textheight]{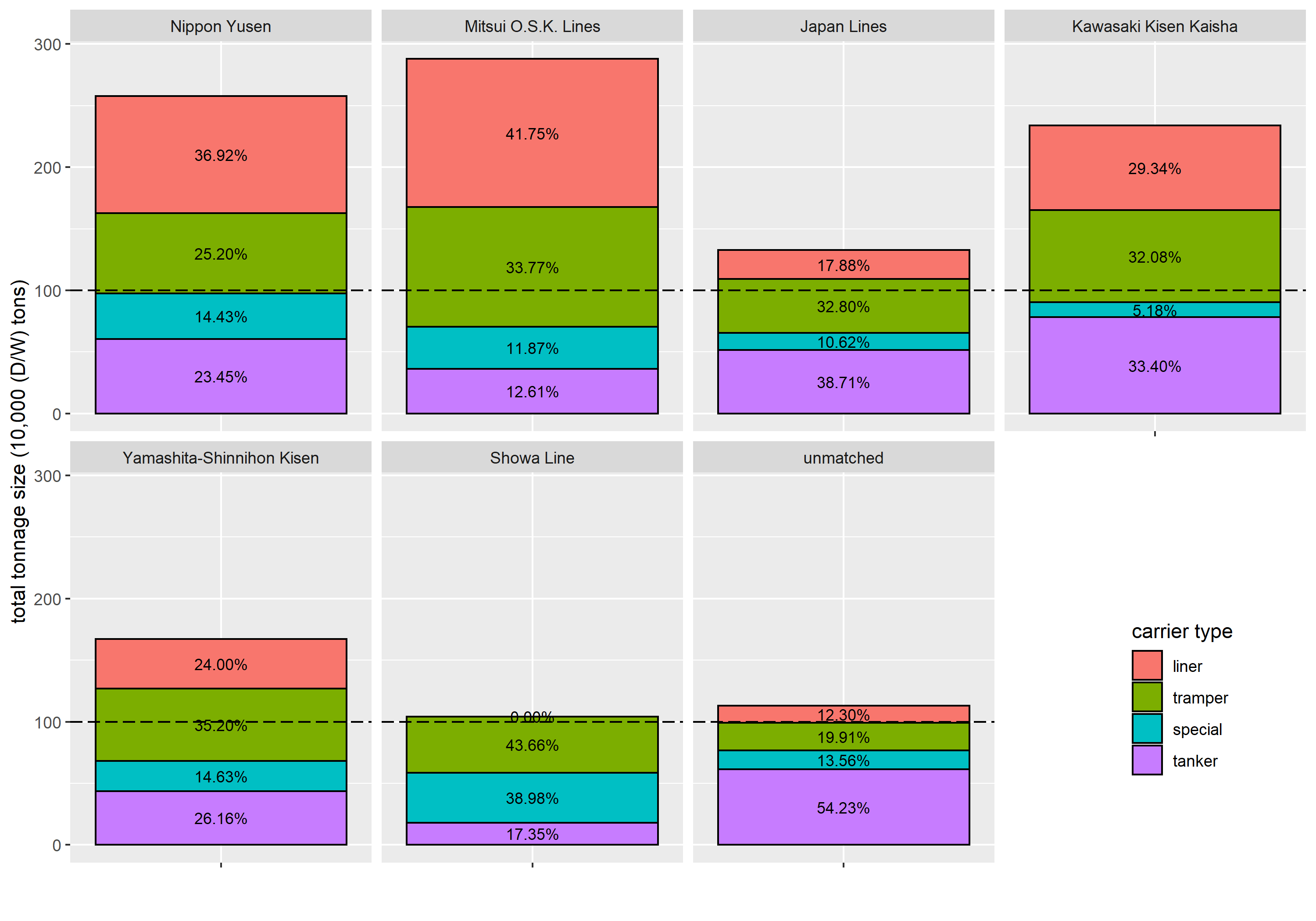}
\end{center}
\caption{Configuration of tonnage for each carrier type.}
\footnotesize
\begin{tablenotes}
\item[a]\textit{Note:} \textit{Source} : \cite{listgaikou} and \cite{nostalgic}. Observation units: Total group-level tonnage for each carrier type after mergers. The dotted horizontal line indicates the subsidy threshold (1 million (D/W) tons).
\end{tablenotes}
\label{fg:carrier_composition_eachgroup}
\end{figure}

To understand the industry-level background, I begin by discussing group-level information. Table \ref{tb:total_tonnage_size_group_table} and Figure \ref{fg:carrier_composition_eachgroup} summarize the final configuration of mergers and consolidations.\footnote{I found that Table \ref{tb:total_tonnage_size_group_table}, which was calculated from \cite{syuuyaku} and \cite{nostalgic} and the corresponding final merger configuration shown in \cite{chida1990japanese} are somewhat different in terms of unknown aggregation, different recorded dates, and some missing firms. For consistency, I continue to use individual firm data collected from the \cite{syuuyaku} and \cite{listgaikou} to compute aggregate measures.} These indicate that all six groups satisfy the subsidy criteria and that unmatched firms have a larger amount of total tonnage than the Showa Line group, which would imply the existence of fixed merger costs. Figure \ref{fg:carrier_composition_eachgroup} also shows post-merger evidence of assortativeness of size and specialization for some groups. For example, Nippon Yusen and Mitsui-O.S.K Lines have balanced portfolios, whereas Showa Line specializes in special shipping. However, it is hard to disentangle the relative importance of post-merger size effects and technological specialization from group-level information.

\begin{table}[!ht]
\caption{Summary of the total tonnage size for each group.}
\begin{center}
\begin{tabular}{llrr}
\toprule
\multicolumn{1}{l}{}&\multicolumn{1}{c}{Firm type}&\multicolumn{1}{c}{Total tonnage}&\multicolumn{1}{c}{Num of firms}\tabularnewline
\midrule
{\itshape Nippon Yusen}&&&\tabularnewline
~~&(1) Main&1.51&2\tabularnewline
~~&(2) Affiliate&0.84&7\tabularnewline
~~&(3) Wholly controlled&0.23&7\tabularnewline
~~&Total&2.58&16\tabularnewline
\midrule
{\itshape Mitsui OSK Line}&&&\tabularnewline
~~&(1) Main&1.92&2\tabularnewline
~~&(2) Affiliate&0.4&5\tabularnewline
~~&(3) Wholly controlled&0.55&20\tabularnewline
~~&Total&2.87&27\tabularnewline
\midrule
{\itshape Japan Line}&&&\tabularnewline
~~&(1) Main&1.12&2\tabularnewline
~~&(2) Affiliate&0.13&1\tabularnewline
~~&(3) Wholly controlled&0.08&4\tabularnewline
~~&Total&1.33&7\tabularnewline
\midrule
{\itshape Kawasaki Kisen Kaisha}&&&\tabularnewline
~~&(1) Main&1.66&2\tabularnewline
~~&(2) Affiliate&0.39&9\tabularnewline
~~&(3) Wholly controlled&0.29&7\tabularnewline
~~&Total&2.34&18\tabularnewline
\midrule
{\itshape Yamashita Shinnihon Kisen}&&&\tabularnewline
~~&(1) Main&0.9&2\tabularnewline
~~&(2) Affiliate&0.6&5\tabularnewline
~~&(3) Wholly controlled&0.17&10\tabularnewline
~~&Total&1.67&17\tabularnewline
\midrule
{\itshape Showa Line}&&&\tabularnewline
~~&(1) Main&0.55&2\tabularnewline
~~&(2) Affiliate&0.46&3\tabularnewline
~~&(3) Wholly controlled&0.03&4\tabularnewline
~~&Total&1.04&9\tabularnewline
\midrule
{\itshape Unmatched}&&&\tabularnewline
~~&Unmatched&1.13&24\tabularnewline
\midrule
{\itshape Total}&&&\tabularnewline
~~&&12.96&118\tabularnewline
\bottomrule
\end{tabular}

\end{center}
\footnotesize
\begin{tablenotes}
\item[a]\textit{Note:} \textit{Source} : \cite{listgaikou} and \cite{nostalgic}. After dropping missing ten firms, the data contain matched 94 firms (12 main firms,  30 affiliates, and 52 wholly controlled firms) and unmatched 24 firms chose to stay unmatched. Total tonnage is measured by million (D/W) tons.
\end{tablenotes}
\label{tb:total_tonnage_size_group_table}
\end{table}

\begin{table}[!ht]
\caption{Summary statistics for independent variables. }
\label{tb:shipping_market_stats_table}
\begin{center}
\begin{tabular}{lrrrrrrrr}
\toprule
\multicolumn{1}{l}{\itshape }&\multicolumn{1}{r}{\itshape N}&\multicolumn{1}{r}{\itshape mean}&\multicolumn{1}{r}{\itshape sd}&\multicolumn{1}{r}{\itshape min}&\multicolumn{1}{r}{\itshape q25}&\multicolumn{1}{r}{\itshape q50}&\multicolumn{1}{r}{\itshape q75}&\multicolumn{1}{r}{\itshape max}\tabularnewline
\midrule
{\itshape Size variables (million (D/W) tons)}&&&&&&&&\tabularnewline
~~total tonnage size&$118$&0.110&0.207&0.000&0.011&0.026&0.091&1.023\tabularnewline
~~total tonnage size of liner&$118$&0.031&0.112&0.000&0.000&0.000&0.000&0.721\tabularnewline
~~total tonnage size of special&$118$&0.015&0.033&0.000&0.000&0.000&0.008&0.177\tabularnewline
~~total tonnage size of tanker&$118$&0.030&0.071&0.000&0.000&0.000&0.024&0.417\tabularnewline
~~total tonnage size of tramper&$118$&0.035&0.055&0.000&0.002&0.013&0.034&0.246\tabularnewline
\midrule
{\itshape Specialization variables (\% share)}&&&&&&&&\tabularnewline
~~share of liner&$118$&0.104&0.235&0.000&0.000&0.000&0.000&1.000\tabularnewline
~~share of special&$118$&0.117&0.244&0.000&0.000&0.000&0.088&1.000\tabularnewline
~~share of tanker&$118$&0.192&0.351&0.000&0.000&0.000&0.208&1.000\tabularnewline
~~share of tramper&$118$&0.587&0.417&0.000&0.135&0.705&1.000&1.000\tabularnewline
~~HHI based on carrier types&$118$&0.815&0.241&0.258&0.584&1.000&1.000&1.000\tabularnewline
\bottomrule
\end{tabular}

\end{center}
\footnotesize
\begin{tablenotes}
\item[a]\textit{Note:} \textit{Source} : \cite{listgaikou} and \cite{nostalgic}. Note that the fact that minimum of some size variable equal to zero implies that the tonnage size is less than 1,000 (D/W) tons. Total tonnage is measured by (D/W) tons.
\end{tablenotes}
\end{table}

To investigate firm-level incentives, I focus on firm-level information. Table \ref{tb:shipping_market_stats_table} shows the summary statistics for the independent variables from all 118 samples. Note that I observe the names of 128 firms but cannot collect variables of 10 firms, so I drop these samples. First, the tonnage of almost all the firms is less than 1 million (D/W) tons, so almost all pre-merger firms have an incentive to be matched with other firms to satisfy the subsidy criteria. Second, the Herfindahl-Hirschman Index (HHI) based on the relative composition of carrier types for each firm is 0.815 on average.\footnote{The Herfindahl-Hirschman Index (HHI) for firm $i$ is calculated by $HHI_i=\sum_{k\in K}(s_{ik})^2.$ where $s_{ik}=\frac{\text{tonnage}_{ik}}{\sum_{k\in K}\text{tonnage}_{ik}}$ and $K = \{Liner,Special,Tramper,Tanker \}$. The index captures the concentration level as a single index measure. This is defined formally in Section \ref{subsec:reduced_form}. } This implies that each firm specialized in one or two shipping categories. For example, the medians of the firm-level tonnage of liners, special vessels, and tankers are zero, which means that more than half of all firms did not have these types of ships. This evidence supports the idea that merger incentives might depend on some measure of post-merger specialization.

\section{Model}\label{sec:model}
\subsection{Setup}

I construct a one-sided one-to-many matching model with complementarities by extending \cite{azevedo2018existence} and \cite{fox2019externality}.
I consider a single large market before and after a subsidy. For convenience, I denote the former pre-merger market by $t=1$ and the latter, post-merger market by $t=2$. Firms $i=1,\cdots,N$ exist at $t=1$. For notational convenience, I use type $i$ firm in a continuum market and individual firm $i$ in a finite market interchangeably.\footnote{The formal discussions are postponed in Section \ref{subsec:formal_definition_of_AH} after introducing the definitions and assumptions.} Let $\mathcal{N}$ be the set of firms at $t=1$. Let $\mathcal{J}$ be a set of all possible coalitions, that is, $\mathcal{J}=\mathcal{P}(\mathcal{N})$ which is the power set of $(\mathcal{N})$. Let $\widehat{J}\in\mathcal{J}$ be a set of realized coalitions. Let $x_i$ be a vector of continuous or discrete observable characteristics for each firm $i$ at $t=1$. 
Suppose that firm $i$ acquires coalition $J\in\mathcal{J}$. Let a finite set of rival's coalitions be denoted as $\widehat{J}_{-J}$ where $J\cup \widehat{J}_{-J}=\widehat{J}$. I specify a vector $X$ of observables in post-merger valuation in the market at $t=2$ as a function of pre-merger observables $x=(x_1,\cdots,x_N)$, 
\begin{align*}
    &X((x_j)_{j\in J}, \widehat{J}_{-J}),
\end{align*}
and $X_i((x_j)_{j\in J}, \widehat{J}_{-J})$ with subscript $i$ indicates observable variables related to firm $i$. This implies that the post-merger valuation is determined by the own coalition and all rival's coalitions.

For simplicity, I assume that the ``main" firm $i$ in coalition $J$ is an acquirer. The ``main" firm, introduced in Section \ref{sec:data_background}, has the right of capital control over the group. I specify the post-merger valuation for coalition $J$ with acquirer $i$ conditional on realized coalitions $\widehat{J}$ as a linear form as follows:
\begin{align*}
F_{i,J}((x_j)_{j\in J}, \widehat{J}_{-J})+\epsilon_{i,J}=X_i((x_j)_{j\in J}, \widehat{J}_{-J})'\theta +\epsilon_{i,J},
\end{align*}
where $F_{i,J}((x_j)_{j\in J}, \widehat{J}_{-J})$ is the production function from coalition matching $(i,J)$, and  $\epsilon_{i,J}$ is an unobservable valuation shock to acquirer $i$ for coalition $J$. Note that $(x_j)_{j\in J}$ includes observed characteristics of firm $i$ belonging to coalition $J$. The post-merger valuation is interpreted as a payoff that is shared with firm $i$ and the members of coalition $J$.

Post-merger profit for coalition $J$ with acquirer $i$ is defined as  follows:
\begin{align*}
F_{i,J}((x_j)_{j\in J}, \widehat{J}_{-J}) +\epsilon_{i,J}- \sum_{j\in J\backslash \{i\}}p_j,
\end{align*}
where $p_j$ is the price for target firm $j\in J\backslash \{i\}$. For target firm $j$, the equilibrium acquisition price $p_j = p_{ij}$ equals the shadow price on the constraint that target firm $j$ may be acquired by at most one acquirer.\footnote{The appendix of \cite{akkus2015ms} considers a similar approach for merger equilibrium price $p_j$. The equilibrium price can be modeled as a bargaining outcome of buyer firm $i$ and target firms within coalition $J$. Since this modeling requires merger price information for both realized and non-realized mergers, this article does not take this approach.} Acquirers demand coalition $J$ to maximize the post-merger profits. To satisfy the sufficient conditions of \cite{fox2018qe} for (set-)identification and rank order property, we assume the following theoretical assumption. 

\begin{assumption}
    \label{as:regular}
    (1) The observations $(i,J)$ are independent and identically distributed (i.i.d.). (2) The term $\epsilon_{i,J}$ has full support in $\mathbb{R}^{\text{dim}(\epsilon_{i,J})}$. (3) The term $\epsilon_{i,J}$ has an exchangeable distribution for each type $i$.\footnote{Denote $\epsilon=\{\epsilon_{i,J}\}_{i\in\mathcal{N},J\in\mathcal{J}}$. Let $\rho$ be a permutation of the elements of $\epsilon.$ An exchangeable distribution satisfies $F\left(\epsilon\mid i, J\right)=F\left(\rho\left(\epsilon\right) \mid i,J\right)$ for all such permutations $\rho .$ This is equivalent to Assumption 2 (iii) in \cite{fox2018qe} when the dimension of unobservable type, $k=1$.} (4) The parameter space of $\theta$ is compact. (5) The cumulative distribution function of unobservable valuation shock conditional of acquirer $i$ for coalition $J$, $F\left(\epsilon_{i,J}|i,J\right)$ has bounded, continuous derivatives.
\end{assumption}

The equilibrium concept of \cite{fox2019externality} is a competitive equilibrium where a price vector with no excess demand for merger target firms exists with externalities. Their full model incorporates strategic interactions across coalitions and allows multiplicity of equilibria. Instead, I impose the following assumption to exclude the possibility of strategic
decisions.

\begin{assumption}
    \label{as:without_externality}
    The post-merger valuation for coalition $J\in\mathcal{J}$ depends on observed characteristics $(x_j)_{j\in J}$ and unobserved characteristics $\epsilon_{i,J}$, but not on characteristics of rival's coalition $\widehat{J}_{-J}$.
\end{assumption}

Assumption \ref{as:without_externality} excludes the possibility of strategic decisions of all firms to participate in coalitional groups, like a static entry game. Although \cite{fox2019externality} and \cite{uetake2019entry} incorporated the strategic decisions into the TU and NTU matching models in different ways, the externality issues are beyond the state of the methodological literature.\footnote{As discussed in \cite{agarwal2021market}, the condition of the existence of the equilibrium with externality is an active area of theoretical research. The conditions of the uniqueness of the equilibrium are not investigated. Also, a global market for each carrier type is characterized differently so that incorporating post-merger competition in the model is difficult in a unified way. For example, tramper bulk carriers operate like taxi drivers, whereas container carriers operate like bus drivers employed by large firms and groups.} Assumption \ref{as:without_externality} allows me to drop $\widehat{J}_{-J}$ and reformulate the post-merger profit into the following:
\begin{align*}
F_{i,J}((x_j)_{j\in J}) +\epsilon_{i,J}- \sum_{j\in J\backslash \{i\}}p_j=X_i((x_j)_{j\in J})'\theta +\epsilon_{i,J}- \sum_{j\in J\backslash \{i\}}p_j,
\end{align*}
where production function $F_{i,J}$ depends only on the pre-merger characteristics of firm $i$ and coalition $J$. Additionally, Assumption \ref{as:without_externality} allows me to implement tractable counterfactual simulations in Section \ref{sec:counter_factual}.

\subsection{Equilibrium Concept}\label{subsec:formal_definition_of_AH}

For convenience, I introduce the formal definition of competitive equilibrium with a continuum of agents according to Section 6 of \cite{azevedo2018existence}. An allocation is a measurable map $A: \mathcal{N} \rightarrow \Delta(\mathcal{P}(\Omega))$ specifying for each type $i\in \mathcal{N}$ a distribution $A_i$ over the power set $\mathcal{P}(\cdot)$ of trade bundles $\Omega$. It allows type $i$ agent to be a single firm. Let $\eta_i$ be a measure of type $i$ firm. The element of $\Omega$ is called a merger bundle in my setting. Let $\mathcal{A}$ be the space of allocations. A price vector $p \in \mathbb{R}^{\Omega}$ assigns a price $p_{\omega}$ for each merger bundle $\omega \in \Omega$. Given a vector of prices $p$,
define expenditure as the vector $e_{p} \in \mathbb{R}^{\mathcal{P}(\Omega) \times \mathcal{P}(\Omega)}$ such that for $\Phi \subseteq \Omega$ and $\Psi \subseteq \Omega,$
$$
e_{p}(\Phi, \Psi)=\sum_{\varphi \in \Phi} p_{\varphi}-\sum_{\psi \in \Psi} p_{\psi},
$$
where $\Phi$ (resp. $\Psi$) is the set of merger bundles in which agents participate as a buyer (resp., a seller).
That is, $e_{p}(\Phi, \Psi)$ is the net transfer paid by an agent buying $\Phi$ and selling $\Psi.$ The
utility of type $i$ agent who buys the set of merger bundles $\Phi \subseteq \Omega$ and sells the set of merger bundles $\Psi \subseteq \Omega$ at prices $p$ is
given as follows:
$$
u^{i}(\Phi, \Psi)-e_{p}(\Phi, \Psi),
$$
where $u^{i}(\Phi, \Psi)$ is the gain from buying the set of merger bundles $\Phi$ and selling the set of merger bundles $\Psi$.

Given allocation $A\in \mathcal{A}$, define the excess demand for each $i\in \mathcal{N}$ and for each merger bundle $\omega \in \Omega$ as follows:
\begin{align*}
    Z_{\omega}^{i}(A) \equiv \sum_{\{\omega\} \subseteq \Phi \subseteq \Omega} A^{i}(\Phi)-\sum_{\{\omega\} \subseteq \Psi \subseteq \Omega} A^{i}(\Phi),
\end{align*}

Define the excess demand for each merger bundle $\omega \in \Omega$
for the entire economy as
\begin{align*}
    Z_{\omega}(A) \equiv \int Z_{\omega}^{i}(A) \mathrm{d} \eta,
\end{align*} 
where $\eta$ is the distribution of the population of full type agents. 
I introduce competitive equilibrium as a solution concept.

\begin{definition}\label{def:competitive_equilibrium_original}
    An arrangement $[A; p]$ is a competitive equilibrium if it satisfies two conditions:
\begin{enumerate}
    \item Incentive Compatibility: Each agent obtains an optimal bundle given prices $p$ for all $i \in \mathcal{N}, A^{i}(\Phi, \Psi)>0$ only if
    \begin{align*}
        (\Phi, \Psi) \in \underset{(\tilde{\Phi}, \tilde{\Psi}) \in \mathcal{P}(\Omega) \times \mathcal{P}(\Omega)}{\arg \max } u^{i}(\tilde{\Phi}, \tilde{\Psi})-e_{p}(\tilde{\Phi}, \tilde{\Psi}).
    \end{align*}
    \item Feasibility: $Z_{\omega}(A)=0$ for all $\omega \in \Omega$.
\end{enumerate}
\end{definition}

\cite{azevedo2018existence} prove the existence of unique competitive
equilibrium in a regular economy with a continuum of agents with a finite set of trades.\footnote{An economy is regular if (1) The integral of absolute values of utility is finite, as long as the agents are not given bundles
for which they have a sufficiently negative utility. That is,
$$
\int_{I} \max _{\Phi, \Psi \subseteq \Omega, u^{i}(\Phi, \Psi) \neq-\infty}\left|u^{i}(\Phi, \Psi)\right| \mathrm{d} \eta<\infty
$$ and (2) Agents can supply any sufficiently small net demand for trades. That is,
$\left\{\mathrm{Y} \in \mathbb{R}^{\Omega}: \exists A \in \mathcal{A}\right.$ such that $\mathrm{Y}=\mathrm{Z}(A)$ and $\left.(u^{i}(\Phi, \Psi)=-\infty \text{, then ,} A^{i}(\Phi, \Psi)=0)\right\}$
contains the neighborhood of 0. The competitive equilibrium is unique if the distribution of full types $\eta$ ensures the uniqueness of the maximizer of the social welfare function.} Also, every competitive equilibrium is efficient. A regular economy where measure $\eta$ has full support has a unique vector of competitive equilibrium prices. \cite{fox2017specifying} discusses the difference in the data between a truly finite market and a continuum market. Following his reasoning, I assume that the observed single large matching market in the data is a finite subset of a continuum of firms and target firm bundles.\footnote{\cite{fox2017specifying} clearly states that working with a continuum of agents is a way of stating an approximation while ensuring the existence of the equilibrium. The assumption on a continuum market is imposed such that the aggregate
matching outcome is a function of the distribution of $\epsilon_{i,J}$ for all $i$ and $J$ but not specific realizations of $\epsilon_{i,J}$. Thus, the market outcome is econometrically deterministic in the aggregate. This is a conceptual assumption for interpreting the matching market(s) in the data.}

These theoretical results allow us to find competitive equilibrium conceptually by discrete assignment linear programming formulation introduced by \cite{shapley1971assignment}. However, the competitive equilibrium above allows each agent to be a seller and buyer over any possible trades in a many-to-many matching environment so that the dimension of $|\Omega| \times |\Omega|$ goes to the size of universe for even small $N$. To overcome the numerical impossibility while applying the fundamental theorems of \cite{azevedo2018existence} to my one-sided one-to-many coalitional matching model, I assume Assumption \ref{as:one_sided}.

\begin{assumption}
    \label{as:one_sided}
    $\mathcal{J}=\mathcal{P}(\mathcal{N}) = \Phi = \Psi$, i.e., the both sets of trades are summarized as the power set of firms. 
\end{assumption}

Assumption \ref{as:one_sided} means that the sets of trades as a seller and buyer are indistinguishable and summarized as the set of possible coalitions $\mathcal{J}$. Also, it explicitly specifies the set of trades $\Omega$ as a set of all possible merger combinations, not arbitrary trades. Assumption \ref{as:one_sided} is innocuous in practical merger situations. The assumption allows us to represent the set of merger bundle $\Omega$ as a list of $N$-dimensional binary digits in the power set $\mathcal{P}(\mathcal{N})$. Also, the power set of $\Omega$ denoted by $\mathcal{P}(\Omega)$ is reduced to the set with a tractable size of dimensions. To understand how the model works, I introduce a simple toy model.

\textbf{Example 1:} Suppose $N=3$, i.e., three firms exist in a market. Each firm can be a buyer and a seller. Then, each firm faces the set of merger bundles described as
\begin{align*}
    \mathcal{P}(\Omega)=\{(000),(100),(010),(001),(110),(101),(011),(111)\},
\end{align*}
where the element $(ijk)$ captures whether firm $i$,$j$, and $k$ are included in the merger bundle. Concretely, merger bundle $(000)$ means a null merger bundle where no firm participates in the merger (i.e., all firms stay unmatched), merger bundle $(100)$ means the merger bundle where only firm 1 participates in the merger, and merger bundle $(110)$ means the merger bundle where firm 1 and 2 participate in the merger. Then, the dimension of $\mathcal{P}(\Omega)$ for each firm is $|\mathcal{P}(\mathcal{N})|=2^N=8$. 

Under the assumptions, I define a competitive equilibrium in the coalitional merger matching game as a simple version of Definition \ref{def:competitive_equilibrium_original}.

\begin{definition}
    \label{def:competitive_equilibrium}
    Under Assumptions \ref{as:regular}, \ref{as:without_externality}, and \ref{as:one_sided}, an allocation is a measurable map $A: \mathcal{N} \rightarrow \Delta(\mathcal{J})$ specifying for each type $i\in \mathcal{N}$ a distribution $A_i$ over the power set of merger bundles $\mathcal{J}$, denoted by $\{A_{i,J}\}_{J\in \mathcal{J}}$. Suppose that the observed matching market comes from a continuum of matching market. I describe the full distribution as $A=\{A_{i,J}\}_{i\in \mathcal{N},J \in \mathcal{J}}$. A merger configuration $[A; p]$ is a competitive equilibrium if it satisfies two conditions:
\begin{enumerate}
    \item Incentive Compatibility: Each firm obtains an optimal merger bundle $J$ given prices $p$ for all $i \in \mathcal{N}, A_{i,J}>0$ only if, for all $i \in \mathcal{N}$, 
    \begin{align*}
        J \in \underset{J' \in \mathcal{J}}{\arg \max }\quad  F_{i,J'}((x_j)_{j\in J'}) +\epsilon_{i,J'}- \sum_{j\in J'\backslash \{i\}}p_j.
    \end{align*}
    \item Feasibility: $Z_{J}(A)=0$ for all $J \in \mathcal{J}$.
\end{enumerate}
\end{definition}

The competitive equilibrium allocation can be computed via the following constrained linear programming problem.
\begin{align}
    \max_{\{A_{i,J}\}_{i\in N,J \in \mathcal{J}}}& \sum_{i\in \mathcal{N}}\sum_{J \in \mathcal{J}} A_{i,J} U_{i,J}\label{eq:LP}\\
    \text{s.t. } U_{i,J}&=F_{i,J}((x_j)_{j\in J}) +\epsilon_{i,J}\nonumber,\\
    A_{i,J} &\in \mathbb{R} \quad \forall i\in \mathcal{N},J \in \mathcal{J}\nonumber,\\
    0\le A_{i,J} & \le \eta_i \quad \forall i\in \mathcal{N},J \in \mathcal{J}\nonumber,\\
    \sum_{J\in \mathcal{J}} A_{i,J} &= \eta_i \quad\forall i\in \mathcal{N}\nonumber,\\
    Z_{J}(A)&=0\quad \forall J \in \mathcal{J}\nonumber,
\end{align}
where the objective function is called a social welfare function, $\eta_i$ is the total population measure over type $i$ firm, and allocation $A_{i,J}$ is allowed to be real-valued. In my empirical application, I assume $\eta_i=1$ for all $i\in \mathcal{N}$ because I interpret the type as firm's identity.\footnote{\cite{azevedo2018existence} use a finite number of trades and a continuum of full agent types. In my empirical application, the number of observable types of firm $i\in\mathcal{N}$ is also finite. I interpret that in mapping the continuum model to the finite data, there is a continuum of firms of  Nippon Yusen's observable type, but I observe only one such firm in the data.} Note that the term $\sum_{j\in J\backslash \{i\}}p_j$ is cancelled out to calculate joint production function $U_{i,J}$ for firm $i$ and coalition $J$.  Except the constraints $Z_{J}(A)=0$, the problem is the natural extension of the classical assignment problem of \cite{shapley1971assignment}. 

\subsection{Specification of feasibility constraint $Z_J(A)=0$}

To compute competitive equilibrium, I need to explicitly specify the feasibility constraints $Z_J(A)=0$ for all $J\in \mathcal{J}$. Recall that the merger bundle $J\in \mathcal{J}$ identifies participants in the merger bundle $J$ by Assumption \ref{as:one_sided}, so I can assign the roles such as buyers and sellers in a merger market to all participants. 
For assignment $A_{i,J}$ for all $i \in \mathcal{N},J\in \mathcal{J}$, type $i$ firm is a seller if merger bundle $J\in \mathcal{J}$ includes only $i$, i.e., only $i$-th element of $J$ is one and is zero otherwise. I denote the set of merger bundles by $J_i^s$ as type $i$ seller firm's bundle. On the other hand, type $i$ firm is a buyer if merger bundle $J\in \mathcal{J}$ includes at least $j$-th ($j\neq i$) element of $J$, where the $j$-th element is one. I denote the set of merger bundles by $J_i^b$ as type $i$ buyer firm's bundle. For assignment $A_{i,J}$ where $J$ consists of only zeros, type $i$ firm chooses to stay autarky. These specifications of assignment $A$ capture all possible merger configurations that are mutually exclusive. \par
Under the specifications, the feasibility constraints $Z_J(A)=0$ for all $J \in \mathcal{J}$ in Definition \ref{def:competitive_equilibrium} are specified as
\begin{align*}
    A_{i,J_i^s}&=\sum_{i'\neq i, i'\in \mathcal{N}} \sum_{J' \in J_{i'}^b}A_{i', J'}\cdot \mathbbm{1}(i\text{-th element of } J' \text{ is } 1), \forall i \in \mathcal{N},
\end{align*}
where $\mathbbm{1}(\cdot)$ is the indicator function which equals to 1 if $(\cdot)$ is satisfied and 0 otherwise, the left-hand side can be interpreted as a supply side for merger bundle $J$, and the right-hand side can be interpreted as a demand side for merger bundle $J$. Example 2 illustrates the simple version of $Z_J(A)=0$.

\textbf{Example 2:} As Example 1, suppose $N=3$, i.e., three types of firms exist in a market. Each type of firm can be a buyer and seller. Then, each type of firm faces the set of merger bundles described as
\begin{align*}
    \mathcal{J}=\{(000),(100),(010),(001),(110),(101),(011),(111)\},
\end{align*}
where the element $(ijk)$ captures whether type $i$, $j$, and $k$ firms are included in the merger bundle. Then, assignment $\{A_{i,J}\}_{i\in \mathcal{N},J \in \mathcal{J}}$ is expressed as $3\times 2^3$-dimensional matrix. For example, $A_{1,(000)}$ means that type 1 firm joins null merger bundle $(000)$, i.e., the firm stays autarky. $A_{1,(100)}$ means that type 1 firm joins merger bundle $(100)$, i.e., type 1 firm sells type 1 firm himself. $A_{1,(011)}$ means that type 1 firm joins merger bundle $(011)$, i.e., type 1 firm buys type 2 and 3 firms.\footnote{Some merger bundles seem unreal for some firms. For example, suppose type 1 firm joins merger bundle $(101)$ which means that the type 1 firm sells the type 1 firm and acquires a type 3 firm. Such an unreal case can be eliminated innocuously by assigning sufficiently large negative values to these assignments. Here, I allow the possibility for completeness of the model description.} Note that the type of firm can be interpreted as an individual firm, but the expression allows for non-integer and probabilistic allocation on $A$. Then, all feasibility constraints $Z_J(A)=0$ for all $J\in \mathcal{J}$ can be described as 
\begin{align*}
    A_{1,(100)}&=A_{2,(100)}+A_{2,(110)}+A_{2,(101)}+A_{2,(111)}+A_{3,(100)}+A_{3,(110)}+A_{3,(101)}+A_{3,(111)}\\
    A_{2,(010)}&=A_{1,(010)}+A_{1,(110)}+A_{1,(011)}+A_{1,(111)}+A_{3,(010)}+A_{3,(110)}+A_{3,(011)}+A_{3,(111)}\\
    A_{3,(001)}&=A_{1,(001)}+A_{1,(101)}+A_{1,(011)}+A_{1,(111)}+A_{2,(001)}+A_{2,(101)}+A_{2,(011)}+A_{2,(111)},
\end{align*}
where left-hand sides of the equalities capture the supply side for each firm and right-hand sides capture the corresponding demand side. 

Under the specification of $Z_J(A)=0$, I can calculate competitive equilibrium numerically by solving (\ref{eq:LP}) via linear programming under equilibrium constraints.

\subsection{Specification of production function $F_{i,J}$}\label{subsec:specification_production_function}

The specification issues about production function $F_{i,J}((x_j)_{j\in J})$, merger cost $C_{i,J}$, and subsidy $s_{i,J}$ remain. These specifications are highly dependent on the data and the researcher's interest, although the specification of $Z_J(A)=0$ is a common simple trick on any one-sided one-to-many matching market. In Sections \ref{subsec:specification_production_function}, \ref{subsec:specification_merger_cost}, and \ref{subsec:specification_subsidy}, I specify $F_{i,J}((x_j)_{j\in J})$, $C_{i,J}$, and $s_{i,J}$ to reflect the matching environment in my data.

To make it consistent with the estimation steps, I specify $F_{i,J}((x_j)_{j\in J})$ as
\begin{align}
    F_{i,J}((x_j)_{j\in J})&=X_i((x_j)_{j\in J})'\theta\nonumber\\
    &=\beta (x_i\circ x_{J}) + \delta s_{i,J} - C_{i,J}(\gamma), \label{eq:utility_specification}
\end{align}
where $X_i((x_j)_{j\in J})'\theta$ is a linear single index of interactions of observed characteristics of firm $i$ and all coalition members $j\in J$ with composite parameter $\theta=[\beta,\delta,\gamma]$ which is a vector of parameters for $X_i((x_j)_{j\in J})$ and subsidy and cost terms, $s_{i,J}$ is the subsidy term, and $C_{i,J}(\gamma)$ is the merger cost term explicitly dependent on parameter $\gamma$ in a flexible way. Following the literature, I incorporate the match-specific interaction term $(x_i \circ x_{J})$ via element-wise product of observed characteristics of firm $i$ and coalition $J$ and the dimension of $\beta$ corresponds to that of the interaction term. The vector of observed characteristics $x_i$ includes firm $i$'s size and share of each carrier type. The vector of $x_{J}$ includes the corresponding observed characteristics of coalition $J$ which summarizes $x_i$ for all $i\in J$.\footnote{In this specification, I can interpret that the production function captures some reduced-form features about ex-post competition under the transportation demand of downstream firms. The post-merger size effects and technological specialization affect quantity competition dynamically in not only the domestic but also the global shipping market, as in \cite{jeon2017learning}. These points are beyond the scope of this article.} If firm $i$ stays unmatched, I assume that $F_{i,\emptyset}((x_j)_{j\in \emptyset})=\beta (x_i\circ x_i)$ which means that the payoff of being unmatched equals to that of matching with a hypothetical firm which has the same composition of carriers.\footnote{In the model, an opt-out action corresponds to the action of selling firms to some buyer and exiting the market, not an unmatched action that incurs operating costs. So, normalizing an unmatched payoff to zero is inappropriate.} If firm $i$ sells itself to some buyer firm, I assume that $F_{i,\{i\}}((x_j)_{j\in \{i\}})=0$ as the payoff of an outside option.

\subsection{Specification of merger cost $C_{i,J}(\gamma)$}\label{subsec:specification_merger_cost}

Both $s_{i,J}$ and $C_{i,J}(\gamma)$ are important terms for eliminating the possibility of a grand coalition (i.e., in a grand coalition, all firms form a single coalitional group). First, as explained in Section \ref{sec:data_background}, no merger occurred before the laws. This implies that if $s_{i,J}=0$ for all $i$ and $J$, no merger occurred. Since I observe a single large matching market before and after mergers, the subsidy term is critical to tilt the playing field before and after the enactment of subsidy laws.

Second, the estimation of fixed costs such as merger costs is critical for policy implications. However, researchers do not observe the proxy variables directly.\footnote{The literature on a static and dynamic entry game provides the point-estimate or bounds of fixed costs regarding a firm's entry decision in the market. The main difference between merger costs in a matching merger market setting and fixed costs in an entry game setting is that merger costs must be determined by a comparison of the payoffs of being unmatched, selling the firm, and buying a single or multiple firms through feasibility condition $Z_J(A)=0$.} In particular, few papers estimate endogenous merger costs. \cite{igami2019mergers} treat the merger cost as a single common constant parameter independent of any observed characteristics, which is incurred only when a firm chooses a merger decision. In the shipping industry allowing a one-to-many matching, this specification is not adequate because the cost of acquiring a single firm is expected to be smaller than that of acquiring many firms.

To reflect the above features, I specify $C_{i,J}(\gamma)$, a merger cost that depends on the number of target firms as follows:
\begin{align*}
    C_{i,J}(\gamma)=\begin{cases}
    0 \quad \quad \quad \quad \quad \quad \quad \quad \quad \quad \quad \quad\quad \quad \quad \quad \text{ if } J\in \{\emptyset,i\},\\
    \gamma \cdot [\text{the number of firms in coalition } J\backslash i ] \quad \text{ otherwise },
    \end{cases}
\end{align*}
where $\gamma$ is a cost parameter that captures merger costs for adding one firm to the coalition $J$, and firm $i$ can remain as an autarky with no merger costs. The case of $J\in \{\emptyset,i\}$ means that firm $i$ is an autarky or selling itself. This specification imposes the condition that $C_{i,J}(\gamma)$ is linearly increasing in the total number of acquired firms in the coalition. The linearity of $C_{i,J}(\gamma)$ might be inadequate because the merger costs of acquiring two firms might not be approximately twice the cost of acquiring a single firm. However, this is unknown to researchers. Thus, I follow and extend the linearly additive separable merger cost of acquiring a single firm in \cite{gowrisankaran1999dynamic}.

\subsection{Specification of subsidy $s_{i,J}$}\label{subsec:specification_subsidy}

The institutional background helps us to specify the subsidy term $s_{i,J}$ explicitly.\footnote{The model has the same spirit of \cite{kawai2011wp} and \cite{weese2015qe}. \cite{kawai2011wp} considered the policy implications of designing procurement auctions, that may yield social gains by inducing investment by firms. He provided empirical evidence on how different subsidy mechanisms affected the incentives of agents to invest using data from plastic recycling auctions in Japan. The main difference between \cite{kawai2011wp} and the present paper is the subsidy requirement. His paper specifies the incentive of firms to invest individually. Conversely, my paper specifies the incentive of acquirers to match targets considering total tonnage.} The subsidy plan depends on the total shipping tonnage of an ocean-going ship owned by coalition $J$ so that $s_{i,J}$ is specified as
\begin{align*}
s_{i,J} = \begin{cases}
M_{i,J} \quad\quad\quad\mbox{ if } [\sum_{i \in J} \text{tonnage}_i >1.0 \text{(million tons)}] \\
0 \quad\quad\quad\mbox{ otherwise }
\end{cases},
\end{align*}
where $M_{i,J}$ denotes the amount of subsidy and the subsidy threshold level (1 million tons) is determined by the government. The key difference from the standard one-to-one matching model is that positive amounts in the joint production function are shared by all members of coalition $J$ including $i$.\footnote{If I reflect the real subsidy criteria, $s_{i,J}$ is written as
\begin{align*}
s_{i,J} = \begin{cases}
M_{i,J} \mbox{ if } [\sum_{i \in J} \text{tonnage}_i >1.000.000] \wedge [\sum_{i \in J, i \text{ is the main firm}} \text{tonnage}_i >500.000] \\
0 \mbox{ otherwise }
\end{cases}.
\end{align*} However, under Assumption \ref{as:main_firm_is_a_buyer}, only 12 main firms can become buyers and Table \ref{tb:total_tonnage_size_group_table} implies that the second subsidy condition is satisfied for almost all merger scenarios if the first subsidy condition is satisfied. Thus, I take the simplification of $s_{i,J}$ for tractability.} Appendix \ref{sec:monte_carlo} shows the alternative subsidy specifications in simulated data.

Importantly, the subsidy amount $M_{i,J}$ was not fixed for all firms because $M_{i,J}$ was dependent on future loans for each group through interest subsidy.\footnote{The details of the subsidy amounts are as follows. The period of the interest subsidy is extended from 5 to 10 years for commercial bank loans and from 5 to 7 years for municipal loans, and the interest rate of the interest subsidy is decreased from 5 years for commercial bank loans to 4 years for shipowners.} Under rational expectation assumption, all firms anticipated all future paths of all possible matchings and investment games, which is infeasible to incorporate in the model. Also, under the data limitation about $M_{i,J}$ for all possible matching pairs $(i,J)$, I assume Assumption \ref{as:subsidy}.

\begin{assumption}
\label{as:subsidy}
Subsidy amount $M_{i,J}$ is fixed, observed by all firms ex-ante, and specified as
$$
M_{i,J}=M,
$$ where $M$ is a positive constant. 
\end{assumption}
Note that, in general, $M_{i,J}$ could be flexibly dependent on the observed characteristics such as the capital size of all possible matching pairs $(i,J)$ if the calculation formula and data were available. However, the subsidy in my data is the interest subsidy involving the future investment cost and benefits. Imposing the ex-ante dependency of the subsidy amount on the observed characteristics will not be innocuous, and the current model could not estimate potential correlation of the unknown subsidy amount with the size of the group. Instead, I focus on the qualification of the subsidy under Assumption \ref{as:subsidy} which is restrictive but the most innocuous. Practically, I assume that $M=1$ so that the subsidy effect term in specification \eqref{eq:utility_specification} is measured by $\delta$.\footnote{Another specification of the subsidy term is shown in Appendix \ref{sec:monte_carlo}. 
} Under Assumption \ref{as:subsidy}, we interpret $\delta$, the coefficient of $M_{i,J}$, as the sensitivity to subsidy qualification. By shifting the level of $M$ from one given a fixed $\delta$, I can simulate the counterfactual scenarios. 

Under Assumptions \ref{as:regular}, \ref{as:without_externality}, \ref{as:one_sided}, and \ref{as:subsidy}, I can characterize the competitive equilibrium fully and compute it numerically.

\section{Estimation}\label{sec:estimation}

\subsection{Matching maximum rank estimator}
The estimation procedure is based on \cite{fox2018qe} and \cite{fox2019externality}. I have the data in a single, large market before and after a subsidy, so I need to use a matching maximum rank estimator instead of a matching maximum score estimator. In this approach, pairwise stability matchings construct inequalities of incentive compatibility conditions for obtaining consistent estimators, even though researchers do not have transfer data.
Recall that incentive compatibility for a type $i$ firm given prices $p$ is described as follows:
\begin{align*}
        J \in \underset{J' \in \mathcal{J}}{\arg \max }\quad  F_{i,J'}((x_j)_{j\in J'}) +\epsilon_{i,J'}- \sum_{j\in J'\backslash \{i\}}p_j.
\end{align*}

I can construct the following five types of inequalities corresponding to incentive compatibility as follows:
\begin{enumerate}
    \item inequalities from two observed coalitions;
    \item inequalities from one observed coalition;
    \item inequalities from one unmatched target;
    \item inequalities from for the individual rationality (IR) conditions of unmatched agents; and
    \item Inequalities from for the IR conditions with and without subsidies.
\end{enumerate}
The last two inequalities are derived from the special features of my data. I explain how the inequalities are constructed in Appendix \ref{subsec:construct_inequalities}. \footnote{To hold these inequalities true, I introduce Assumption \ref{as:rank_order} known as the rank order conditions \citep{fox2010qe}.} 

I estimate and simulate using all or a sub-sample of the data containing at most 118 firms. If all firms can be either buyers, sellers, or unmatched firms, I need to construct inequalities for all possible combinations of $\mathcal{N}=\{1,\cdots,118\}$ and $|\mathcal{J}|=2^{118}$, which is infeasible to compute an equilibrium even once. To make the model tractable in the estimation step, I additionally impose Assumption \ref{as:main_firm_is_a_buyer} for eliminating unreasonable merger scenarios.

\begin{assumption}
    \label{as:main_firm_is_a_buyer}
    Only the main firms defined in Section \ref{sec:data_background} are eligible to be buyer firms. In other words, affiliated or wholly controlled firms that arise after the merger cannot act as buyer firms.
\end{assumption}

Note that I require Assumption \ref{as:main_firm_is_a_buyer} only for estimation with the full sample. Although Assumption \ref{as:main_firm_is_a_buyer} seems natural since small and medium-sized firms could not acquire much larger and many firms in general due to financial constraints, Assumption \ref{as:main_firm_is_a_buyer} eliminates the possibility that many small and medium-sized firms could form a single coalitional group with tonnage that would be comparable to that of the six groups. Assumption \ref{as:main_firm_is_a_buyer} also ignores the possibility of the profitable deviation of affiliated and wholly controlled firms from belonging to a coalition to joining the other coalitions or staying unmatched. 

Based on these assumptions, I construct the  objective function as follows: 
\begin{align}
    Q(\theta)=\sum_{g \in G_N} 1\left[Z_{g}^{\prime} \theta \geq 0\right],\label{eq:obj}
    \end{align}
where $G_N$ is the set of observed inequalities and $Z_{g}^{\prime} \theta$ is the corresponding inequality of matching pairs $g$. The set of maximizers of $Q(\theta)$ is a vector of point-estimated parameters.

\subsection{Identification, consistency, and inference}\label{subsec:identification_consistency_and_inference}

The present paper follows \cite{fox2018qe} and \cite{fox2013aej} who both applied the subsampling approach to the matching maximum score and rank estimator, based on the literature developed by \cite{manski1975maximum}, \cite{han1987non}, and \cite{kim1990cube}.\footnote{The alternative approach is a set identification method. The subsampling method developed by \cite{romano2008inference} is used in a single-agent maximum score and rank estimator \citep{bajari2008evaluating} and a many-to-many matching maximum score estimator \citep{fox2018qe}.}

I repeat the notion of pairwise stability in matches only from \cite{fox2010qe} and \cite{fox2013aej}. I assume that the econometrician observes some finite number of recorded firms from a continuum matching market, as in \cite{azevedo2018existence}. This means that I introduce the fiction that the real-life matching market with $N$ firms is a subset of some very large matching market. I formally assume a rank order property without price data as in \cite{fox2018qe}:
\begin{assumption}
    \label{as:rank_order}
    Suppose that type $i$ firm acquires firms $J_i$ in the data, including firm $h\in J_i$, whereas type $l$ firm acquires firms $J_l$ in the data, including firm $k\in J_l$. Define swapped coalitions as $\overline{J_{i}}=J_{i} \cup\{k\}-\{h\},\overline{J_{l}}=J_{l} \cup\{h\}-\{k\},$. Then, 
\begin{align}
    \text{Pr}[\text{observed matching}|X]&\geq \text{Pr}[\text{swapped counterfactual matching}|X],\nonumber
\end{align}
if and only if 
\begin{align*}
X_i\left(\left(x_{j}\right)_{j \in J_{i}} \right)^{\prime} \theta+X_l\left(\left(x_{j}\right)_{j \in J_{l}} \right)^{\prime}\theta &\geq X_i\left(\left(x_{j}\right)_{j \in \overline{J_{i}}} \right)^{\prime} \theta+X_l\left(\left(x_{j}\right)_{j \in \bar{J}_{l}} \right)^{\prime}\theta. 
\end{align*}
\end{assumption}

Assumption \ref{as:rank_order} says that if the exchange of a subset of merger members produces a lower sum of deterministic valuations,
then the frequency of observed matching pairs with the same characteristics as the exchange of a subset of merger members must be lower than that of observed matching pairs with
characteristics that yield higher valuations.\footnote{As shown in Appendix \ref{subsec:construct_inequalities}, Assumption \ref{as:rank_order} corresponds to inequality \eqref{ineq1}, and the corresponding rank order conditions with inequalities \eqref{ineq2}, \eqref{ineq3}, \eqref{ineq4}, and \eqref{ineq5} hold analogously.}  

Given additional assumptions on the support of $\beta$ and $x,$ \cite{fox2010qe} shows that the set of all maximizers of $Q(\theta)$ are identified as true parameters and the matching maximum score estimator is consistent. Continuous support also ensures that the maximum is unique. \cite{sherman1993limiting} demonstrated that the maximum rank correlation estimator proposed by \cite{han1987non} is $\sqrt{N}$ -consistent and asymptotically normal if the parameter is point identified. The asymptotic variance matrix in \cite{sherman1993limiting} is complex to implement in that it requires additional nonparametric estimates of the components that appear in the variance matrix. To avoid this complexity, \cite{fox2013aej} proposed a subsampling approach \citep{politis1994large}, which is consistent under weak conditions. 

For the point estimate, I use all possible pairwise inequalities to construct the objective function $Q(\theta)$ and numerically find a maximizer $\hat{\theta}$. For inference, I compute the confidence intervals via bootstrap. An advantage of bootstrap over subsampling is that researchers do not need to tune the subsampled size. \cite{subbotin2007asymptotic} showed that a nonparametric bootstrap is valid for inference under point-identification.\footnote{I confirm that both bootstrap and subsampling approaches with one-third of the full sample generated similar confidence intervals with the same signs using my data, although this is not a theoretical finding.} To construct a meaningful 95 \% confidence interval, I calibrate subsidy sensitivity $\delta$ because Appendix \ref{sec:monte_carlo} finds that a dummy variable like the subsidy term can fail its identification because the identification set may be unbounded on one side.\footnote{As shown in Appendix \ref{subsec:construct_inequalities}, since parameter $\delta$ mainly evaluates inequality \eqref{ineq5} before and after the subsidy period, the value of calibrated $\delta$ can take a positive infinity to achieve inequality \eqref{ineq5}. So, as the most insensitive scenario to the subsidy, I calibrate $\delta$ to the lower bound, achieving the best fitting, i.e., the most correct prediction of pairwise inequalities. For the government, this scenario is the most expensive and worrisome case.}  Thus, for the main estimations, I estimate nine parameters: coefficient parameter $\beta$ for eight-dimensional observed characteristics and merger cost parameter $\gamma$. The estimation details are provided in the footnote of the estimation result tables.

\section{Results}\label{sec:results}
\subsection{Estimation results on two observed characteristics}\label{subsec:estimation_two_variables}

Table \ref{tb:score_results_two_variables} reports the estimation results with two observed characteristics and merger cost.
\footnote{In Appendix \ref{subsec:estimation_HHI_only}, I investigate an alternative simple specification of aggregate-level technological specialization (i.e., carrier-type HHIs of buyer firms and coalitions in Table \ref{tb:shipping_market_stats_table}). The results are consistent with the results in Table \ref{tb:score_results_two_variables}.} The numbers in parentheses are 95 \% confidence intervals via bootstrap. Estimation details are summarized in the footnote of Table \ref{tb:score_results_two_variables}. I normalized $\beta_0$ to $+1$ for identification. I estimate the other parameters separately under the normalization and pick the vector with the highest number of satisfied inequalities. The number of inequalities for the full sample is $17,864$. As found in Section \ref{subsec:identification_monte_carlo}, the identification of the upper bound of $\delta$ is difficult. To deal with this problem, I find in the preliminary point estimation that $\delta=20$ achieves the maximum score as the integer lower bound of $\delta$ so I calibrate $\delta=20$ for all specifications. Thus, I assume the worst case regarding $\delta$ from the perspective of a government. 

Now, I will discuss the estimated signs in Table \ref{tb:score_results_two_variables}. First, all specifications report the robust result that the estimated coefficient of merger cost $\gamma$ is positive, which means that a merger incurs significant costs. Second, in Columns 1 through 4, the point-estimated coefficients of assortativeness of scale for $\beta_1$ to  $\beta_4$ are positive. This implies that assortativeness of scale exists. In Columns 5 through 8, the estimated coefficients of assortativeness of specialization are negative for $\beta_5$ to $\beta_8$. This means that technological diversification on carrier types contributes to merger incentives. The 95 \% confidence intervals of these parameters indicate that the signs are robust.

\begin{landscape}
{
\begin{table}[!htbp] \centering 
  \caption{Results of matching maximum rank estimator with two observed variables}
  \label{tb:score_results_two_variables} 
  \begin{tabular}{lccccccccc}
\toprule 
 &  & (1) & (2) & (3) & (4) & (5) & (6) & (7) & (8) \\
 &  & Point Est & Point & Point & Point & Point & Point & Point & Point \\
 &  & [95\% CI] & [95\% CI] & [95\% CI] & [95\% CI] & [95\% CI] & [95\% CI] & [95\% CI] & [95\% CI] \\
\midrule 
Scale variables &  &  &  &  &  &  &  \\
total$_{b}$ $\times$ total$_{t}$ & $\beta_0$ & +1 & +1 & +1 & +1 & +1 & +1 & +1 & +1 \\
 &  & (S) & (S) & (S) & (S) & (S) & (S) & (S) & (S) \\
liner$_{b}$ $\times$ liner$_{t}$ & $\beta_1$ & 93.61 &  &  &  &  &  &  &  \\
 &  & [41.7, 294.7] &  &  &  &  &  &  &  \\
tramper$_{b}$ $\times$ tramper$_{t}$ & $\beta_2$ &  & 34.71 &  &  &  &  &  &  \\
 &  &  & [13.5, 295.0] &  &  &  &  &  &  \\
special$_{b}$ $\times$ special$_{t}$ & $\beta_3$ &  &  & 227.29 &  &  &  &  &  \\
 &  &  &  & [39.9, 297.2] &  &  &  &  &  \\
tanker$_{b}$ $\times$ tanker$_{t}$ & $\beta_4$ &  &  &  & 34.71 &  &  &  &  \\
 &  &  &  &  & [10.0, 289.6] &  &  &  &  \\
Share variables &  &  &  &  &  &  &  &  &  \\
liner$_{b}$ $\times$ liner$_{t}$ & $\beta_5$ &  &  &  &  & -90.68 &  &  &  \\
 &  &  &  &  &  & [-280.6, -41.1] &  &  &  \\
tramper$_{b}$ $\times$ tramper$_{t}$ & $\beta_6$ &  &  &  &  &  & -285.49 &  &  \\
 &  &  &  &  &  &  & [-296.0, -91.5] &  &  \\
special$_{b}$ $\times$ special$_{t}$ & $\beta_7$ &  &  &  &  &  &  & -261.56 &  \\
 &  &  &  &  &  &  &  & [-296.7, -86.6] &  \\
tanker$_{b}$ $\times$ tanker$_{t}$ & $\beta_8$ &  &  &  &  &  &  &  & -240.9 \\
 &  &  &  &  &  &  &  &  & [-294.5, -125.8] \\
 &  &  &  &  &  &  &  &  &  \\
 &  &  &  &  &  &  &  &  &  \\
merger cost & -$\gamma$ & -1.24 & -0.94 & -2.05 & -0.94 & -1.35 & -0.84 & -1.14 & -0.89 \\
 &  & [-2.5, -0.9] & [-2.2, -0.7] & [-3.0, -0.9] & [-3.0, -0.8] & [-2.7, -0.6] & [-12.0, -0.5] & [-9.0, -0.3] & [-5.7, -0.2] \\
subsidy sensitivity & $\delta$ & 20 & 20 & 20 & 20 & 20 & 20 & 20 & 20 \\
 &  &  &  &  &  &  &  &  &  \\
\hline 
$\sharp$ Inequalities (Point) &  & 17864 & 17864 & 17864 & 17864 & 17864 & 17864 & 17864 & 17864 \\
\% Inequalities &  & 0.8226 & 0.798 & 0.8465 & 0.8032 & 0.9126 & 0.9102 & 0.9267 & 0.9079 \\
\bottomrule 
\end{tabular}

   \footnotesize
   \begin{tablenotes}
\item[a]\textit{Note:} The objective function was numerically maximized using differential evolution (DE) algorithm in \texttt{BlackBoxOptim.jl} package. For DE algorithm, I require setting the domain of parameters and the number of population seeds so that I fix the former to $[-20, 10]$ for $\gamma$ and $[-300,300]$ for other parameters, and the latter to 100. For point estimation, 100 runs were performed for all specifications. The reported point estimates are the best-found maxima. The parentheses are 95 \% confidence intervals which are computed by taking the 2.5th percentile
and the 97.5th percentile of the empirical sampling distribution of estimated parameters. Bootstrap uses 200 replications with 200 population seeds and 118 firms (all 12 main firms and 106 non-main firms sampled with replacement out of 106 non-main firms) per replication. Confidence intervals are not necessarily symmetric around the point estimate. Parameters that can take on only a finite number of values (here +1) converge at an arbitrarily fast rate, then they are superconsistent (denoted by (S)). The unit of measure of assortativeness of scale is a million (D/W) tonnage tons. The sub-indices $b$ and $t$ mean the buyer's and target coalition's covariates. 
   \end{tablenotes}
\end{table} 
}\end{landscape}

\begin{table}[!htbp] \centering 
  \caption{Comparable results of matching maximum rank estimator in Table \ref{tb:score_results_two_variables}} 
  \label{tb:ratio_score_results_two_variables} 
  \begin{tabular}{@{\extracolsep{5pt}}lcccccccc}
\toprule 
$m$ & $x_0$ & $\beta_m$ & $\bar{x}_m$ & $1SD(x_m)$ & $-\gamma$ & (1) $V(\bar{x}_m)$ & (2) $V(\bar{x}_m+1SD(x_m))$ & (2) - (1) \\
\midrule 
1 & 0.11 & 93.61 & 0.031 & 0.112 & -1.24 & -1.138 & 0.686 & 1.824 \\
 &  &  &  &  &  &  &  &  \\
2 & 0.11 & 34.71 & 0.015 & 0.033 & -0.94 & -0.92 & -0.848 & 0.072 \\
 &  &  &  &  &  &  &  &  \\
3 & 0.11 & 227.29 & 0.03 & 0.071 & -2.05 & -1.833 & 0.281 & 2.114 \\
 &  &  &  &  &  &  &  &  \\
4 & 0.11 & 34.71 & 0.035 & 0.055 & -0.94 & -0.885 & -0.647 & 0.238 \\
 &  &  &  &  &  &  &  &  \\
 &  &  &  &  &  &  &  &  \\
5 & 0.11 & -90.68 & 0.104 & 0.235 & -1.35 & -2.319 & -11.759 & -9.44 \\
 &  &  &  &  &  &  &  &  \\
6 & 0.11 & -285.49 & 0.117 & 0.244 & -0.84 & -4.736 & -38.033 & -33.297 \\
 &  &  &  &  &  &  &  &  \\
7 & 0.11 & -261.56 & 0.192 & 0.351 & -1.14 & -10.77 & -78.249 & -67.479 \\
 &  &  &  &  &  &  &  &  \\
8 & 0.11 & -240.9 & 0.587 & 0.417 & -0.89 & -83.885 & -243.709 & -159.824 \\
 &  &  &  &  &  &  &  &  \\
\bottomrule 
\end{tabular}

\footnotesize
\begin{tablenotes}
\item[a]\textit{Note:} $V(\bar{x}_m)=x_0 x_0 + \beta_m \bar{x}_m \bar{x}_m-\gamma$. The parameters are shown in Table \ref{tb:score_results_two_variables}. Models $m=1,\cdots,8$ corresond with columns in Table \ref{tb:score_results_two_variables}. I denote the mean and one standard deviation of observed characterisitics $x_m$ as $\bar{x}_m$ and $1SD(x_m)$ shown in Table \ref{tb:shipping_market_stats_table}.
\end{tablenotes}
\end{table}

To obtain the intuition of the relative importance of each variable, Table \ref{tb:ratio_score_results_two_variables} reports the value of the joint production function for each specification with the corresponding units of observed characteristics. To pin down the situation, suppose that there is a single firm that has a mean observed characteristic. Then, the value of acquiring a single firm with the same characteristic is shown in Column 6 as a benchmark. Column 7 shows the hypothetical value when their observed characteristics change by one standard deviation.\footnote{The matching outcomes under different specifications are not comparable because the competitive equilibrium in the market is determined by these joint production functions for all possible matching configurations under equilibrium conditions, as shown in equation \eqref{eq:LP}. Recall that the value of the joint production function of staying unmatched that is dependent on the specification varies, whereas the value of selling itself is invariant and normalized to zero. Thus, direct comparison of the effect of the parameters on matching outcomes is inadequate in my estimation. See Appendix \ref{subsec:estimation_multivariate} for details of the further comparison.} 

As comparable results, Column 8 summarizes the changes in the values. This illustrates that the size of a liner is 30 times more important than the size of a tramper. In addition, the share of a tramper is twice as important as the share of a liner. Although it would not be adequate to compare the results based on size and share because of crucial differences in the units of the measures, I would interpret that the size of a liner is 5.5 times more important than the share of a liner. In Appendix \ref{subsec:estimation_multivariate}, I investigate this comparison further.

\subsection{Estimation results on two observed characteristics for 12 main firms}\label{subsec:estimation_two_variables_main_firms_only}

Finally, I report the estimation results on two observed characteristics by focusing only on 12 main firms, each of which can be either a buyer, a seller, or stay unmatched. This is the most flexible way to model matching behaviors instead of using a small subsample. The 12 main firms hold about 60 \% of the total tonnage in the industry, and are the main merger decision makers. Ideally, one would solve a matching equilibrium for the whole sample with $N=118$. Computational constraints, however, preclude this approach. Instead, I show that the estimation procedure is not only applicable to market(s) in which large firms compete even in a single market but also implementable for counterfactual analysis in a tractable size market as shown in Section \ref{sec:counter_factual}. I find the lower bound of $\delta$ in the preliminary point estimation and fix $\delta$ to the value as a calibrated parameter ($\delta=400$).\footnote{The value is much greater than the previous value in Table \ref{tb:score_results_two_variables}. This is because the subsample includes only matched groups that consist of much fewer firms, and the interaction terms of the explanatory variables are much less than the full sample. Thus, to explain the difference between pre-subsidy and post-subsidy matching outcomes, the subsidy term will play a more important role. Robustness checks are provided in Appendix \ref{sec:robustness_check}.} The goodness of fit in Table \ref{tb:score_results_two_variables_main_firms_only} is similar to the results presented in Table \ref{tb:score_results_two_variables}.

Table \ref{tb:score_results_two_variables_main_firms_only} reports the maximum rank estimation results for the subsample. Since I use 12 main firms in this section, I could not construct a 95 \% confidence interval with enough subsampled data. Instead, the numbers in brackets indicate the lower and upper bounds of a set of maximizers of the objective function. The estimated signs of parameters such as $\beta_1$ and $\beta_2$ are consistent with Table \ref{tb:score_results_two_variables}. On the other hand, as different points, other parameters such as $\beta_3$ and $\beta_4$ show negative signs, which implies that regarding special and tanker ships, the main firms consider the size of their tonnage relative to the total tonnage a disincentive to merge. Columns 5 to 8 show that all parameters regarding the assortativeness of specialization have positive signs, which implies that main firms aim for technological specialization. These results indicate that a few main firms specializing in special and tanker shipping want to match with each other, whereas many non-specialized main firms considered special and tanker tonnage as a disincentive to merger.\footnote{The results look consistent with some institutional facts. For example, the main firms in the Showa Line group specialized in special shipping. Furthermore, the main firms in the Japan Lines group and the Kawasaki Kisen Kaisha group specialized in tankers.}

Combining the results in Table \ref{tb:score_results_two_variables_main_firms_only} with those in Table \ref{tb:score_results_two_variables}, I find that the main firms improve the scale effects and specialization by merging main large firms while aiming for total diversification by merging small and middle-sized non-main firms. This captures well the actual post-merger configuration, in particular, regarding the special and tanker tonnage of small firms for each group.\footnote{See Figure \ref{fg:carrier_dist_eachgroup} in Appendix \ref{sec:institutional_background}.} The decomposition of merger incentives illustrates how and why mergers occur at each firm size level.

Finally, some parameters such as $\beta_1$, $\beta_4$, and $\beta_8$ are point-identified in the small subsample. This is a desirable feature for solving matching outcomes. In Section \ref{sec:counter_factual}, I simulate counterfactual analyses of 12 main firms based on the estimated parameters in Table \ref{tb:score_results_two_variables_main_firms_only} to answer several policy questions regarding the effects of subsidies on merger configurations and matching outcomes.

\begin{landscape}
{
\begin{table}[!htbp] \centering 
  \caption{Results of matching maximum rank estimator with two observed variables for 12 main firms. } 
  \label{tb:score_results_two_variables_main_firms_only} 
  \begin{tabular}{lccccccccc}
\toprule 
 &  & (1) & (2) & (3) & (4) & (5) & (6) & (7) & (8) \\
 &  & [LB, UB] & [LB, UB] & [LB, UB] & [LB, UB] & [LB, UB] & [LB, UB] & [LB, UB] & [LB, UB] \\
\midrule 
Scale variables &  &  &  &  &  &  &  \\
total$_{b}$ $\times$ total$_{t}$ & $\beta_0$ & +1 & +1 & +1 & +1 & +1 & +1 & +1 & +1 \\
 &  & (S) & (S) & (S) & (S) & (S) & (S) & (S) & (S) \\
liner$_{b}$ $\times$ liner$_{t}$ & $\beta_1$ & [124.0, 124.0] &  &  &  &  &  &  &  \\
tramper$_{b}$ $\times$ tramper$_{t}$ & $\beta_2$ &  & [54.1, 293.4] &  &  &  &  &  &  \\
special$_{b}$ $\times$ special$_{t}$ & $\beta_3$ &  &  & [-151.8, -42.6] &  &  &  &  &  \\
tanker$_{b}$ $\times$ tanker$_{t}$ & $\beta_4$ &  &  &  & [-220.8, -220.8] &  &  &  &  \\
Share variables &  &  &  &  &  &  &  &  &  \\
liner$_{b}$ $\times$ liner$_{t}$ & $\beta_5$ &  &  &  &  & [33.9, 37.4] &  &  &  \\
tramper$_{b}$ $\times$ tramper$_{t}$ & $\beta_6$ &  &  &  &  &  & [241.9, 279.9] &  &  \\
special$_{b}$ $\times$ special$_{t}$ & $\beta_7$ &  &  &  &  &  &  & [278.9, 292.6] &  \\
tanker$_{b}$ $\times$ tanker$_{t}$ & $\beta_8$ &  &  &  &  &  &  &  & [286.4, 286.4] \\
 &  &  &  &  &  &  &  &  &  \\
 &  &  &  &  &  &  &  &  &  \\
merger cost & -$\gamma$ & [-15.4, -15.4] & [-10.3, -5.2] & [-9.1, -1.8] & [-8.4, -8.4] & [-13.6, -3.5] & [-9.1, -3.8] & [-7.8, -5.2] & [-8.9, -8.9] \\
subsidy sensitivity & $\delta$ & 400 & 400 & 400 & 400 & 400 & 400 & 400 & 400 \\
 &  &  &  &  &  &  &  &  &  \\
\hline 
$\sharp$ Inequalities (Point) &  & 330 & 330 & 330 & 330 & 330 & 330 & 330 & 330 \\
\% Inequalities &  & 0.9636 & 0.9271 & 0.9394 & 0.9152 & 0.9756 & 0.9514 & 0.9635 & 0.9515 \\
\bottomrule 
\end{tabular}

\footnotesize
\begin{tablenotes}
\item[a]\textit{Note:} The numbers in parentheses are lower and upper bounds of a set of maximizers of the maximum rank estimator. The implementation details are the same as the footnote of Table \ref{tb:score_results_two_variables}.
\end{tablenotes}
\end{table} 
}
\end{landscape}

\section{Counterfactual simulations}\label{sec:counter_factual}

Could the government have stimulated endogenous merger incentives more efficiently? How would a merger configuration have changed in such possible scenarios? Given the estimated parameters in Section \ref{sec:results}, I exercise two counterfactual simulations to answer these questions. First, I investigate the effect of a change in the subsidy threshold on matching formation outcomes. Second, I investigate the effect of shifting the subsidy amount on matching outcomes. These simulations evaluate the efficiency of an actual subsidy provision. Finally, I investigate how counterfactual merger configurations would have been formed under different subsidy thresholds and amounts by focusing on the most frequently observed simulated matching outcomes. The numerical comparative statics experiments with the simulated data are shown in Appendix \ref{sec:comparative_statics}.

To make the counterfactual situation clear, I specify the objective function of the government as follows.\footnote{To characterize the data generating process for all economic environments in which all firms and the government behave rationally, I ideally need to solve the government problem and obtain the optimal solution $M^{*}_{J}$ for all $J$ and $\kappa^*$. I do not take this direction because $M^{*}_{J}$ in my data was the form of an interest subsidy not measured as a lump sum amount. In addition, subsidy threshold $\kappa^*$ in my data was equal to 1 million tons, which seemed to be roughly determined by the government in the 1960s without an exact calculation, as some institutional evidence such as \cite{book4} has described.} Given data about all observed firm's characteristics $X$, parameters $\theta$, and thresholds $\underline{\mathcal{J}},\overline{\mathcal{J}} \in \mathbb{N}$ such that $\underline{\mathcal{J}}\le\overline{\mathcal{J}}$, the government solves the expenditure minimization problem as follows:
\begin{align}
    \min_{\{M_J\}_{J \in \mathcal{J}} \in \mathbb{R}_{+}^{|\mathcal{J}|},\kappa\in \mathbb{R}_{+}} &\sum_{J\in \mathcal{J}(M,\kappa,\epsilon) } M_J, \label{eq:expenditure_minimization}\\
    \text{s.t. }& \underline{\mathcal{J}} \le |\mathcal{J}(M,\kappa,\epsilon)| \le \overline{\mathcal{J}},\nonumber\\
    &\mathcal{J}(M,\kappa,\epsilon) = \Gamma(M,\kappa,\epsilon|\theta,X),\nonumber\\
    &\epsilon=\{\epsilon_{i,J}\}_{i \in \mathcal{N},J \in \mathcal{J}},\quad \epsilon_{i,J}\sim_{iid} F_{\epsilon},\nonumber
\end{align}
where $M_J$ is the subsidy amount paid to group $J$ from the government, $\kappa$ is the predetermined subsidy threshold, $\overline{\mathcal{J}}$ and $\underline{\mathcal{J}}$ are the predetermined maximum and minimum numbers of groups that the government must satisfy with its subsidy design, $\Gamma(M,\kappa,\epsilon|\theta,X)$ is the operator that determines the matching allocation $\mathcal{J}(M,\kappa,\epsilon)$ conditional on $M$ and $\kappa$ by equation \eqref{eq:LP} given parameters $\theta$, observed characteristics $X$, and random term $\epsilon$ drawn from the distribution $F_{\epsilon}$. The computation of a competitive equilibrium as a central planner's problem is nested in $\mathcal{J}(M,\kappa,\epsilon)$. The analytical or approximate solution to equation \eqref{eq:expenditure_minimization} is impossible to obtain because $\Gamma$ involves a random term $\epsilon$ in the equilibrium computation and the domain of the control variables, $\mathbb{R}_{+}^{|\mathcal{J}|}$ varies via the computation of $\Gamma$. Instead, I focus on how the equilibrium number of groups, i.e.,  $|\mathcal{J}(M,\kappa,\epsilon)|$ would change if the government changed $M=M_J$ for all $J\in \mathcal{J}$ and $\kappa$ given the $\epsilon$ drawn.

For subsequent counterfactual simulations, I focus on 12 main firms (i.e., $N=12$) in six groups for computational feasibility, as in Section \ref{subsec:estimation_two_variables_main_firms_only}. I recompute equation \eqref{eq:LP} 20 times based on the observed characteristics in the data under the estimated parameters in Table \ref{tb:score_results_two_variables_main_firms_only}.\footnote{I confirmed from 20 simulated equilibria that increasing the number of simulations would not affect the simulated results because the error term is relatively small in the specification. In addition, unlike dynamic game papers computing the equilibrium paths such as a conditional choice probability under 1000 simulated draws for each scenario, my model needs to compute the matching allocations for 60 scenario points (i.e., 6 thresholds $\times$ 10 amounts) under one simulated draw. Thus, the cost of a single equilibrium computation of my model is very low, but the cost of computing a counterfactual ``path" is high, depending on the number of counterfactual scenarios. This is a common feature of a matching model.} For each simulation, I draw $\varepsilon_{i,J}\sim_{iid}N(0,5)$.

One aspect of our matching model requires fine-tuning. This aspect is the credibility of the estimated parameters in Section \ref{sec:estimation}.\footnote{Another potential aspect is the level of matched payoffs and unmatched payoffs. The estimated parameters in Table \ref{tb:score_results_two_variables_main_firms_only} are based on 12 main matched firms, which implies that these estimated parameters do not reflect any information about unmatched firms. To correct this for counterfactual equilibrium computation, researchers may need to add a positive common constant to each matched payoff that fits the market environment in the data but does not affect the evaluation of pairwise inequalities. This fine-tuning will enable them to generate an actual matching outcome before a merger, which is consistent with the fact that no mergers occurred without subsidies. In this paper, the estimated model generates consistent outcomes before and after mergers without the correction.}  Appendix \ref{sec:monte_carlo} suggests that in a small matching market, the bounds of parameters such as merger costs are correctly identified, but those of others may not be identified. Fortunately, I can use point-identified estimates in Column 1 of Table \ref{tb:score_results_two_variables_main_firms_only}, which achieves the second-highest score of the objective function and the scale variable is easy to interpret. Concretely, I use ($(\beta_1,-\gamma)=(124.0,-15.4)$) as a middle scenario.\footnote{As another alternative for the counterfactual simulations, it is possible to simulate equilibrium outcomes based on the lower and upper bounds of the estimates for the cheapest and most expensive cases, like in \cite{ciliberto2009market}.}

\subsection{Changes in subsidy threshold ($\kappa$) and subsidy level ($M$)}

First, I consider the effect of the magnitude of the subsidy threshold $\kappa$ on the matching outcome, especially the number of groups. \cite{igami2019mergers} showed that the optimal policy should stop mergers when five or fewer firms exist, highlighting a dynamic welfare trade-off between ex-post pro-competitive effects and ex-ante value-destroying side effects.\footnote{The applications of dynamic games to a merger incorporate different merger incentives. \cite{stahl2011dynamic} modeled a firm's station acquisition as a portfolio choice to maximize the discounted expected future profit. \cite{igami2019mergers} modeled six mergers in their data as the proposer firm's discrete choice to maximize expected future profit. These dynamic merger models are too complex to explain a firm's merger incentive since the papers involve constructing value functions built on state and action variables with industry-specific features. Instead of dropping the firm's dynamic consideration, matching models such as in \cite{fox2019externality} and \cite{akkus2015ms} capture the merger incentive explicitly and straightforwardly unlike the dynamic merger model.} In addition, \cite{book_kiseki} recorded a detailed discussion about how the number of groups after consolidation would be favorable. For example, some council members insisted that three or four groups were the best number of groups. In fact, until 1999, six groups had been consolidated into three groups. Given these suggestions, I recomputed the matching outcome under different subsidy thresholds conditional on fixed subsidy amounts. This simulation explores the optimal subsidy issue and disentangles the incentives of endogenous matching.

\begin{figure}[htbp]
 \begin{minipage}{0.5\hsize}
  \begin{center}
   \includegraphics[width=80mm]{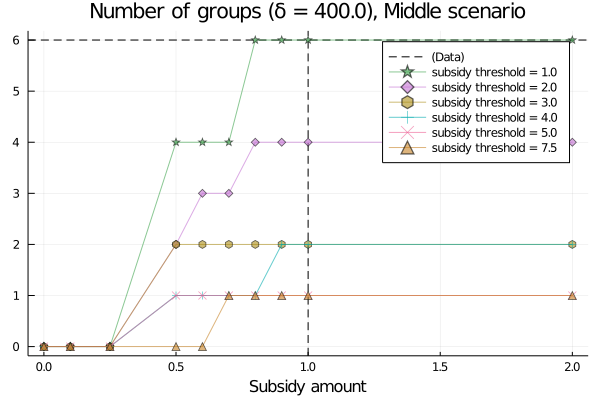}
  \end{center}
 \end{minipage}
 \begin{minipage}{0.5\hsize}
  \begin{center}
   \includegraphics[width=80mm]{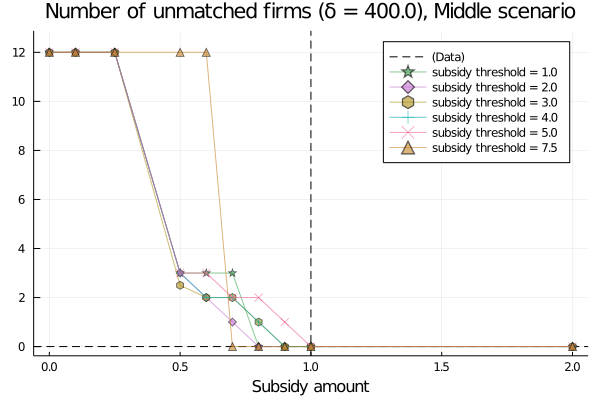}
  \end{center}
 \end{minipage}\\
 \begin{minipage}{0.5\hsize}
  \begin{center}
   \includegraphics[width=80mm]{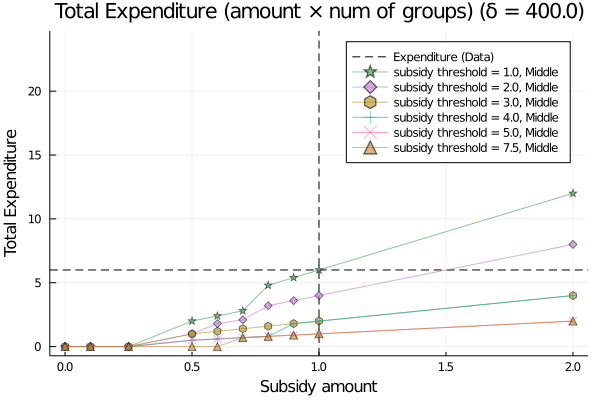}
  \end{center}
 \end{minipage}
 \begin{minipage}{0.5\hsize}
  \begin{center}
   \includegraphics[width=80mm]{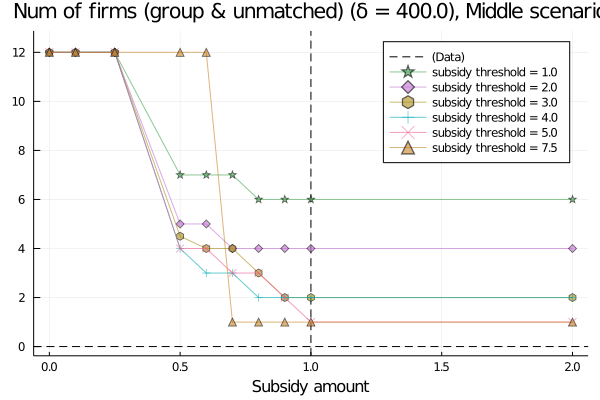}
  \end{center}
 \end{minipage}
 \caption{Simulation results of the number of groups under different subsidy thresholds and amounts.}
 \label{fg:counterfactual_different_subsidy_amount_threshold}
 \begin{tablenotes}
 \footnotesize
\item[a]\textit{Note:} I set grid points of the subsidy amount in the X-axis to $[0,0.1,0.25,0.5,0.6,0.7,0.8,0.9,1.0,2.0]$. The threshold level of 7.5 million tons can be satisfied if and only if all firms form a single grand coalition. I use the estimated parameters in Column 1 of Table \ref{tb:score_results_two_variables_main_firms_only}, for each scenario point. If the solution is a non-integer, I add small perturbations drawn from a standard normal distribution to the allocation matrix and then pick up the maximum one as a probabilistic matching. I simulate 20 times and take the median of the number of groups and unmatched firms.
\end{tablenotes}
\end{figure}

Figure \ref{fg:counterfactual_different_subsidy_amount_threshold} reports in the top panels the number of groups and unmatched firms under different subsidy thresholds.
First, I confirm that if the subsidy amount had been zero, all firms would have remained unmatched, which is consistent with institutional fact. Second, the top-left panel shows that shifting the subsidy threshold would have significantly changed the number of groups given the benchmark subsidy amount level (the black dotted vertical line). For example, if the subsidy threshold had been set to 2 million tons, the number of groups would have been four. This implies that a higher subsidy threshold would have encouraged firms to form larger groups. Thus, the government would have been able to control the number of groups by setting the subsidy threshold under a fixed budget. The right panel also indicates that if the subsidy threshold had been high, mergers would have been relatively less sensitive to an increase in the subsidy amount because it might be more difficult to compensate for merger costs by forming large groups under small subsidy amounts. 

To find an alternative way to control the merger outcome, I investigate the effect of subsidy amount level $M$ on the matching outcome. Figure \ref{fg:counterfactual_different_subsidy_amount_threshold} reports in the top left panel that the level of subsidy provision would have changed the number of groups. First, if matched firms have satisfied the subsidy requirement at the benchmark subsidy amount (black dotted vertical line) and the subsidy threshold (1 million), increasing the subsidy amount would not have affected the matching outcomes because each group had no incentive for additional mergers, which induces splitting the fixed amount of the subsidy with more group firms. Second, the number of groups would have increased if the amount of subsidies had increased. The case of a subsidy threshold of two million tons (purple line) illustrates that increasing the subsidy amount from 0.5 million to 0.8 million would have split two large groups into four mid-sized groups because the split groups would have obtained a larger amount of the subsidy shared with fewer group firms and the mid-sized groups could have compensated for merger costs. Third, the upper right panel shows that increasing the subsidy amount would have reduced the number of unmatched firms.

\subsection{Efficiency of the actual merger-induced subsidy}

As a by-product, the above counterfactual simulations reveal the efficiency of the government subsidy. The efficient subsidy generates a targeted matching outcome with minimum expenditure. First, I evaluate the efficiency of the actual subsidy provided in 1964. Suppose that the government set the target equilibrium number of firms to six, as in the data. The case of a subsidy threshold of 1 million, as shown by the green line in the top left panel of Figure \ref{fg:counterfactual_different_subsidy_amount_threshold} shows that the government could have cut 20 \% of the expenditures to achieve the same matching outcome. The cost-savings amount would have been equivalent to the joint matching production that a hypothetical firm with 2 million tons in a liner would have generated by acquiring a firm having 2 million tons in a liner.\footnote{The calculation of the total production function in \eqref{eq:LP} is as follows. In the matching production function, $20$ \% cost savings is equal to 80 when $\delta=400$ (i.e., $400\times 0.2=80$). The data observed six groups, so the total cost savings is 480 (i.e., $80\times 6=480$). Since $\beta_1=124.0$ and $-\gamma=-15.4$, adding a firm with 2 million tons in a liner to a group that has 2 million tons in a liner yields about 480 (i.e., $124.0\times(2.0\times2.0)-15.4\times 1 \approx 480$). Thus, the cost-savings amount is equivalent to the matching production generated by the hypothetical matching of a firm having 2 million tons in a liner with another firm with 2 million tons in a liner. Note that this cost-savings amount cannot be converted into a monetary amount such as JPY.} Concretely, the impact is approximately equivalent to that of the two biggest mergers (i.e., the Mitsui-OSK Line merger and the Kawasaki Kisen Kaisha merger) shown in Table \ref{tb:total_tonnage_size_group_table}. The cost-savings amount is negligible for the government.

In contrast, suppose that the government set the target equilibrium number of firms to four, which was achieved 25 years later in the industry. The case of a subsidy threshold of 1 million, as shown by the green line in the top left panel of Figure \ref{fg:counterfactual_different_subsidy_amount_threshold} shows that the hypothetical target equilibrium number of firms could have been achieved in 1964, given a half of the total expenditure.\footnote{The actual expenditure is $6=1\times 6(\text{groups})$ and the hypothetical expenditure is $2.0=0.5\times 4(\text{groups})$.}

Second, I explore the feasible target equilibrium number of firms given the same government budget constraint. The bottom left panel in Figure \ref{fg:counterfactual_different_subsidy_amount_threshold} shows the total expenditure incurred by a policy mix combining shifting the subsidy threshold with the subsidy amount. This shows that under the same government's budget level (i.e., the dotted horizontal line), the government could have changed the equilibrium number of firms to between one and six using a policy mix. For example, per the top left panel, fixing the subsidy threshold at two and the subsidy amount at 0.9 would have generated four groups, while fixing the subsidy threshold at three and a subsidy amount at 0.5 would have generated two groups. The bottom-right panel summarizes the number of post-merger firms in the market for each scenario. Therefore, subsidy design is crucial to inducing endogenous mergers for any targeted number of groups.



\subsection{Investigation of a counterfactual merger configuration}

\begin{figure}[htbp]
 \begin{minipage}{\hsize}
  \begin{center}
   \includegraphics[width=170mm]{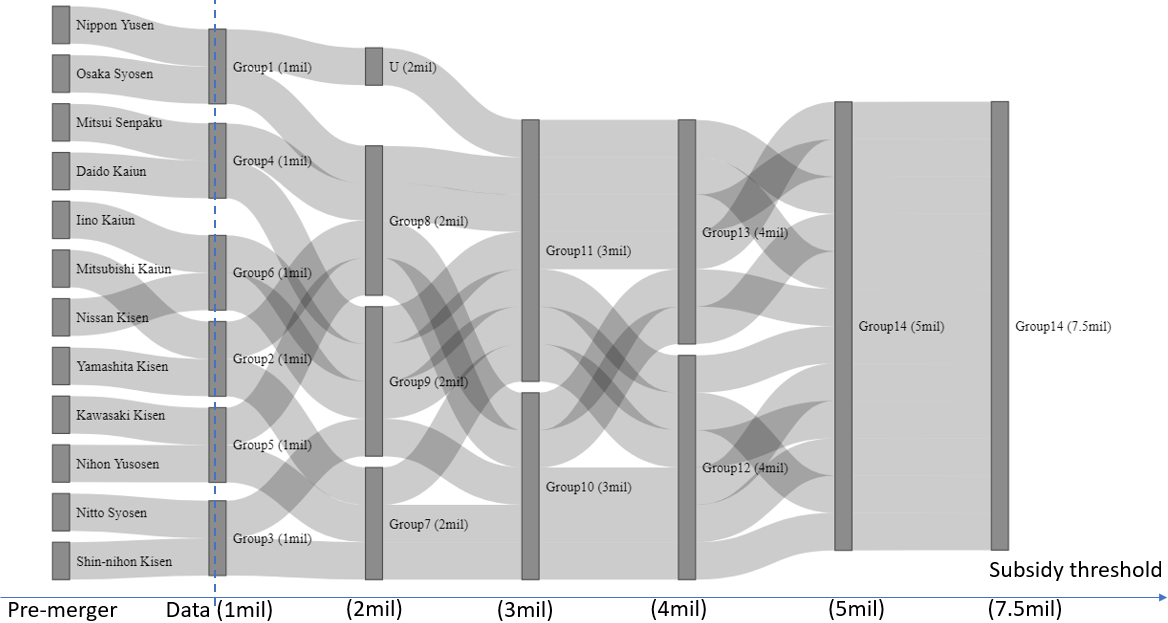}
  \end{center}
 \end{minipage}\\
 \begin{minipage}{\hsize}
  \begin{center}
   \includegraphics[width=170mm]{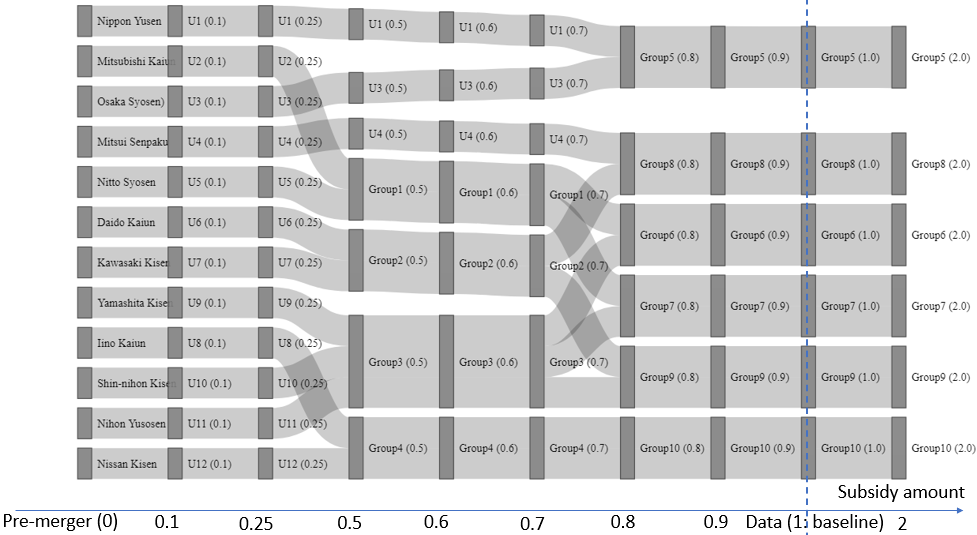}
  \end{center}
 \end{minipage}
 \caption{Merger simulation across different subsidy thresholds and amounts. }
 \begin{tablenotes}
 \footnotesize
\item[a]\textit{Note:} Given my data, the estimated parameters in Column 1 of Table \ref{tb:score_results_two_variables_main_firms_only}, and the match-specific errors drawn, the upper panel picks up the most frequently observed counterfactual mergers under different subsidy thresholds, and the lower panel picks up those most frequently observed under different subsidy amounts. To show the Sankey diagrams, if the simulated matching outcome includes non-integer allocations, I assign firms in the allocations to the existing groups equally, as in a probabilistic matching. The names shown in the figures indicate the firm's identity and hypothetical independent group's ID. I denote an unmatched firm by $U$.
\end{tablenotes}
\label{fg:sankey_diagram_merger}
\end{figure}

Finally, I investigate how a counterfactual merger configuration would have been constructed under different subsidy thresholds and amounts by focusing on the most frequently observed simulated matching outcomes. Figure \ref{fg:sankey_diagram_merger} illustrates counterfactual merger configuration under different subsidy thresholds and amounts. 

First, it would be desirable for the merger configurations, such as merger pairs with firms' identities, to be perfectly predicted by the estimated parameters in the full sample. However, given the data and estimated parameters, the merger configuration in the subsample may not reflect the entire configuration because the merger configuration uses a tiny fraction of all inequalities. In addition, the estimation procedure utilizes pairwise stability based on observed characteristics without firms' identities so that simulated stable matching does not need to fit the actual merger configuration with the firm's identities. Second, the merger configuration varies only if the match-specific error $\epsilon_{i, J}$ varies and turns over pairwise inequalities. Given the estimated parameters in Column 1 of Table \ref{tb:score_results_two_variables_main_firms_only}, the error does not affect the inequalities significantly because the terms of the observed characteristics are much larger than $\epsilon_{i,J}$ in the matching production function.

Figure \ref{fg:sankey_diagram_merger} shows the most frequently observed simulated matching configurations. First, the upper panel of Figure \ref{fg:sankey_diagram_merger} demonstrates that if the subsidy threshold had increased from 1 million tons to 2 million tons, the six existing groups would have formed a few large groups, and ultimately, an oligopoly market with three large groups and one unmatched firm would have resulted. As the subsidy threshold had increased, the market would have become close to a monopoly market.

Second, the lower panel of Figure \ref{fg:sankey_diagram_merger} shows that even if the subsidy amount had increased from the baseline level, the merger configurations would not have varied because the six existing groups had already satisfied the subsidy threshold. On the other hand, if the subsidy amount had decreased from the baseline level, the merger configuration would have consisted of some unmatched firms and four groups because only four firms could have overcome the high merger costs while satisfying the subsidy threshold. Although these hypothetical simulated merger scenarios would have been blocked by competition policy authority, the simulations provide a cut-off point regarding whether endogenous mergers could have occurred innocuously or harmfully.

\section{Conclusions}\label{sec:conclusion}

This paper proposes a structural model to investigate the effect of subsidy thresholds and amounts on endogenous coalition mergers by constructing a one-sided one-to-many matching model with complementarities, an extension of \cite{azevedo2018existence}. I apply the model to shipping mergers and consolidations and the related subsidy laws enacted between 1960 and 1964 in Japan. I find that the assortativeness of both size and technological specialization contributes to merger incentives. The assortativeness of technological specialization varies across carrier types, which implies that the importance of technological diversification and specialization varies across technological types. The counterfactual simulations under the estimated parameters indicate that the real subsidy provision was far from optimal to achieve the same matching outcome in terms of cost minimization. It also shows that a policy mix of shifting the subsidy threshold and amount under a fixed budget constraint could have controlled matching outcomes. 

Finally, I should mention possible extensions as well as some shortcomings of this study. First, this paper focuses on disentangling the endogenous merger incentives under the unique and specific subsidy provision policy while ignoring future competition in the market due to data limitations. Thus, a welfare evaluation of the post-merger market was not investigated. Combining a firm's strategic interactions with estimations of demand and supply sides with the endogenous matching merger model remains a challenging and open research question in the field of industrial organization \citep{agarwal2021market}. Second, this paper does not consider the potential effect of global competition in a foreign transportation market. My current data could not address this potentially interesting question. Third, my matching model does not incorporate unobserved heterogeneity which is identified nonparametrically \citep{fox2018jpe}. Pursuing this direction will require a different econometric approach such as the simulated method of moments in \cite{fox2018jpe} and multiple market data. Fourth, because the subsidy law did not specify subsidy amount flexibly, I could not incorporate potentially heterogeneous subsidy effect although the theoretical model in itself allows it. Further empirical research in this direction will be fruitful.

\bibliographystyle{aer}
\bibliography{00ship_merger}

\newpage
\appendix
\noindent{\LARGE{\textbf{Appendix (For Online Publication)} }}
\onehalfspacing

This Appendix is prepared for online publication and contains supplementary materials for Sections \ref{sec:data_background} through \ref{sec:counter_factual}. Table \ref{tb:contents_of_appendix} lists the contents.

\begin{table}[hbt]
    \centering
    {\footnotesize{}\begin{tabular}{clcc}
        Appendix & Contents &Corresponding main text  \\\hline
        \ref{sec:institutional_background} & Institutional background   &\ref{sec:introduction} and \ref{sec:data_background}\\
        \ref{subsec:supplemental_figures} & Supplemental figures &\\
        \ref{subsec:institutional_facts} & Institutional analysis: what mergers occurred and why these mergers occurred.   &\\
        \ref{subsec:reduced_form} & Preliminary analysis:  Merger patterns   &\\
        \ref{subsec:preliminary_regressions} & Preliminary regression analysis: Who matches whom? &\\
        \hline
        \ref{sec:algorithm} & Estimation algorithm   &\ref{sec:estimation}\\
        \ref{subsec:construct_inequalities} & Construction of inequalities based on pairwise stability &\\
        \ref{subsec:score_function_detail} & Mathematical representation of the score function & \\
        \ref{subsec:score_evaluation} & Procedures for the score function evaluation &\\
        \ref{subsec:romano_shaikh_not_works} & Implementations for inference &\\\hline
        \ref{sec:monte_carlo} & Monte Carlo simulation  &\ref{sec:estimation} and \ref{sec:counter_factual}\\
        \ref{subsec:setup_monte_carlo} & Data generating process & \\
        \ref{subsec:identification_monte_carlo} &  Identification results for a single market  &\\
        \ref{subsec:estimation_monte_carlo}  & Estimation results for a single market & \\
        \ref{subsec:monte_carlo_no_merger}   & Identification power with a no-merger market & \\
        \ref{subsec:monte_carlo_small_n} & Identification power with a small $N$ market & \\
        \ref{subsec:monte_carlo_large_firms}& Identification and Estimation results for a single market with only large firms & \\\hline
        \ref{sec:comparative_statics} & Comparative statics  &\ref{sec:estimation} and \ref{sec:counter_factual}\\
        \ref{subsec:comparative_statics_benchmark_markets}& Equilibrium properties in a benchmark market &\\
        \ref{subsec:comparative_statics_large_firms}& Equilibrium properties in an oligopoly market with only large firms & \\\hline
        \ref{sec:supplemental_estimation}& Supplemental estimation results & \ref{sec:results}\\
        \ref{subsec:estimation_HHI_only} &  Estimation results based on assortativeness of HHI levels &\\
        \ref{subsec:estimation_multivariate} & Estimation results on multivariate observed characteristics &\\
        \hline
        \ref{sec:robustness_check} & Robustness Checks & \ref{sec:results}\\
        \ref{subsec:robustness_check_delta}& Different calibrated $\delta$ & 
    \end{tabular}
    }
    \caption{Contents of the Appendix}
    \label{tb:contents_of_appendix}
\end{table}

\section{Institutional background}\label{sec:institutional_background}

\subsection{Supplemental figures}\label{subsec:supplemental_figures}

\begin{figure}[!ht]
\begin{center}
\includegraphics[height = 0.45\textheight]{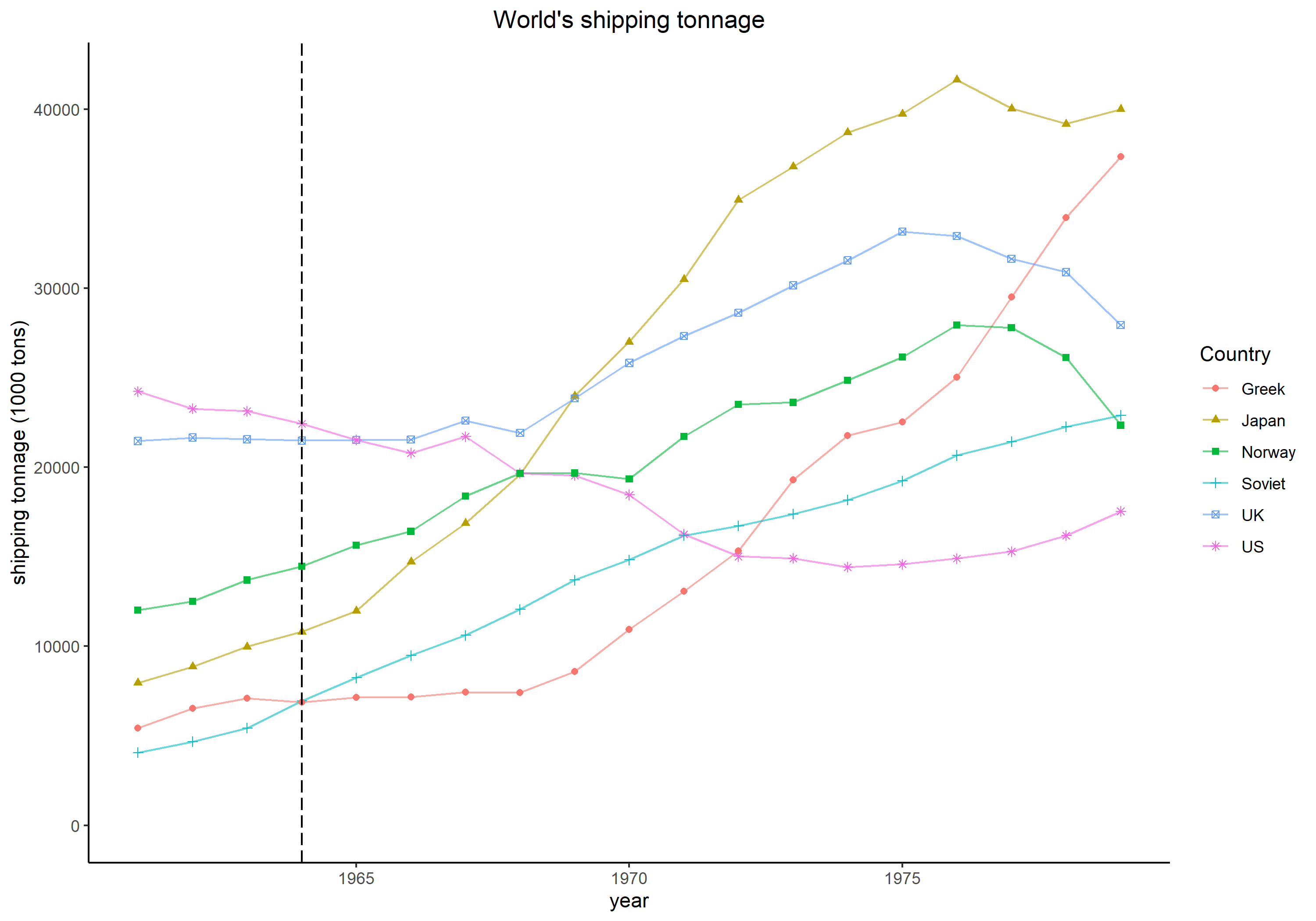}
\end{center}
\caption{The trend of the world's shipping gross tonnage (1000 gross weight tons). }
\footnotesize
\begin{tablenotes}
\item[a]\textit{Note:} \textit{Source}: \cite{syuuyaku} which borrows data from the Statistical Tables in \textit{Lloyd's Statistics}. The data contain ships with tonnage sizes of at least 100 tons and include fishing vessels. The dotted vertical line indicates the period before and after mergers. I omit Liberia. This is because Liberia had the most shipping tonnage through flag of convenience ships, but this feature did not reflect actual economic activity in the shipping industry.
\end{tablenotes}
\label{fg:shippingtonnage}
\end{figure}

Figure \ref{fg:shippingtonnage} depicts the  country-level trends of total tonnage, which is referred in the footnote in Section \ref{sec:introduction}. Table \ref{tb:industry_history} documents the major events related to my sample. 

\begin{table}[ht!]
    \centering
    \begin{tabular}{cll}
      Timing / Period & Event & Japanese shipping industry\\\hline
      1945 (Aug) & The end of WW2 & lost 80\% of tonnage\\
      1950 -- 1953 & Korean War & shipbuilding boom\\
      1956 -- 1957 & Suez Crisis & shipbuilding boom\\
      1960 -- 1971 & Income-Doubling Plan & targeted as a subsidized industry \\
      1961 (Dec) & The 1st bill of the subsidy laws was submitted  & shared information with all firms \\
      1962 (Dec) & The 2nd bill of the subsidy laws was submitted  & shared information with all firms \\
      1963 (July) & The subsidy laws were enforced & prepared for merger plans\\
      1964 (April) & The laws were implemented & mergers and coalition formations
    \end{tabular}
    \caption{The industry historical background}
    \label{tb:industry_history}
\end{table}

\subsection{Institutional analysis: What mergers occurred and why these mergers occurred. }\label{subsec:institutional_facts}

Before quantitatively investigating merger patterns, I provide institutional facts based on \cite{book_kiseki} who recorded implicit communications between some firms and the government's detailed plans for merger processes. 
First, \cite{book_kiseki} reported that although newspaper companies predicted matching pairs for the main firms based on shareholder information before the laws were enacted, the actual matching pairs differed somewhat from their predictions.\footnote{For example, the birth of Mitsui O.S.K Lines formed by Osaka Merchant Ship and Mitsui Lines was not predicted based on shareholder information. \cite{book_kiseki} mentioned that Osaka Merchant Ship parted from ex-ante Japan Lines and found unmatched Mitsui Lines sequentially.} \cite{book_kiseki} also recorded talks by the JSA and some corporate executives that the main firms considered post-merger size effects and technological specialization important on different levels. Therefore, the matching outcome is not easy to predict via simple reasoning.

Second, the actual matching processes were determined sequentially and certainly involved sequential private bargaining and communication, which were essentially not reported. However, \cite{book_kiseki} recorded that the newspapers reported many specific private and public meetings of corporate executives of many firms between 1960 and July 1, 1964. These reports about sharing information among firms for finite but sufficiently long periods helped me to approximate the sequential matching process to the static matching process in a single large market with complete information. While these institutional reports enable us to infer the qualitative reasons why some specific firms were matched to another specific firm, it highlights the significance, from the government's perspective, of quantitative reasoning of matching outcomes and evaluation of the subsidy design. 

Shipping firms have a substantial incentive for post-merger size effects and technological specialization, which was a general feature of the shipping industry that was not specific to the Japanese at that time.\footnote{See \cite{branch2013maritime} for industry-specific background information as a general reference. } \cite{notteboom2004container} maintained that the economic rationality for mergers and acquisitions is rooted in the objectives of economies of scale and power: to gain instant access to markets and distribution networks in the liner shipping industry. The industry-specific features support the above institutional narratives.

\subsection{Preliminary analysis: Merger patterns}\label{subsec:reduced_form}

Based on the institutional provided in Appendix \ref{subsec:institutional_facts}, I illustrate some quantitative features of the ex-post matching results in the following figures. Due to shipping mergers, 27 main firms and 68 affiliate and wholly controlled firms formed six mutually exclusive groups through mergers and acquisitions, while 33 firms chose to stay unmatched. I was unable to collect variables for ten firms, so the final sample size is 118. Explicitly, the six groups are Nippon Yusen, Mitsui-O.S.K Lines, Japan Lines, Kawasaki Kisen Kaisha, Yamashita-Shinnihon Kisen, and Showa Line. Preliminary regression results are shown in Appendix \ref{subsec:preliminary_regressions}.

\begin{figure}[!ht]
\begin{center}
\includegraphics[height = 0.45\textheight]{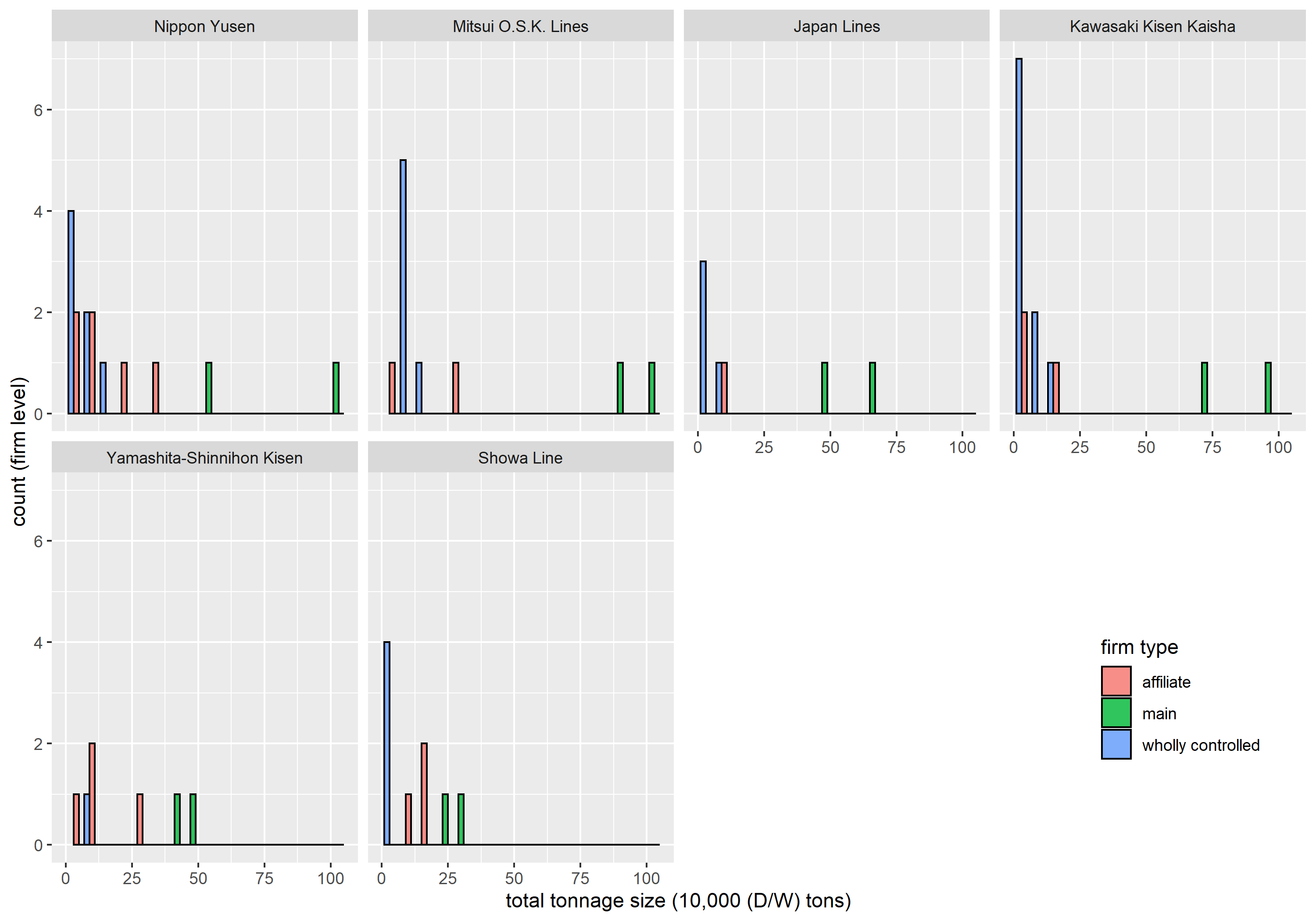}
\end{center}
\caption{Distribution of tonnage for each firm type.}
\footnotesize
\begin{tablenotes}
\item[a]\textit{Note:}  Observation units: Firm-level tonnage size for each firm type of each group after mergers. I omit the observations that have less than 10,000 (D/W) tons in total tonnage.
\end{tablenotes}
\label{fg:type_dist_eachgroup}
\end{figure}

Figure \ref{fg:type_dist_eachgroup} reports the post-merger distribution of the firm-level tonnage for each firm type. This distribution is evidence that all the groups have two main large firms that played the role the main merger decision-makers. Some groups also have many wholly controlled firms, whereas others do not. Finally, unmatched firms are not significantly small. These facts support the idea that the main firms in each group and the unmatched firms might have different capacities for marginal merger costs. 

\begin{figure}[!ht]
\begin{center}
\includegraphics[height = 0.45\textheight]{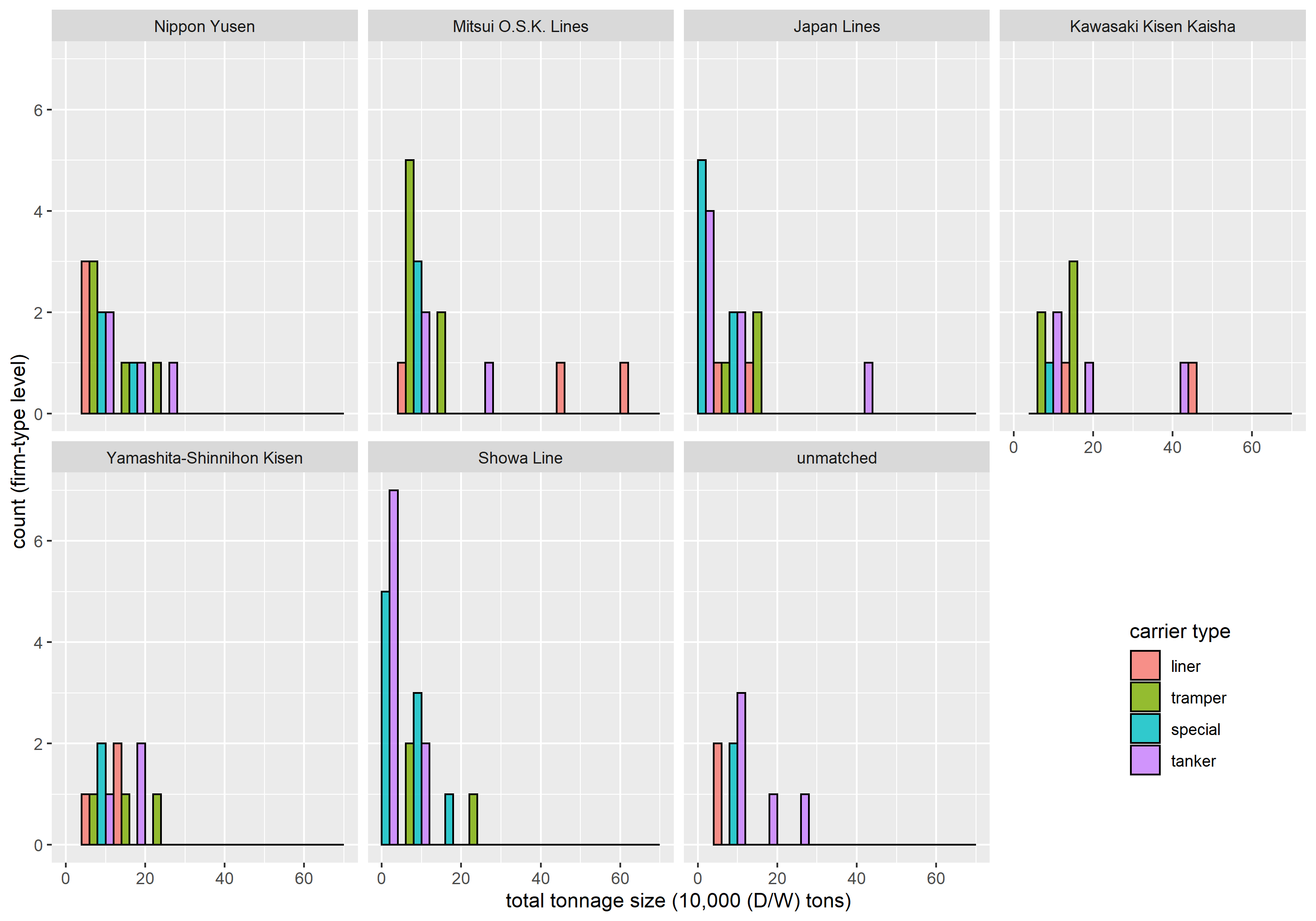}
\end{center}
\caption{Distribution of tonnage for each carrier type.}
\footnotesize
\begin{tablenotes}
\item[a]\textit{Note:} \textit{Source} : \cite{listgaikou} and \cite{nostalgic}. Observation units: Firm-type-level tonnage size of each group after mergers. I omit observations which have less than 10,000 (D/W) tons in the total tonnage.
\end{tablenotes}
\label{fg:carrier_dist_eachgroup}
\end{figure}

\begin{figure}[!ht]
\begin{center}
\includegraphics[height = 0.45\textheight]{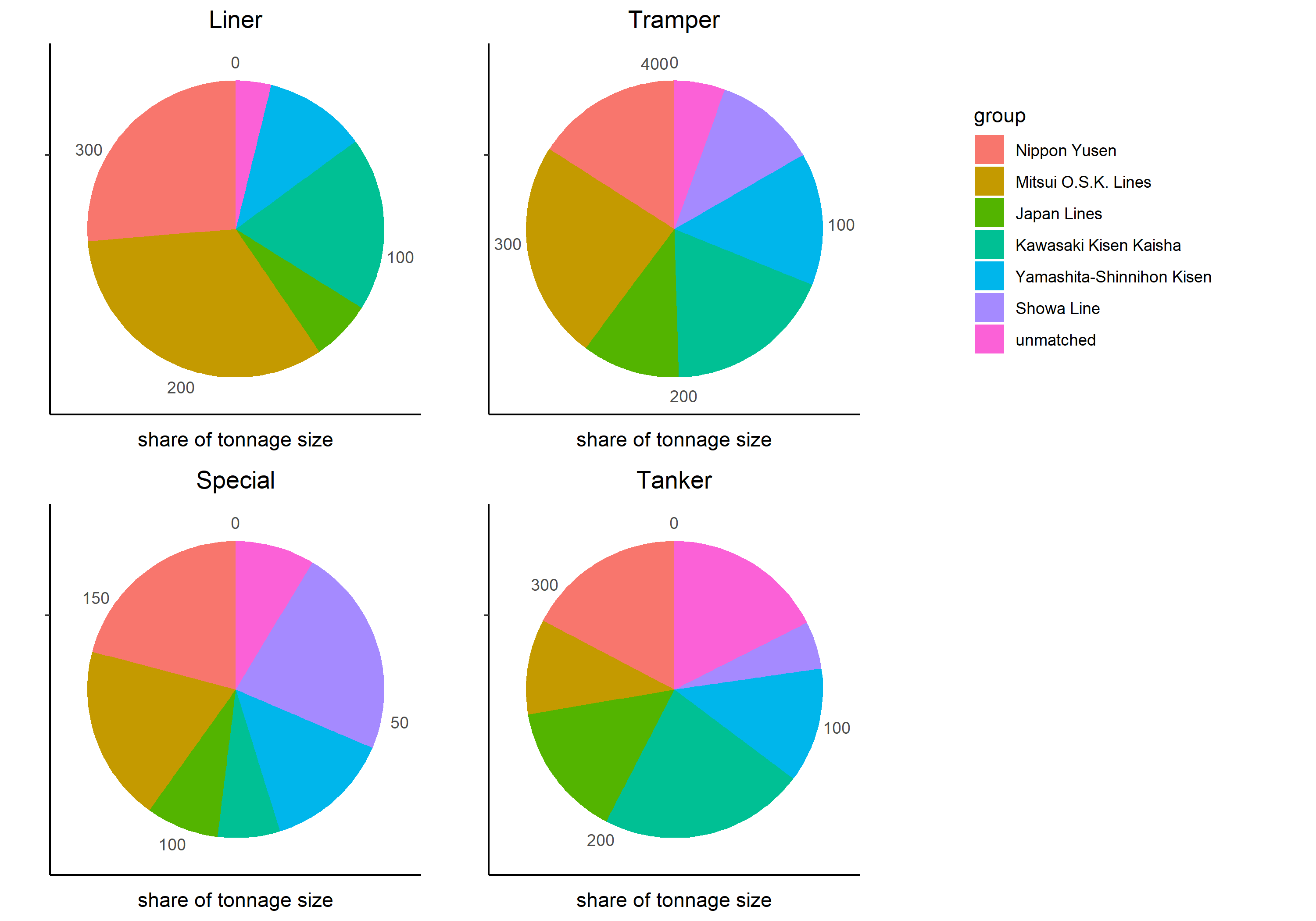}
\end{center}
\caption{Shares of each carrier type and each group.}
\begin{tablenotes}
\item[a]\textit{Note:} \textit{Source} : \cite{listgaikou} and \cite{nostalgic}. Observation units: The total tonnage for each carrier type of each group after mergers. 
\end{tablenotes}
\label{fg:carrier_share_eachgroup}
\end{figure}

Figure \ref{fg:carrier_dist_eachgroup} reports the post-merger distribution of firm-level tonnage for each carrier type. In particular, this illustrates group-specific features about the configuration of carrier types. The remarkable finding is that many large unmatched firms own special vessels or tankers. In addition, Showa Line has many small special vessels and tankers. This implies that the preferences for the assortativeness of specialization are heterogeneous at the group level. Figure \ref{fg:carrier_share_eachgroup} illustrates the post-merger tonnage share of segregated shipping markets. It indicates that more than half of the liner ships were owned by two groups, whereas the other carrier types seem to be more competitive. The findings imply that with these carrier types, concentration seems to be important for potential economies of scale.

\subsection{Preliminary regression analysis: Who matches with whom}\label{subsec:preliminary_regressions}
To obtain some intuition about the matches that occurred quantitatively, I run two preliminary regression models for predicting pairwise match probability: (1) the logit model, and (2) the linear probability model. Here, I simplify the coalitional relationships into pairwise matchings. If two firms belonged to the same group, I interpret that these two firms were matched. I observe $N=118$ firms, so I can construct $13,806$ combinations as possible pairwise matchings. For notational convenience, I assign the index $b$ (buyer) and $t$ (target) to each agent in all possible matching combinations. As a dependent variable, define the indicator variables $Y_{bt}$, that is, $Y_{bt}=1$ if buyer $b$ and target $t$ are matched in the data and $Y_{bt}=0$ otherwise. 

For each match, I identify several observed characteristics $X_{bt}$. First, I use the interaction terms of buyer $b$ and target $t$ 's log of the tonnage of each carrier type, which captures the economies of scale at the pairwise level. 

Second, to capture the assortativeness of specialization, I include the interaction terms of buyer $b$ and target $t$ 's specialization index for each carrier type which is defined as follows:
$$
\text{share}_{ik}=\frac{\text{ton}_{ik}}{\sum_{k\in K}\text{ton}_{ik}},\quad K = \{Liner,Special,Tramper,Tanker \},
$$
where $\text{ton}_{ik}$ is the tonnage of carrier type $k$ owned by firm $i$. 

To measure total diversification, I also include the interaction terms of buyer $b$ and target $t$ 's Herfindahl-Hirschman-Index (HHI), which is defined as follows:
$$
    HHI_i=\sum_{k\in K}(\text{share}_{ik})^2.
    $$
Third, I can observe banking relationships that provide information on capital efficiency, encouraging mergers.\footnote{I can observe the banking relationships for only 55 samples which includes almost all the main and wholly controlled firms. I assume that the bank coverage similarity ratio is zero for missing samples. }\footnote{I confirmed that the information about major shareholders and the main banks overlapped because at that time, large banks were operated by ``Zaibatsu" (a financial conglomerate). In this sense, the bank coverage similarity ratio implicitly captures the effects of shareholders.} Based on the variables, I calculate the bank coverage similarity ratio between buyer $b$ and target $t$ as follows:
$$
    \sum_{z\in Z}(\frac{\mathbbm{1}[b \text{ and } t \text{ have the relationship with bank } z]}{|Z|})
    $$
where $Z$ is the set of banks, $\mathbbm{1}(\cdot)$ is an indicator function, and $|\cdot|$ is a cardinality number. I observe 18 banks in the data, so $|Z|=18$. The ratio also captures the geographic proximity of the head offices because the contracted local banks were located in regions such as Tokyo, Osaka, and Kobe, which were close to the head offices. I also include the dummy variable that indicates whether buyer $b$ and target $t$ are from the same firm type.

\begin{table}[!ht]
\footnotesize
\caption{Preliminary regression results for predicting matchings. }
\begin{center}

\begin{tabular}{@{\extracolsep{5pt}}lccccc} 
\\[-1.8ex]\hline 
\hline \\[-1.8ex] 
 & \multicolumn{5}{c}{\textit{Dependent variable:}} \\ 
\cline{2-6} 
\\[-1.8ex] & \multicolumn{5}{c}{1(match)} \\ 
\\[-1.8ex] & (1) & (2) & (3) & (4) & (5)\\ 
\hline \\[-1.8ex] 
 log(liner$_{b}$ *liner$_{t}$+1) & $-$0.002 &  & $-$0.013 & $-$0.029 & $-$0.003 \\ 
  & (0.006) &  & (0.009) & (0.010) & (0.001) \\ 
  & & & & & \\ 
 log(tramper$_{b}$ *tramper$_{t}$+1) & 0.005 &  & 0.004 & 0.018 & 0.002 \\ 
  & (0.002) &  & (0.005) & (0.006) & (0.001) \\ 
  & & & & & \\ 
 log(special$_{b}$ *special$_{t}$+1) & $-$0.009 &  & $-$0.002 & $-$0.017 & $-$0.002 \\ 
  & (0.004) &  & (0.006) & (0.006) & (0.001) \\ 
  & & & & & \\ 
 log(tanker$_{b}$ *tanker$_{t}$+1) & $-$0.003 &  & $-$0.017 & $-$0.026 & $-$0.003 \\ 
  & (0.004) &  & (0.007) & (0.007) & (0.001) \\ 
  & & & & & \\ 
 log(total$_{b}$ *total$_{t}$+1) & $-$0.021 &  & $-$0.010 & 0.049 & 0.006 \\ 
  & (0.013) &  & (0.017) & (0.018) & (0.002) \\ 
  & & & & & \\ 
 bank coverage similarity ratio &  & 1.598 & 2.052 & 0.649 & 0.088 \\ 
  &  & (0.525) & (0.575) & (0.617) & (0.076) \\ 
  & & & & & \\ 
 log(HHI$_{b}$ *HHI$_{t}$+1) &  & 0.525 & 0.372 & $-$0.123 & $-$0.019 \\ 
  &  & (0.148) & (0.221) & (0.231) & (0.028) \\ 
  & & & & & \\ 
 log(share of liner$_{b}$ *share of liner$_{t}$+1) &  & 0.334 & 1.159 & 2.140 & 0.253 \\ 
  &  & (0.473) & (0.739) & (0.788) & (0.096) \\ 
  & & & & & \\ 
 log(share of special$_{b}$ *share of special$_{t}$+1) &  & $-$0.996 & $-$0.990 & $-$0.529 & $-$0.041 \\ 
  &  & (0.519) & (0.667) & (0.694) & (0.072) \\ 
  & & & & & \\ 
 log(share of tramper$_{b}$ *share of tramper$_{t}$+1) &  & 0.308 & 0.165 & $-$0.558 & $-$0.058 \\ 
  &  & (0.091) & (0.188) & (0.200) & (0.024) \\ 
  & & & & & \\ 
 log(share of tanker$_{b}$ *share of tanker$_{t}$+1) &  & 0.257 & 0.992 & 1.311 & 0.158 \\ 
  &  & (0.210) & (0.335) & (0.354) & (0.043) \\ 
  & & & & & \\ 
 same type &  &  &  & 1.600 & 0.229 \\ 
  &  &  &  & (0.052) & (0.007) \\ 
  & & & & & \\ 
 Intercept & $-$1.271 & $-$2.041 & $-$1.748 & $-$3.312 & $-$0.033 \\ 
  & (0.260) & (0.083) & (0.395) & (0.425) & (0.051) \\ 
  & & & & & \\ 
\hline \\[-1.8ex] 
Model & Logit & Logit & Logit & Logit & OLS \\ 
Observations & 13,806 & 13,806 & 13,806 & 13,806 & 13,806 \\ 
Akaike Inf. Crit. & 12,056.510 & 12,037.120 & 12,034.740 & 11,053.080 & 10,230.180 \\ 
\hline 
\hline \\[-1.8ex] 
\end{tabular} 

\end{center}
\footnotesize
\begin{tablenotes}
\item[a]\textit{Note:} Observation unit: a one-to-one matching pair. The sample size is determined by all possible matching pairs from 118 firms in my data set. Parentheses show standard errors.
\end{tablenotes}
\label{tb:regression_matching}
\end{table}

Table \ref{tb:regression_matching} reports the preliminary regression results. I find robust results for predicting matches. First, if the firms have the same main banks and are of the same type, then they are likely to belong to the same group. Second, the coefficients of assortativeness of size in trampers (resp. in liners and tankers) are positive (resp. negative). The coefficients of economies of scope also have similar positive signs. These findings imply the heterogeneity of matching incentives across carrier types.

Although the preceding reduced-form analysis is useful as a first
cut of the data, the estimated signs and significance levels are ambiguous and not robust. 
As \cite{fox2010qe} mentioned, the matching outcomes are determined by considering the possible matching patterns and error distributions, which essentially involves a high-dimensional numerical integration, so the simple logit regression is not adequate in a structural sense. Specifically, the reduced-form analysis ignores important features such as one-to-many matches with complementarities, the one-sidedness of matches, and trade-offs between merger costs and subsidies.

To understand the trade-offs and merger patterns, a structural model is necessary to quantify matching production and quantitatively investigate the balance of subsidies and costs. In addition, this enables us to simulate counterfactual policies. This is the purpose of the structural model described in Section \ref{sec:model}.
\newpage
\section{Estimation algorithm for evaluating the score function}\label{sec:algorithm}

\subsection{Construction of inequalities based on pairwise stability}\label{subsec:construct_inequalities}

Suppose that type $i$ firm acquires firms $J_i$ in the data, including $h\in J_i$, whereas firm $l$ acquires firms $J_l$ in the data, including $k\in J_l$. For notational convenience, I define swapped coalitions as follows:
\begin{align*}
        \begin{array}{l}
\overline{J_{i}}=(J_{i} \cup\{k\})\backslash\{h\}, \\
\overline{J_{l}}=(J_{l} \cup\{h\})\backslash\{k\}.
\end{array}
    \end{align*}
The following inequality constructions are derived based on the rank order property formally introduced as Assumption \ref{as:rank_order} in Section \ref{subsec:identification_consistency_and_inference}.

First, I construct a matching maximum score inequality for two acquiring firms without price data. Through the rank order property, 
$$
\text{Pr}[\text{observed matching}|X]\geq \text{Pr}[\text{swapped counterfactual matching}|X]
$$ if and only if
\begin{align}
    X_i\left(\left(x_{j}\right)_{j \in J_{i}} \right)^{\prime} \theta+X_l\left(\left(x_{j}\right)_{j \in J_{l}} \right)^{\prime}\theta &\geq X_i\left(\left(x_{j}\right)_{j \in \overline{J_{i}}} \right)^{\prime} \theta+X_l\left(\left(x_{j}\right)_{j \in \bar{J}_{l}} \right)^{\prime}\theta. \label{ineq1}
\end{align}
which corresponds to the standard pairwise inequalities between observed and swapped matches.

Second, I construct inequalities from an observed coalition. Through the rank order property,  $$
\text{Pr}[\text{observed matching}|X]\geq \text{Pr}[\text{matching excluding matched $h$}|X]
$$ if and only if
\begin{align}
    X_i\left(\left(x_{j}\right)_{j \in J_i}\right)^{\prime} \theta &\geq X_i\left(\left(x_{j}\right)_{j \in J_i\backslash\{h\}}\right)^{\prime} \theta+\underbrace{X_h\left(x_{h}\right)^{\prime} \theta}_{\text{unmatched } h},\forall i\in\mathcal{N}\label{ineq2}
\end{align}
where $h \in J_i$ and $X_h\left(x_{h}\right)^{\prime} \theta$ is deviated firm $h$'s unmatched payoff. This corresponds with the pairwise inequalities between observed and hypothetical matches. This captures a balance between additional matched payoffs and matched costs regarding adding firm $h$ to group $J_i$.

Third, I construct inequalities from an unmatched target. Through the rank order property,  
$$
\text{Pr}[\text{observed matching}|X]\geq \text{Pr}[\text{matching including unmatched $h$}|X]
$$ if and only if
\begin{align}
    X_i\left(\left(x_{j}\right)_{j \in J_i\backslash\{h\}} \right)^{\prime} \theta+\underbrace{X_h\left(x_{h}\right)^{\prime} \theta}_{\text{unmatched } h} &\geq X_i\left(\left(x_{j}\right)_{j \in J_i}\right)^{\prime} \theta. \label{ineq3}
\end{align}
which is analogous to the second inequality. The right-hand side can be a composition-swapped term such as $X_i\left(\left(x_{j}\right)_{j \in (J_i\cup \{h\})\backslash \{k\}}\right)^{\prime} \theta+X_k(x_k)'\theta$ for any $k\in J_i$.

Fourth, I construct inequalities from the IR condition for matched buyer $i$ of unmatched agents. Through the rank order property, I can derive the following:
\begin{align}
        \underbrace{X_i\left(\left(x_{i}\right) \right)'\theta}_{\text{unmatched}} \geq X_i\left(\left(x_{j}\right)_{j \in \tilde{J}_i} \right)'\theta, \forall \tilde{J}_i, \label{ineq4}
    \end{align}
where coalition $\tilde{J}_i$ consists of type $i$ firm and any combination of unmatched firms. 

Fifth, I construct inequalities from the IR condition with and without subsidies. Through the rank order property, 
\begin{align*}
    \text{Pr}[\text{observed matching with subsidy}|X]&\geq \text{Pr}[\text{hypothetical unmatched without subsidy}|X]\nonumber\\
    &\geq \text{Pr}[\text{observed matching without subsidy}|X]\nonumber
\end{align*}
if and only if
\begin{align}
    X_i\left(\left(x_{j}\right)_{j \in J_i}\right)^{\prime} \theta|_{M=1}&\geq\underbrace{X_i\left(\left(x_{i}\right) \right)'\theta}_{\text{unmatched}}\nonumber\\
    &\geq X_i\left(\left(x_{j}\right)_{j \in J_i}\right)^{\prime} \theta |_{M=0},\label{ineq5}
\end{align}
where the left- and right-hand sides mean matched payoffs conditional on subsidy level $M=1$ (data) or $M=0$. Inequality \eqref{ineq5} reflects the fact that no firms did not merge before the subsidy laws. For notational convenience, let $\chi (i)$ be a indicator function regarding whether inequality \eqref{ineq5} is satisfied for the buyer of a type $i$ firm.

Finally, I construct the matching maximum score objective function \eqref{eq:obj} as follows: 
\begin{align*}
    Q(\theta)=\sum_{g \in G_N} 1\left[Z_{g}^{\prime} \theta \geq 0\right]
    \end{align*}
where $G_N$ is the set of observed inequalities and $Z_{g}^{\prime} \theta$ is the corresponding inequality of matching pairs $g$ based on inequalities (\ref{ineq1}), (\ref{ineq2}), (\ref{ineq3}), (\ref{ineq4}), and \eqref{ineq5}.\footnote{In general, maximizing $Q(\theta)$ over $G_N$ is impossible for a large $N$ so the researcher needs to replace $G_N$ with $\hat{G}_N$ based on subsampled inequalities, as proposed by \cite{fox2018qe}. In this paper, I use only 118 firms so that I can construct $G_N$ fully for point estimates.} Since $Q(\theta)$ is discontinuous over $\theta$, I need to use a global optimization algorithm. Following the literature, I use a differential evolution (DE) algorithm \citep{storn1997differential} to find a maximizer $\hat{\theta}$ of $Q(\theta)$. 

In Appendices \ref{sec:algorithm} and \ref{sec:monte_carlo}, I show how the objective function is evaluated and how the proposed method works for identification and estimation.

\subsection{Mathematical representation of the score function}\label{subsec:score_function_detail}

Several pairwise inequalities are introduced in Section \ref{sec:estimation}. Under Assumption \ref{as:main_firm_is_a_buyer}, each firm is a buyer, seller, or an unmatched firm. Then, the possible pairwise inequalities are classified into nine types. The objective function (\ref{eq:obj}) of matching maximum rank estimation is explicitly constructed as
{\scriptsize{}
\begin{align}
    Q(\theta)&=\frac{2}{N(N-1)}\sum_{a = 1}^{N-1}\sum_{b = 2}^{N} \Bigg[\label{eq:score_function_detail}\\
    &1[\text{firm $a$ is a buyer and firm $b$ is a buyer.}]\cdot \nonumber\\
    &\Bigg(\left( \sum_{k \in J_a} \sum_{h \in J_b} 1\left[X_a\left(\left(x_{j}\right)_{j \in J_{a}} \right)^{\prime} \theta+X_b\left(\left(x_{j}\right)_{j \in J_{b}} \right)^{\prime}\theta \geq X_a\left(\left(x_{j}\right)_{j \in (J_{a} \cup\{h\})\backslash\{k\}} \right)^{\prime} \theta+X_b\left(\left(x_{j}\right)_{j \in (J_{b} \cup\{k\})\backslash\{h\}} \right)^{\prime}\theta\right]\right)\nonumber\\
    & \quad \quad  \quad \quad  \quad \quad +\chi(a)+\chi(b)\Bigg)\nonumber\\
    &+1[\text{firm $a$ is a buyer and firm $b$ is a seller.}]\cdot \nonumber\\
    &\left(\left( \sum_{k \in J_a}  1\left[X_a\left(\left(x_{j}\right)_{j \in J_{a}} \right)^{\prime} \theta \geq X_a\left(\left(x_{j}\right)_{j \in J_{a}\backslash\{k\}} \right)^{\prime} \theta+\underbrace{X_k\left(x_{k}\right)^{\prime} \theta}_{\text{unmatched } k}\right]\right)+\chi(a)\right)\nonumber\\
    &+1[\text{firm $a$ is a seller and firm $b$ is a buyer.}]\cdot \nonumber\\
    &\left(\left( \sum_{h \in J_b}  1\left[X_b\left(\left(x_{j}\right)_{j \in J_{b}} \right)^{\prime} \theta \geq X_b\left(\left(x_{j}\right)_{j \in J_{b}\backslash\{h\}} \right)^{\prime} \theta+\underbrace{X_h\left(x_{h}\right)^{\prime} \theta}_{\text{unmatched } h}\right]\right)+\chi(b)\right)\nonumber\\
    &+1[\text{firm $a$ is a buyer and firm $b$ is unmatched.}]\cdot\nonumber\\
    &\left(\left( \sum_{k \in J_a}  1\left[X_a\left(\left(x_{j}\right)_{j \in J_{a}} \right)^{\prime} \theta+\underbrace{X_b\left(x_{b}\right)^{\prime} \theta}_{\text{unmatched } b} \geq X_a\left(\left(x_{j}\right)_{j \in (J_{a} \cup\{b\})\backslash\{k\}} \right)^{\prime} \theta+\underbrace{X_k\left(x_{k}\right)^{\prime} \theta}_{\text{unmatched } k}\right]\right)+\chi(a)\right)\nonumber\\
    &+1[\text{firm $a$ is unmatched and firm $b$ is a buyer.}]\cdot\nonumber\\
    &\left(\left( \sum_{h \in J_b}  1\left[\underbrace{X_a\left(x_{a}\right)^{\prime} \theta}_{\text{unmatched } a}+X_b\left(\left(x_{j}\right)_{j \in J_{b}} \right)^{\prime} \theta \geq X_a\left(\left(x_{j}\right)_{j \in (J_{a} \cup\{a\})\backslash\{h\}} \right)^{\prime} \theta+\underbrace{X_h\left(x_{h}\right)^{\prime} \theta}_{\text{unmatched } h}\right]\right)+\chi(b)\right)\nonumber\\
    &+1[\text{firm $a$ is unmatched and firm $b$ is unmatched.}]\cdot\nonumber\\
    &\left( \sum_{h \in J_b}  1\left[\underbrace{X_a\left(x_{a}\right)^{\prime} \theta}_{\text{unmatched } a}+\underbrace{X_b\left(x_{b}\right)^{\prime} \theta}_{\text{unmatched } b} \geq X\left(\left(x_{j}\right)_{j \in \{a\} \cup\{h\}} \right)^{\prime} \theta \right]\right)\nonumber\\
    &+1[\text{firm $a$ is a seller and firm $b$ is unmatched.}]\cdot 0 \nonumber\\
    &+1[\text{firm $a$ is a seller and firm $b$ is a seller.}]\cdot 0\nonumber\\
    &+1[\text{firm $a$ is unmatched and firm $b$ is a seller.}]\cdot 0 \Bigg],\nonumber
\end{align}
}
where the first term is based on inequality \eqref{ineq1}, the second and third terms are based on inequality \eqref{ineq2}, the fourth and fifth terms are based on inequality \eqref{ineq3}, the sixth term is based on inequality \eqref{ineq4}, the last three terms are evaluated to zero to avoid affecting the score because of Assumption \ref{as:main_firm_is_a_buyer}, and all terms consisting of buyer firms are evaluated by inequality \eqref{ineq5}. 

\subsection{Procedures for the score function evaluation}\label{subsec:score_evaluation}

For each candidate parameter value, the score function \eqref{eq:score_function_detail} is evaluated as follows:
\begin{itemize}
    \item Find all possible pairs of firms, $k=1,2,\cdots,K$ from the bootstrap data $\mathcal{K}$ generated from $\mathcal{N}$ ($N=118$ samples). I fix $|\mathcal{K}|=118$ (i.e., 12 main firms and 106 randomly chosen other types of firms) so that $K=118\times 117=13,806$ for each sub-sampled data $\mathcal{K}$.
    \item For $k=1,2,\cdots,K$, given the specification of production function $F_{i,J}$ and candidate $\theta$,
    \begin{enumerate}
    \item Denote each of the firms in pair $k$ as firms $1$ and $2$
    \item Extract information of individual covariates of firms $1$ and $2$ as buyer's covariates, denoted by $X_1$ and $X_2$
    \item Identify the types of firms 1 and 2:
    \begin{enumerate}[(I)]
        \item If both firms 1 and 2 are buyers: 
        \begin{enumerate}[(i)]
            \item Pick up all target firms for firms 1 and 2 and the sets are denoted by $T_1$ and $T_2$
            \item Construct the target's covariates at coalition level for firms 1 and 2 based on $T_1$ and $T_2$, denoted by $X_{T_1}$ and $X_{T_2}$
            \item Construct the observed production function based on candidate parameters and covariates $X_1$ and $X_{T_1}$ for firm 1 and covariates $X_2$ and $X_{T_2}$ for firm 2
            \item Construct a swapped production function based on candidate parameters and covariates $X_1$ and $X_{T_2}$ for firm 1 and covariates $X_2$ and $X_{T_1}$ for firm 2
            \item Construct inequalities by the observed production function and swapped production function
        \end{enumerate}
        \item If firm 1 is a buyer and firm 2 is a seller:
        \begin{enumerate}[(i)]
            \item Pick up all target firms for firm 1 and the sets are denoted by $T_1$
            \item Pick up a deviated target single firm $\tilde{t}_1\in T_1$. This results in constructing $|T_1|$ inequalities for all possible deviated firms in this step.
            \item Construct the target's covariates at coalition level for firm 1 based on $T_1$, denoted by $X_{T_1}$. Additionally, construct own covariates $X_{\tilde{t}_1}$ for the deviated firm $\tilde{t}_1$
            \item Construct the observed production function based on candidate parameters and covariates $X_1$ and $X_{T_1}$ for firm 1
            \item Construct swapped production function based on candidate parameters and covariates $X_1$ dropping $X_{\tilde{t}_1}$ for firm 1
            \item Construct inequalities by the observed production function and swapped production function
        \end{enumerate}
        \item If firm 1 is a seller and firm 2 is a buyer: 
        \begin{itemize}
            \item Same as (II)
        \end{itemize}
        \item If firm 1 is a buyer and firm 2 is unmatched:
        \begin{itemize}
            \item Pick up all target firms for firm 1 and the sets are denoted by $T_1$
            \item Construct the target's covariates at coalition level for firm 1 based on $T_1$, denoted by $X_{T_1}$. 
            \item Construct the observed production function based on candidate parameters and covariates $X_1$ and $X_{T_1}$ for firm 1
            \item Construct a swapped production function based on candidate parameters and covariates $X_1$ and $X_{2}$ for firm 1
            \item Construct inequalities by the observed production function and swapped production function
        \end{itemize}
        \item If firm 1 is unmatched and firm 2 is a buyer:
        \begin{itemize}
            \item Same as (IV)
        \end{itemize}
        \item If firms 1 and 2 are unmatched:
        \begin{itemize}
            \item Construct observed production function functions based on candidate parameters and covariates $X_1$ for firm 1 and covariates $X_2$ for firm 2
            \item Construct swapped production function based on candidate parameters and covariates $X_1$ and $X_{2}$ for merged group formed by firms 1 and 2
            \item Construct inequalities by the observed production function and swapped production function
        \end{itemize}
        \item If firm 1 and 2 are sellers:
        \begin{itemize}
            \item Assign zero to the score function
        \end{itemize}
        \item If firm 1 is a seller and firm 2 is unmatched:
        \begin{itemize}
            \item Assign zero to the score function
        \end{itemize}
        \item If firm 1 is an unmatched and firm 2 is a seller:
        \begin{itemize}
            \item Assign zero to the score function
        \end{itemize}
    \end{enumerate}
\end{enumerate}
\end{itemize}
\begin{itemize}
    \item Summing up the scores over $k=1,2,\cdots,K$ and get $Q(\theta)$
    \item Choose parameters which maximize $Q(\theta)$ over $\theta$ for sub-sampled data $\mathcal{K}$
\end{itemize}

Following the literature, I use differential evolution (DE) algorithm \citep{storn1997differential} to find maximizers of $Q(\theta)$. The point estimation procedure is analogously implemented by replacing $\mathcal{K}$ with $\mathcal{N}$. The search domains of DE algorithm are found by preliminary estimation and fixed to all estimations. 

\subsection{Implementations for inference}\label{subsec:romano_shaikh_not_works}

For the maximum rank estimator, I construct 95 \% confidence intervals via nonparametric bootstrap proposed by \cite{subbotin2007asymptotic}. The implementation procedure is the same as the point estimation in Appendix \ref{subsec:score_evaluation}, except for using subsample $\mathcal{K}$ via sampling the full sample $\mathcal{N}$ with replacement.

There are some remarks. First, under Assumption \ref{as:main_firm_is_a_buyer}, I keep the main 12 firms as buyer firms across all subsamples because the buyer firms are the most informative for constructing inequalities in Equation \eqref{eq:score_function_detail}. Second, the number of all possible inequalities for each subsampled set $\mathcal{K}\neq \mathcal{N}$ is different from $\mathcal{K}'\neq \mathcal{K}$ because of the existence of pairwise inequalities regarding composition swaps. Although these inequalities are important to capture the trade-off between merger costs and subsidy effects, instead, the property that achievable maximum score on $\mathcal{K}$ must be different from the set $\mathcal{K}'$ holds and makes it impossible to conduct inference \citep{romano2008inference} based on partial identification analysis straightforwardly, as I explain below.

Ideally, as in Appendix of \cite{fox2018qe}, confidence regions should be reported as projections onto the axes of the confidence regions proposed by \cite{romano2008inference} which is valid under both set and point identification. For a practical purpose, I show that the method constructing confidence regions proposed by \cite{romano2008inference} is not applicable to the case that the total number of pairwise inequalities is different for each bootstrapped or subsampled sample.

To follow the notations of \cite{romano2008inference}, I redefine the objective function $Q(\theta)$ as $-Q(\theta)-\left(-\max _{\theta} Q(\theta)\right)$ in equation \eqref{eq:obj} and find the minimizer of $Q(\theta)$. Let $\mathcal{C}$ be the confidence regions for identifiable parameters that are pointwise consistent in level. The individual null hypotheses as $\mathrm{H}_{\theta}: Q(\theta)=0$ for $ \theta \in \Theta$. Suppose that I construct a critical value for the test by subsampling. Let $\hat{Q}_n(\theta)$ be a sample analog of $Q(\theta)$ for a full sample of size $n$. Let $b=b_{n}<n$ be a sequence of positive integers tending to infinity, but satisfying $b / n \rightarrow 0$. Let 
$B_n$ be a positive sequence of numbers tending to $\infty$ as $n \rightarrow \infty$ and let $I_{1}, \ldots, I_{B_{n}}$ be chosen randomly with or without replacement from the numbers $1, \ldots, N_{n} =\left(\begin{array}{l}
n \\
b
\end{array}\right).$ Let $a_n$ and $a_b$ be constants such that $a_b/a_n \rightarrow 0$. Let $\hat{Q}_{n, b, i}(\theta)$ denote the statistic $\hat{Q}_{n}(\theta)$ evaluated at the $i$ th subset of data of size $b$ from the $n$ observations. For $\alpha \in(0,1)$, define a critical value as
\begin{align*}
    \tilde{d}_{n}(\theta, 1-\alpha)=\inf \left\{x: \frac{1}{B_{n}} \sum_{1 \le i \le B_{n}} I\left\{a_{b} \hat{Q}_{n, b, I_{i}}(\theta)\le x\right\} \ge 1-\alpha\right\},
\end{align*}
where the construction details are shown in remarks 3.2 and 3.3 of \cite{romano2008inference}. Then, the confidence regions are characterized as follows:
\begin{align*}
    \mathcal{C}_{n}=\left\{\theta \in \Theta: a_{n} \hat{Q}_{n}(\theta) \le \tilde{d}_{n}(\theta, 1-\alpha)\right\}
\end{align*}

In the standard setting listed in \cite{romano2008inference}, the maximum achievable number of $\hat{Q}_{n,b,I_i}(\theta)$ is the same for all $I_i$, given the same $n$ and $b$. Then, the bracket term of $\tilde{d}_{n}(\theta, 1-\alpha)$ in the indicator function works properly for common level $x$. On the other hand, in the setting such as the present study and \cite{fox2013aej}, the maximum achievable number of $\hat{Q}_{n,b,I_i}(\theta)$ can be different for each $I_i$. For example, in the setting in this study, the set of target firm bundles for each subsample can be different. Then, the bracket term of $\tilde{d}_{n}(\theta, 1-\alpha)$ in the indicator function does not work under common level $x$ and $a_b$.

\newpage
\section{Monte Carlo simulations}\label{sec:monte_carlo}

This section shows how the proposed method works for identification and estimation. Particularly, the method might not be able to identify the upper bound of subsidy sensitivity $\delta$ in a single market. Therefore, two practical single-market settings are considered. In the first case, two or three large firms, two or three mid-size firms, and three small-sized firms exist in a market. In the second case, there are eight large firms in a market, as shown in Appendices \ref{subsec:monte_carlo_large_firms} and \ref{subsec:comparative_statics_large_firms}. 

For tractability and time efficiency, the number of firms is set as $N=8$, so the simulation focuses on a small, single matching market as a minimal setting. It should be emphasized that generating data in a large market is infeasible because the model must solve a large linear programming problem with equilibrium constraints. For example, if $N=20$, the model must solve $20\times 2^{20}\approx 20,000,000$-dimensional linear programming problem under much larger equilibrium constraints.\footnote{I confirm that $N=8$ is a feasible size for repeated simulation. Fortunately, $N=12$ which is the number of main firms in my data used in Section \ref{subsec:estimation_two_variables_main_firms_only} is a computational limit size of repeated counterfactual simulations. Thus, I can simulate counterfactual scenarios as the full solution approach.} Instead, the subsequent Monte Carlo simulations investigate a finite small sample property of identification and estimation.

The results in independent many markets setting are analogously derived from distributions of estimated parameters for each single market because the objective function in many markets setting evaluates the summation of the score for each single independent market. Figure \ref{fg:histogram_beta_delta_gamma_to_buyer_shared} shows that if we observe 1000 one-to-many markets, $\beta$ and $\gamma$ will be consistently estimated because the most frequently observed and estimated values are close to the true value through ignoring a few unidentified markets. In this study, the property in many markets setting is not checked because the evaluation step of the objective function in many simulated markets for each random seed is computationally burdensome.

\subsection{Data generating process}\label{subsec:setup_monte_carlo}

To generate samples based on \eqref{eq:LP}, the payoff function of matching $(i,J)\in \mathcal{N}\times \mathcal{P}(\mathcal{N})$ is specified as
\begin{align*}
    F_{i,J}&=\bar{\beta} (x_i\circ x_{J}) + \delta s_{i,J} - C_{i,J}\\
    &=\bar{\beta}  (x_i\circ x_{J}) + \delta s_{i,J} - \gamma \cdot[\text{the total number of firms in coalition } J\backslash i ]
\end{align*}
where $\bar{\beta} =(\beta_0,\beta)=(1.0,0.0)$, $\beta_0$ must be normalized to 1 for identification, $\delta=1.0$ and $\gamma=1.0$ as true parameters, match-specific errors $\epsilon_{i,J}$ are drawn i.i.d. from the standard normal distribution, two types of tonnage covariates $\text{ton}_{i1},\text{ton}_{i2}$ are drawn i.i.d. from $\text{Log-Normal}[2,1]$ and divided by 100 for scale standardization, and $x_i\circ x_{J}$ means the elemental product of $x_i$ and $x_J$ defined later. The simulated tonnage sizes approximate features of a single group, on average, which consists of two or three large firms, two or three mid-sized firms, and three small-sized firms in my study. Another important case, such as a market consisting of all eight large-sized firms as in the counterfactual simulations in Section \ref{sec:counter_factual}, is explored in Sections \ref{subsec:monte_carlo_large_firms} and \ref{subsec:comparative_statics_large_firms}. 

The covariates $x_{i}$ and $x_{J}$ are specified as:
\begin{align*}
    x_{i1}&=\text{ton}_{i1}, \quad x_{J,1}=\sum_{j\in J}\text{ton}_{j1},\\
    x_{i2}&=\frac{\text{ton}_{i1}}{\text{ton}_{i1}+\text{ton}_{i2}}, \quad x_{J,2}=\frac{1}{|J|}\sum_{j\in J}x_{j2},
\end{align*}
where the first line of the variables captures the scale for type 1 carrier and the second line of the variables captures the specialization on type 1 carrier. Two types of subsidy specifications, that is, $s_{i,J}^{\text{to buyer}}$ and $s_{i,J}^{\text{shared}}$ are considered.
These subsidy specifications are defined as follows:
\begin{align*}
s_{i,J}^{\text{to buyer}} &= \begin{cases}
1 \quad \mbox{ if } [\sum_{i \in J} (\text{ton}_{i1}+\text{ton}_{i2}) >1] \\
0 \quad \mbox{ otherwise }
\end{cases},\\
s_{i,J}^{\text{shared}} &= \begin{cases}
\frac{1}{|J|} \quad \mbox{ if } [\sum_{i \in J} (\text{ton}_{i1}+\text{ton}_{i2}) >1] \\
0 \quad \mbox{ otherwise }
\end{cases},
\end{align*}
where $|J|$ denotes the number of firms in coalition $J$, and $s_{i,J}^{\text{to buyer}}$ can be interpreted as a match-specific dummy variable. The subsidy specification $s_{i,J}^{\text{to buyer}}$ implies a lump-sum form of subsidy to the joint production function $F_{i,J}$, whereas the subsidy specification $s_{i,J}^{\text{shared}}$ captures that the subsidy is shared equally by group members. Note that I do not need to specify how the joint production would be allocated to each group member. Moreover, the levels of subsidy amount and subsidy threshold ($1$ and $1$) affect matching outcomes, but these depend on the parameters and data generating process. Finally, type $i$ firm's payoff of selling type $i$ firm is normalized to zero, while there is a need to specify the payoff of staying unmatched as $F_{i\{i\}}=\bar{\beta} (x_i\circ x_i)$. The benchmark settings are summarized in Table \ref{tb:monte_carlo_parameters_list}.

This study considers mergers in a generic single market, but it might be adequate in practice that the market is defined as an industry-level market consisting of some firms. Therefore, this study focuses on a single one-to-many market setting in the subsequent simulation and estimation results. Additionally, under the size limitation of the data generation process, Monte Carlo simulations cannot formally verify the asymptotic property because the maximum score is at most 30 which is too far from the asymptotic size.\footnote{In recent econometrics literature on the maximum score estimator, there are several novel methods with improved properties. As the most applicable method to this study, \cite{rosen2019finite} showed a finite sample inference method of a semiparametric binary response model under a conditional median restriction and investigated its finite sample performance with sample size $N=250$ and two parameters. If we consider the many cross-sectional market setting, their approach can be used.}
However, I need to stress that all properties investigated here hold even in a very small sample.  

\begin{table}[!htbp] \centering 
  \caption{Base Case Parameters of Model. } 
  \label{tb:monte_carlo_parameters_list} 
  \scriptsize{}
  \begin{tabular}{@{\extracolsep{5pt}}lllccc}
\toprule 
 &  &    & Section \ref{subsec:estimation_monte_carlo} (benchmark), \ref{subsec:monte_carlo_no_merger} & Section \ref{subsec:monte_carlo_small_n} & Section \ref{subsec:monte_carlo_large_firms} \\
  &  &    & Section \ref{subsec:comparative_statics_benchmark_markets} && Section \ref{subsec:comparative_statics_large_firms}\\\hline
Parameter &  &    &  &&\\
$\beta_0$  & & the coefficient of $x_{i1}$ with $x_{J1}$   & 1.0 &1.0& 1.0\\
 &  &    &&  &\\
$\beta$& & the coefficient of $x_{i2}$ with $x_{J2}$  & 0.0 &0.0& 0.0\\
 &  &    &&  &\\
$\delta$  & & the coefficient of subsidy $s_{i,j}$  & 1.0 &1.0& 1.0\\
 &  &    && & \\
$\gamma$  & & the coefficient of merger cost $C_{i,j}$  & 1.0 \text{ or } 5.0 &1.0& 1.0\\
 &  &    &&  &\\
$N$ &  &  the number of firms & 8 &2,3,4,5,6,7& 8\\
 &  &   &&  &\\
Specification &  &    &&  &\\
 &  &    &&  &\\
$\text{ton}_1$,$\text{ton}_2$ &  &  tonnage sizes & \text{Log-Normal}[2,1] &\text{Log-Normal}[2,1]&\text{Uniform}[20,80]\\
 &  &   &&  &\\
$s_{i,J}$ &  &  subsidy term & $s_{i,J}^{\text{shared}}$ or $s_{i,J}^{\text{to buyer}}$ &benchmark& benchmark\\
 &  &  subsidy amount & 1 (= dummy variable)&benchmark&benchmark\\
 &  &  subsidy threshold & 1 (i.e., 1 million tons) &benchmark&benchmark\\
 & & &&&\\
$C_{i,J}$ &  & merger cost term & $[\text{the number of firms in coalition } J \backslash i ]$ &benchmark& benchmark\\
\bottomrule 
\end{tabular}
\end{table} 

\subsection{Identification results for a single market}\label{subsec:identification_monte_carlo}

I construct simulated data based on the aforementioned data generating process and focus on integer matching outcomes.\footnote{Although the model of \cite{azevedo2018existence} allows non-integer matching allocations, I do not take into account for non-integer solutions for deriving bounds of estimators as in \cite{ciliberto2009market}. They stated that if a game does not have an equilibrium in pure strategies for some realization of the errors in one market, then they do not consider that particular realization of the errors when they construct the lower and upper bounds for simplicity. See Appendix of \cite{ciliberto2009market} for details of the Monte Carlo simulation.} Then, I construct possible pairwise inequalities as in Appendix \ref{sec:algorithm} and estimate the three parameters $\beta$, $\delta$, and $\gamma$ while fixing $\beta_0=1$ for normalization.\footnote{Unlike the empirical application in Section \ref{sec:estimation}, the evaluations based on inequality \eqref{ineq5} are not included in Monte Carlo simulations because this is an institutional-specific inequality that uses an equilibrium before and after subsidies.} To illustrate the comparable figures and the shape of the objective function, I start to discuss the results of an illustrative one of the simulated market.

\begin{figure}[htbp]
 \begin{minipage}{0.33\hsize}
  \begin{center}
   \includegraphics[width=50mm]{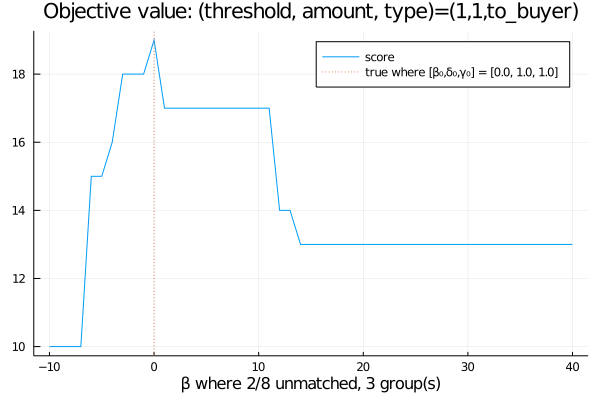}
  \end{center}
 \end{minipage}
 \begin{minipage}{0.33\hsize}
  \begin{center}
   \includegraphics[width=50mm]{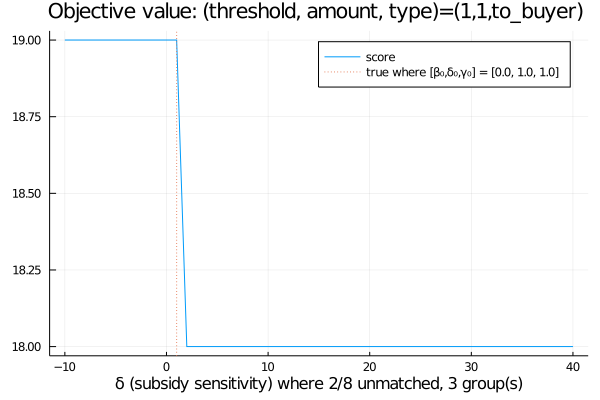}
  \end{center}
 \end{minipage}
 \begin{minipage}{0.33\hsize}
  \begin{center}
   \includegraphics[width=50mm]{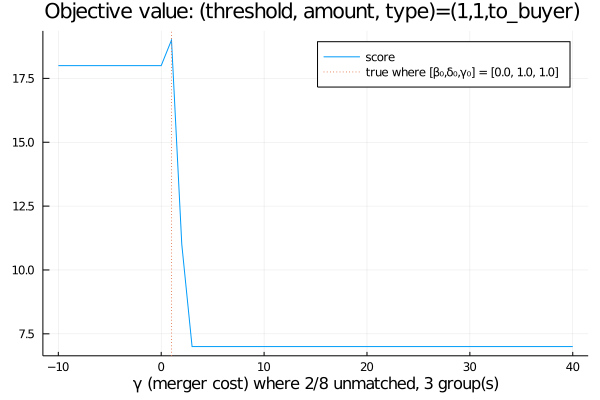}
  \end{center}
 \end{minipage}\\
 \begin{minipage}{0.33\hsize}
  \begin{center}
   \includegraphics[width=50mm]{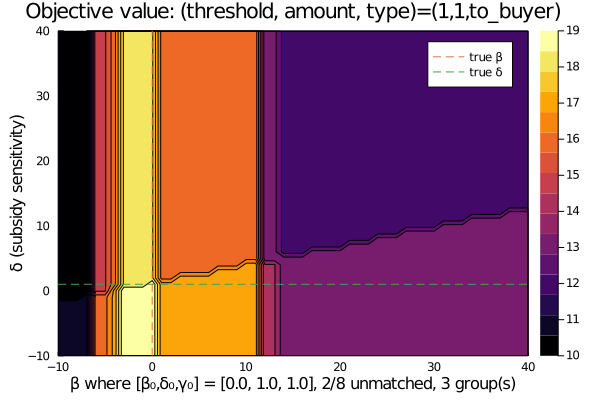}
  \end{center}
 \end{minipage}
 \begin{minipage}{0.33\hsize}
  \begin{center}
   \includegraphics[width=50mm]{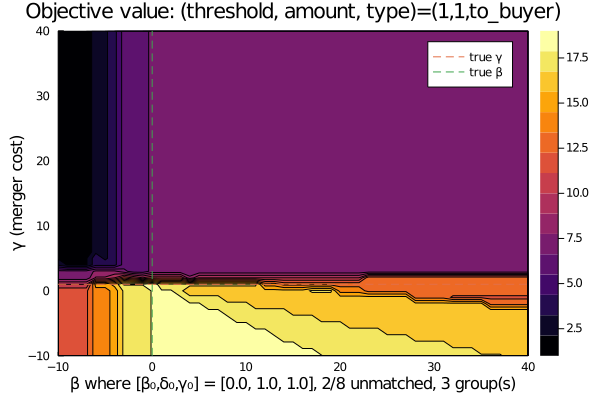}
  \end{center}
 \end{minipage}
 \begin{minipage}{0.33\hsize}
  \begin{center}
   \includegraphics[width=50mm]{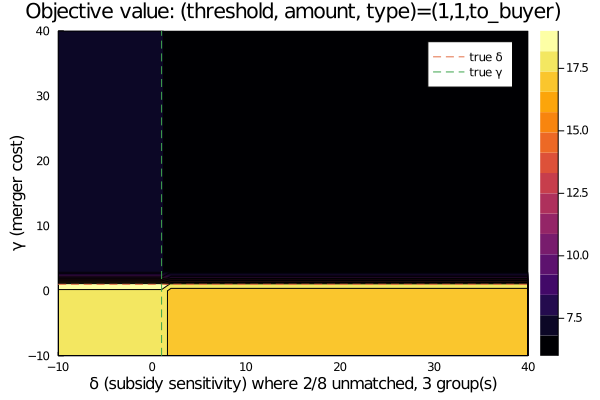}
  \end{center}
 \end{minipage}
 \caption{The maximum rank objective function across different values of a parameter fixing the other parameters to true values (one-dimensional and two-dimensional plots in the upper and lower panels). The subsidy is specified as $s_{i,J}^{\text{to buyer}}$. The matching outcome consists of three unmatched firms, a single one-to-two matching, and a single one-to-one matching. The number of inequalities is 19.}
 \label{fg:plot_beta_delta_gamma_to_buyer}
\end{figure}

First, the case of subsidy specification $s_{i,J}^{\text{to buyer}}$ is considered. The subsidy specification generates the matching outcome in which two firms stay unmatched, and three one-to-one matching occur. The number of inequalities is 19. The upper panels in Figure \ref{fg:plot_beta_delta_gamma_to_buyer} show the plots of the value of the discontinuous objective function based on one of the simulated matching data. The lower panels in Figure \ref{fg:plot_beta_delta_gamma_to_buyer} also show the contour maps of the value of the objective function. I find that parameters $\beta$ and $\gamma$ are identified. However, the coefficients of $\delta$ provide only the upper bound.

\begin{figure}[htbp]
 \begin{minipage}{0.33\hsize}
  \begin{center}
   \includegraphics[width=50mm]{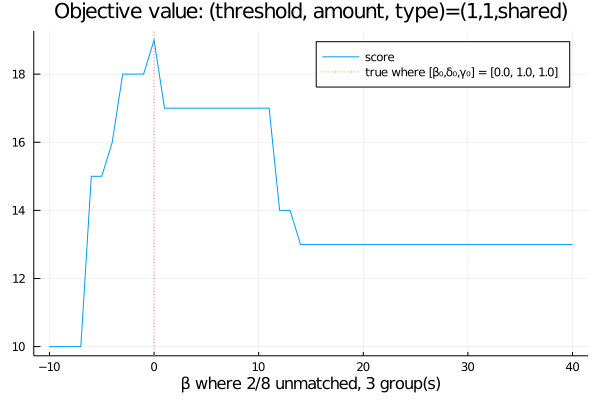}
  \end{center}
 \end{minipage}
 \begin{minipage}{0.33\hsize}
  \begin{center}
   \includegraphics[width=50mm]{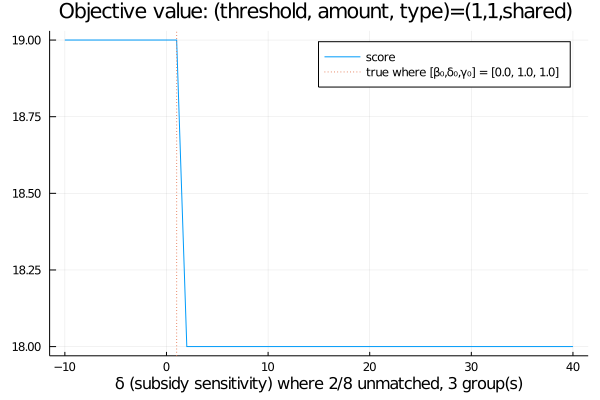}
  \end{center}
 \end{minipage}
 \begin{minipage}{0.33\hsize}
  \begin{center}
   \includegraphics[width=50mm]{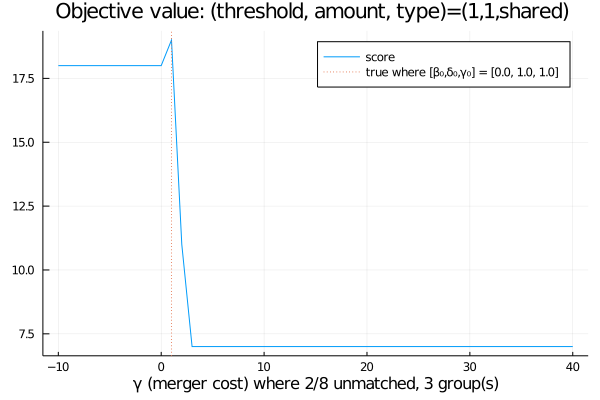}
  \end{center}
 \end{minipage}\\
 \begin{minipage}{0.33\hsize}
  \begin{center}
   \includegraphics[width=50mm]{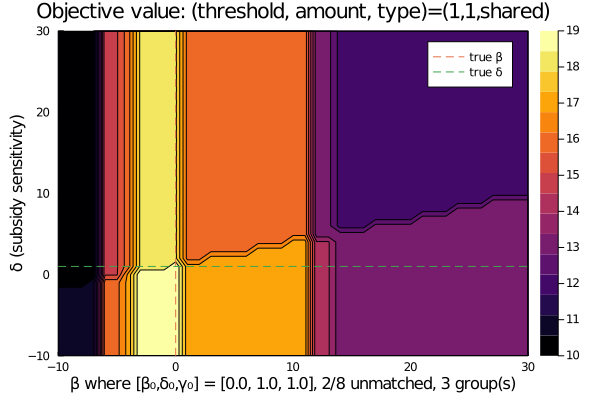}
  \end{center}
 \end{minipage}
 \begin{minipage}{0.33\hsize}
  \begin{center}
   \includegraphics[width=50mm]{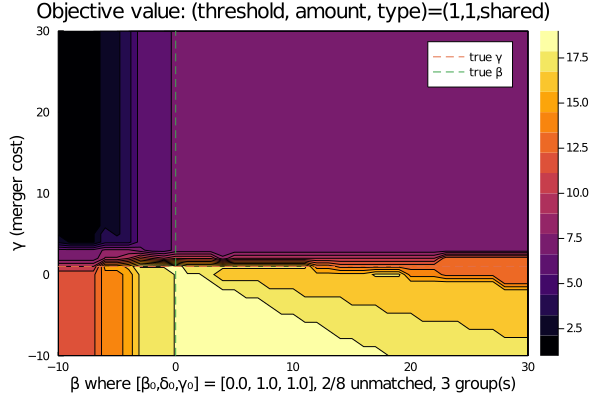}
  \end{center}
 \end{minipage}
 \begin{minipage}{0.33\hsize}
  \begin{center}
   \includegraphics[width=50mm]{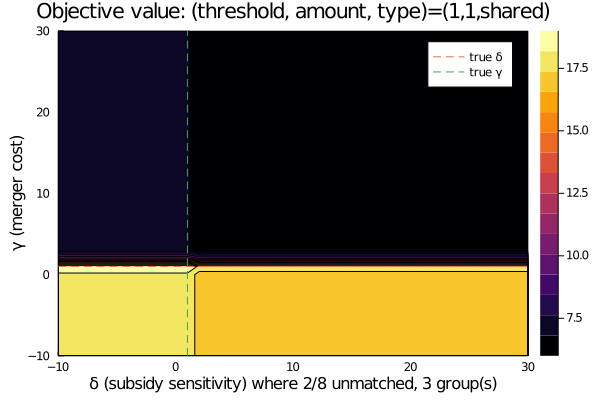}
  \end{center}
 \end{minipage}
 \caption{The maximum rank objective function across different values of a parameter fixing the other parameters to true values (one-dimensional and two-dimensional in the upper and lower panels). The subsidy is specified as $s_{i,J}^{\text{shared}}$. The matching outcome consists of four unmatched firms and two one-to-one matchings. The number of inequalities is 19. }
 \label{fg:plot_beta_delta_gamma_shared}
\end{figure}

Second, the case of subsidy specification $s_{i,J}^{\text{shared}}$ is examined. Under the same data and random variables, the subsidy specification generates the same matching outcome. The number of inequalities is 19.  
Figure \ref{fg:plot_beta_delta_gamma_shared} correspond to Figure \ref{fg:plot_beta_delta_gamma_to_buyer}. The specification does not significantly affect the shape of the objective function.

The aforementioned identification results, in particular, of $\delta$, in a small matching market are natural in the one-to-many one-sided matching model allowing complementarities. Suppose that the coefficient of the subsidy $\delta$ goes to infinity in a market consisting of some matched firms and unmatched small firms. Then, buyer firms would not enjoy the benefits of additional mergers because they already satisfy the requirement of subsidy before additional mergers. In competitive equilibrium, additional matching of a matched group and another matched group induces dividing the constant subsidy amount into more members in the new group, which must be less profitable in the joint production functions for two matched groups. On the other hand, unmatched firms also would not deviate to any matching with the other unmatched firms because the total tonnage size of any coalition does not satisfy the requirement of subsidy. Thus, the increase of $\delta$ affects the firm's matching behavior only when the subsidy can make firms satisfy the subsidy requirement.\footnote{To obtain the identified set of subsidy sensitivity $\delta$, a researcher needs the data of a variety of subsidy thresholds and amounts from many markets and, thus, excluded variables from the joint production function. The model is analogous to the sample selection model. For merger analysis in a single market, it is difficult to apply the method. 
} 

In summary, the experiments show that even in a small single market the model has identification power for $\beta$ and merger cost $\gamma$. Particularly, the lower bounds of merger cost are informative even under the small market setting and can be utilized for deriving optimal subsidy amount under the scenarios where firms are at most insensitive to subsidy and at most sensitive to merger cost. The scenarios give us the most expensive, that is, the worst case for the government expenditure in merger policy.

However, the bound at one side of subsidy sensitivity $\delta$ is impossible to obtain. It suggests that, for counterfactual simulation, the bounds of estimated subsidy sensitivity in Table \ref{tb:score_results_two_variables_main_firms_only} are not so reliable that users should not use these for counterfactual simulation as in Section \ref{sec:counter_factual}. In Section \ref{sec:estimation}, the subsidy sensitivity $\delta$ is calibrated to proceed reliably and safely.

\subsection{Estimation results for a single market}\label{subsec:estimation_monte_carlo}

\begin{figure}[htbp]
 \begin{minipage}{0.33\hsize}
  \begin{center}
   \includegraphics[width=50mm]{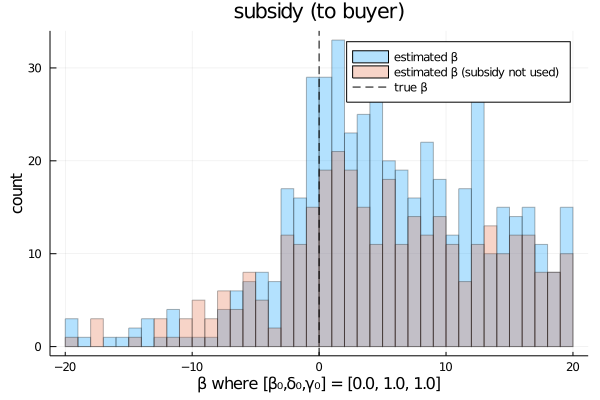}
  \end{center}
 \end{minipage}
 \begin{minipage}{0.33\hsize}
  \begin{center}
   \includegraphics[width=50mm]{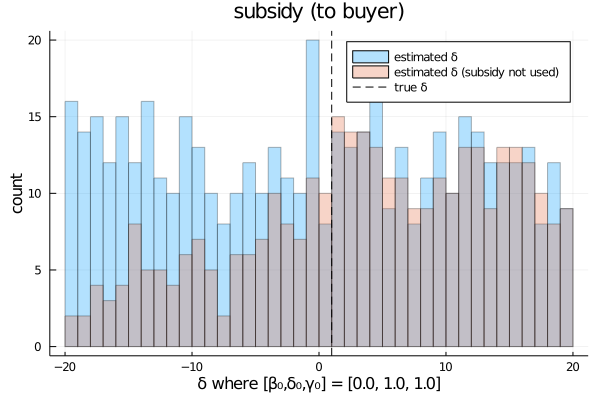}
  \end{center}
 \end{minipage}
 \begin{minipage}{0.33\hsize}
  \begin{center}
   \includegraphics[width=50mm]{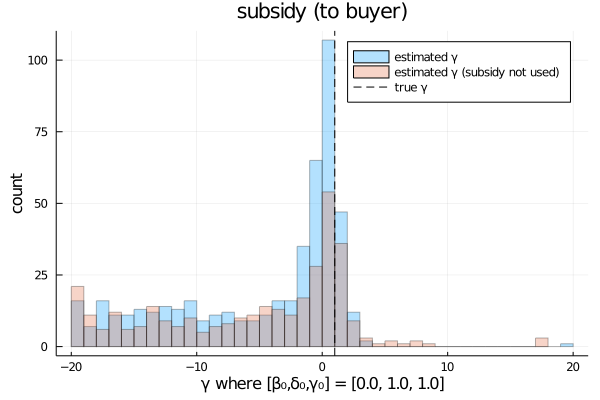}
  \end{center}
 \end{minipage}\\
 \begin{minipage}{0.33\hsize}
  \begin{center}
   \includegraphics[width=50mm]{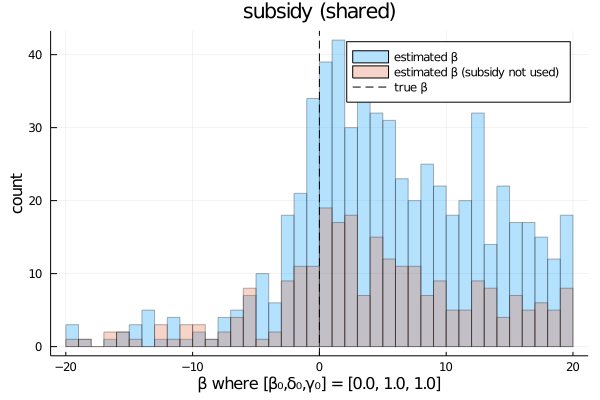}
  \end{center}
 \end{minipage}
 \begin{minipage}{0.33\hsize}
  \begin{center}
   \includegraphics[width=50mm]{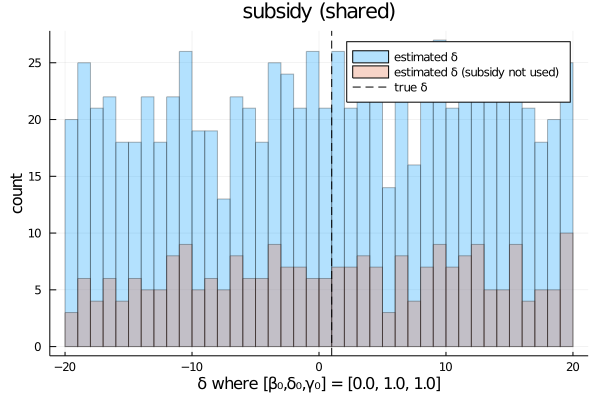}
  \end{center}
 \end{minipage}
 \begin{minipage}{0.33\hsize}
  \begin{center}
   \includegraphics[width=50mm]{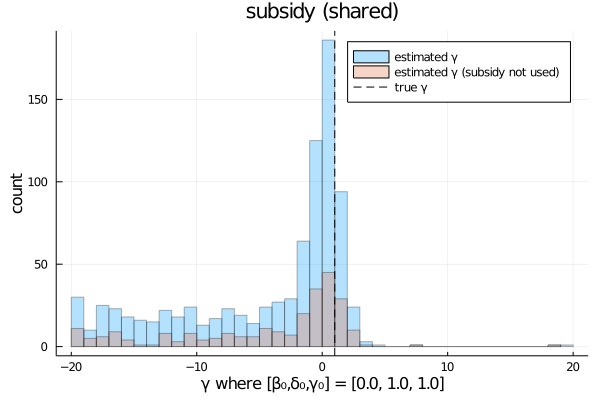}
  \end{center}
 \end{minipage}
 \caption{Distribution of the estimated parameters for 1000 simulated data dropping non-integer matching outcomes. For DE algorithm with 50 steps, the search domain is restricted to $[-20,20]$. The upper panels use $s_{i,J}^{\text{to buyer}}$ and the lower panels use $s_{i,J}^{\text{shared}}$.}
 \label{fg:histogram_beta_delta_gamma_to_buyer_shared}
\end{figure}

\begin{table}[!ht]
\caption{\textbf{Maximum score Monte Carlo results}. The true values are set at $\beta=0$ and $\gamma=1$. I calculate the median bias as $\text{Med}(\hat{\theta_s}-\theta)$ and RMSE as $\sqrt{\frac{1}{n_s}\sum_{s=1}^{n_s}(\hat{\theta_s}-\theta)^2}$ where $n_s$ is the number of samples using subsidy and $\hat{\theta}_s$ is estimated parameters for each set of simulated data $s$.}
\begin{center}
\begin{tabular}{@{\extracolsep{5pt}}ccccc}
\toprule 
Num of firms & subsidy type & parameter & median bias & RMSE \\
\midrule 
8 & to buyer & $\beta$ & 4.78 & 9.45 \\
8 & to buyer & $\gamma$ & -2.16 & 8.89 \\
8 & shared & $\beta$ & 4.99 & 9.49 \\
8 & shared & $\gamma$ & -1.84 & 8.49 \\
\bottomrule 
\end{tabular}

\end{center}
\label{tb:monte_carlo_results_bias_RMSE}
\end{table}

Keeping the same settings as in Appendix \ref{subsec:identification_monte_carlo}, I move on to investigate the properties of estimated parameters via 1000 simulations. The estimation procedure follows Appendix \ref{sec:algorithm} as in Section \ref{sec:estimation}. Figure \ref{fg:histogram_beta_delta_gamma_to_buyer_shared} reports the estimation results for $N=8$ under $s_{i,J}^{\text{to buyer}}$ and $s_{i,J}^{\text{shared}}$. For each simulation, only the random seed for generating $x_{i,1}, x_{i,2}$ and $\epsilon_{i,J}$ is changed. Recall that if the parameters are completely not-identified, the distribution of estimated parameters must appear a uniform distribution according to the property of DE algorithm, which randomly searches global maxima on $[-20,20]$. 

Figure \ref{fg:histogram_beta_delta_gamma_to_buyer_shared} reports the distribution of the estimated parameters in each simulated single market. First, I focus on the case where any group does not satisfy the subsidy requirement (orange histograms). This case naturally occurs if the merger cost is low. The orange histogram in the upper middle panel shows that estimated $\delta$ has an ambiguous lower bound but no upper bound, which implies that the coefficients of dummy variables will be only identified at the lower bound. Specifically, increasing subsidy sensitivity changes the matching outcome, then the evaluation of pairwise inequalities varies, and the variation is detected correctly for parameter search. The orange histograms in the other panels show results similar to the blue histograms.

Second, I move on to discuss the case that at least one group satisfies the subsidy requirement (blue histogram).
The bounds are identified for $\beta$ and $\gamma$ but not for $\delta$. Because of the DE algorithm, not-identified parameters such as $\delta$ take arbitrary values uniformly on $[-20,20]$. Also, estimated $\beta$ might fall in the tail because the upper bound is hard to trace out in the specific data. These results are consistent with the identification results in Figures \ref{fg:plot_beta_delta_gamma_to_buyer} and \ref{fg:plot_beta_delta_gamma_shared}. 

Table \ref{tb:monte_carlo_results_bias_RMSE} reports the finite-sample median bias and RMSE for simulated data using subsidy scheme (blue samples in Figure \ref{fg:histogram_beta_delta_gamma_to_buyer_shared}). As the identification results in Section \ref{subsec:identification_monte_carlo} suggested, I find that $\beta$ is upward biased and $\gamma$ is downward biased because some parameters for some simulated data are not identified and take extreme values. This suggests that point-identification may be misleading at least in a small market setting ($N=8$), although unidentified simulated data is easily detected. In addition, bootstrapped confidence intervals might be affected by non-identified cases via sampling, so that the confidence intervals can be wider.\footnote{As \cite{fox2013aej} suggested, if the correctly-specified equilibrium transfer data are available, the estimator would be much more accurate even in a single market setting.} 

In conclusion, I find that even in a small matching market ($N=8$) where there are a few mergers and at most 30 pairwise inequalities are observed, the proposed estimator can estimate $\beta$ and merger cost $\gamma$ with bias.\footnote{It seems desirable to construct confidence region as \cite{romano2008inference}. However, this fails in the setting of this study. To construct confidence regions, the researcher must check whether the objective function for each subsample has the common achievable maximum score. In the specification focusing on a specific subsidy mechanism, the achievable maximum score depends on subsamples so that confidence regions cannot be constructed as shown in Appendix \ref{subsec:romano_shaikh_not_works}.} However, I find that the coefficient of subsidy sensitivity is not identified in a small market setting. The estimation in a large market setting as the empirical application ($N=118$) uses more than 10 thousand pairwise inequalities, which provides much information to improve the identification power. However, Appendices \ref{subsec:identification_monte_carlo} and \ref{subsec:estimation_monte_carlo} suggest me calibrate the parameter $\delta$ on several candidate values as a safe and reliable way.

\subsection{Identification power with a no-merger market}\label{subsec:monte_carlo_no_merger}

The antitrust agency may be interested in estimating the merger fixed costs prior to mergers. In particular, the lower bound of a merger cost would be informative as a simple device to detect the possibility of mergers within a market. Also, it provides a comparable single index measure across markets or industries. The proposed method uses information from unmatched firms and inequalities \eqref{ineq4} induced by the incentive compatibility condition in \eqref{def:competitive_equilibrium}. This subsection investigates the identification power of $\beta$ and merger cost $\gamma$ when the market does not observe any mergers. 

For consistency, I use the same data generating process with $N=8$ in Section \ref{subsec:identification_monte_carlo} and \ref{subsec:estimation_monte_carlo}, except that the value of merger cost $\gamma$ to generate no-merger markets, that is, $\gamma=5$. In this experiment, I want to trace out the lower bound of $\gamma$ because no merger would occur if $\gamma$ went to infinity from above a certain threshold. Specifically, given a certain threshold $\bar{\gamma}$, the matching outcomes under any $\gamma \ge \bar{\gamma}$ are observationally equivalent.

\begin{figure}[htbp]
 \begin{minipage}{0.30\hsize}
  \begin{center}
   \includegraphics[width=50mm]{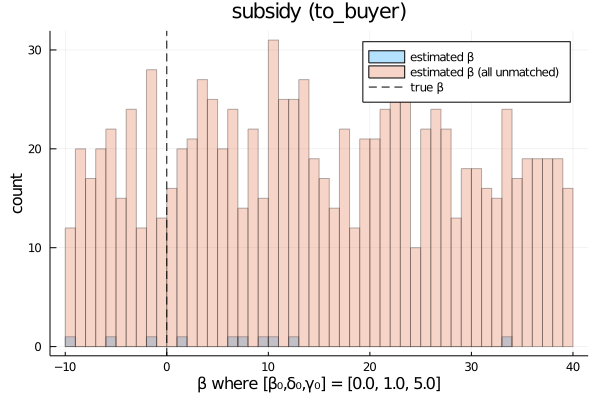}
  \end{center}
 \end{minipage}
 \begin{minipage}{0.30\hsize}
  \begin{center}
   \includegraphics[width=50mm]{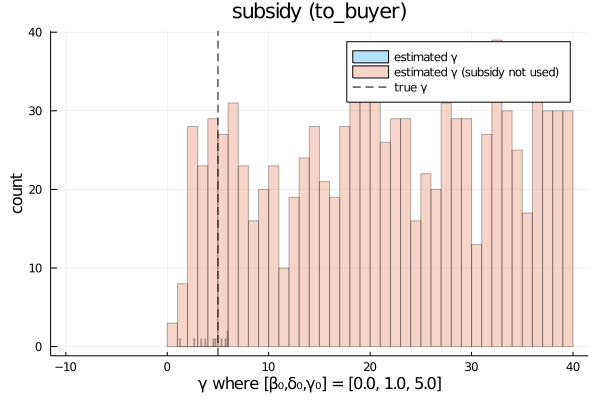}
  \end{center}
 \end{minipage}
 \begin{minipage}{0.30\hsize}
  \begin{center}
   \includegraphics[width=50mm]{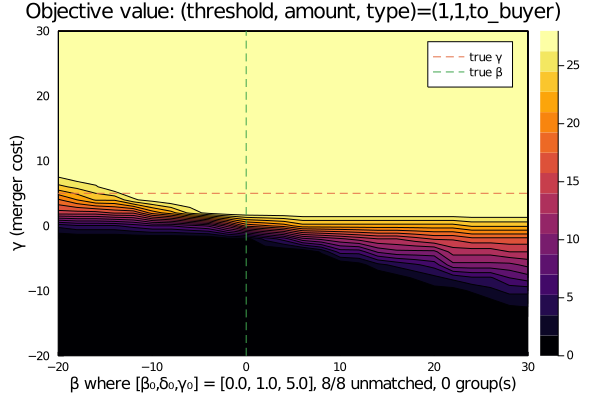}
  \end{center}
 \end{minipage}
 \caption{Distribution of the estimated parameters (left and middle panels) for 1000 simulated market data dropping non-integer matching outcomes. For DE algorithm with 50 steps, the search domain is restricted to $[-10,40]$. Contour map of the objective function for chosen simulated data (right panel) where the search domain is restricted to $[-20,30]$. The panels use $s_{i,J}^{\text{to buyer}}$ and I omit the cases of $s_{i,J}^{\text{shared}}$ because the subsidy has no effect in no-merger markets so that it generates the same plots. }
 \label{fg:histogram_beta_gamma_to_buyer_shared_no_merger}
\end{figure}

Figure \ref{fg:histogram_beta_gamma_to_buyer_shared_no_merger} reports that the lower bound of merger cost $\gamma$ is detected but $\beta$ is not identified. This is because $\beta$ can take an arbitrary value given higher $\gamma$ in the evaluation of the objective function. Interestingly, the right panel shows that if matching shows a negative assortative feature in some parameter ($\beta<0$), the lower bound of the merger cost would shrink. In addition, as expected, the identified set of $\gamma$ is unbounded on the positive side. The experiment shows that even a single ``no-merger" market can provide an informative lower bound of merger cost $\gamma$ without any additional restrictions.

The results are similar to the identification bounds of some weak equilibrium concepts such as the set of rationalizable strategies in a static complete information game \citep{aradillas2008identification} and supermodular game \citep{uetake2013estimating}. In particular, \cite{aradillas2008identification} showed that one side of the identification set of a rationalizable strategy in a binary entry game with complete information is unbounded.



\subsection{Identification power with a small $N$ market}\label{subsec:monte_carlo_small_n}

The antitrust agency may also have some practical interest in applying the proposed method to a single duopoly or oligopoly market (i.e., $N$ is very small). In this case, the number of pairwise inequalities is so much smaller than the case of $N=8$ that the identification power would decrease. For illustration, one of the simulated markets as in Section \ref{subsec:identification_monte_carlo} is chosen to investigate the limit of the applicability.

\begin{figure}[htbp]
 \begin{minipage}{0.30\hsize}
  \begin{center}
   \includegraphics[width=50mm]{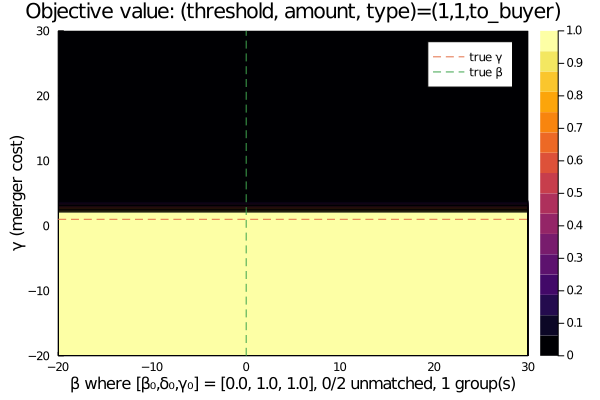}
  \end{center}
 \end{minipage}
 \begin{minipage}{0.30\hsize}
  \begin{center}
   \includegraphics[width=50mm]{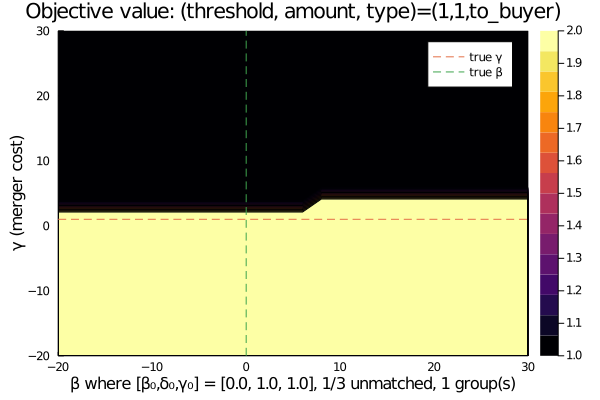}
  \end{center}
 \end{minipage}
 \begin{minipage}{0.30\hsize}
  \begin{center}
   \includegraphics[width=50mm]{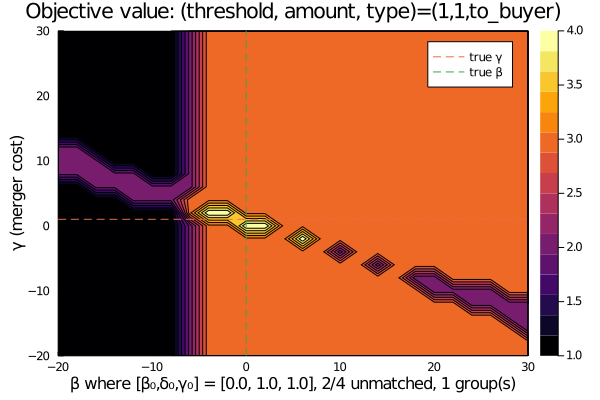}
  \end{center}
 \end{minipage}\\
 \begin{minipage}{0.30\hsize}
  \begin{center}
   \includegraphics[width=50mm]{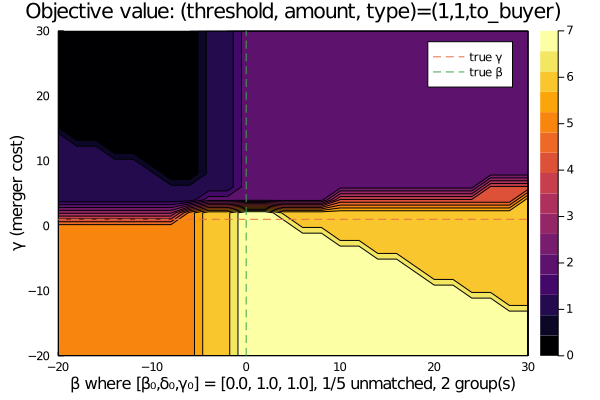}
  \end{center}
 \end{minipage}
 \begin{minipage}{0.30\hsize}
  \begin{center}
   \includegraphics[width=50mm]{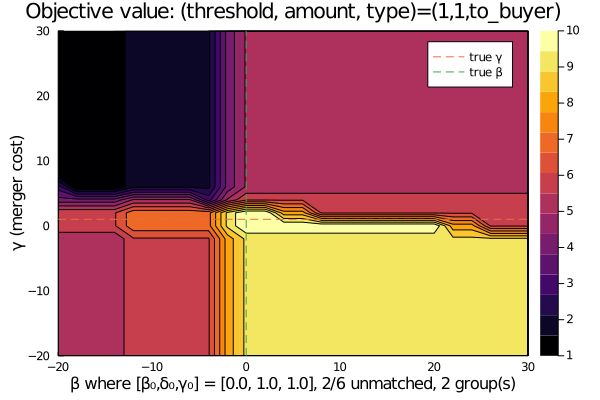}
  \end{center}
 \end{minipage}
 \begin{minipage}{0.30\hsize}
  \begin{center}
   \includegraphics[width=50mm]{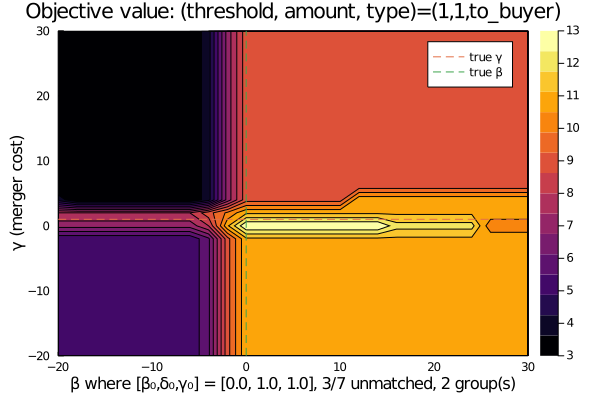}
  \end{center}
 \end{minipage}
 \caption{Contour maps of the objective function for the chosen simulated data under different $N$ where the search domain is restricted to $[-20,30]$. The panels use $s_{i,J}^{\text{to buyer}}$ and I omit the cases of $s_{i,J}^{\text{shared}}$ because the subsidy has no effect in no-merger markets so that it generates the same plots. }
 \label{fg:histogram_beta_gamma_to_buyer_shared_different_N}
\end{figure}

Figure \ref{fg:histogram_beta_gamma_to_buyer_shared_different_N} shows how the identification power decreases as the number of firms $N$ decreases. If $N=2$ and $3$, only one and two pairwise inequalities can be constructed, which then cannot obtain a bounded identification set. The result is robust in other simulated markets. However, even a market with $N=4$ can generate identified sets for $\beta$ and $\gamma$ using only four pairwise inequalities in some markets. Note that the bounds shown in Figure \ref{fg:histogram_beta_gamma_to_buyer_shared_different_N} are generated in one of the simulated markets and the bounds may not be obtained for other large simulated markets as in the case of $N=5$. However, the result is helpful that the limit of identification possibility is $N=4$.

\subsection{Identification and Estimation results for a single market with only large firms}\label{subsec:monte_carlo_large_firms}

So far, I simulate the market where there are two or three large firms and six or five mid-sized or small-sized firms. Section \ref{subsec:monte_carlo_large_firms} considers the market where $N=8$ firms are relatively large firms. The market structure is different from the market in Sections \ref{subsec:estimation_two_variables_main_firms_only} and \ref{sec:counter_factual}. The only modification from the previous sections is that two types of tonnage covariates $\text{ton}_{i1},\text{ton}_{i2}$ are drawn from i.i.d.  $\text{Uniform}[20,80]$. 

\begin{figure}[htbp]
 \begin{minipage}{0.33\hsize}
  \begin{center}
   \includegraphics[width=50mm]{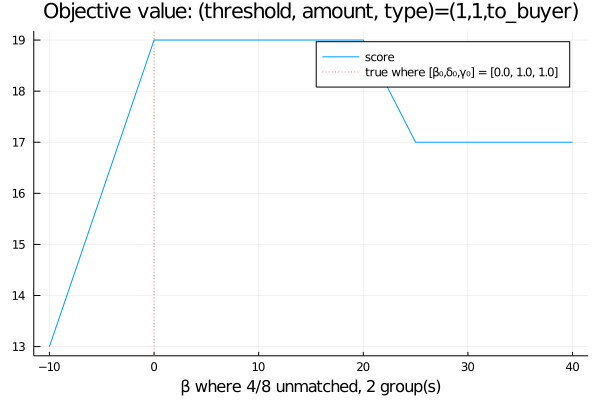}
  \end{center}
 \end{minipage}
 \begin{minipage}{0.33\hsize}
  \begin{center}
   \includegraphics[width=50mm]{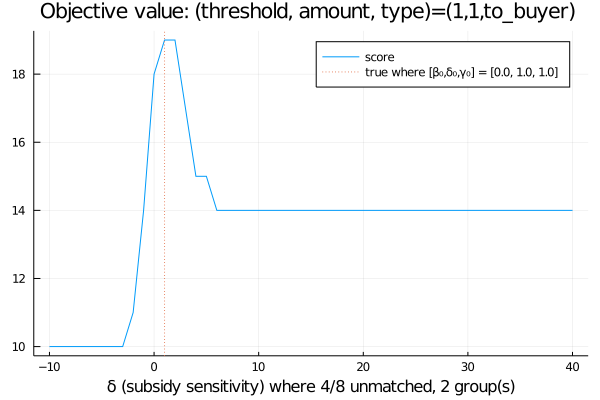}
  \end{center}
 \end{minipage}
 \begin{minipage}{0.33\hsize}
  \begin{center}
   \includegraphics[width=50mm]{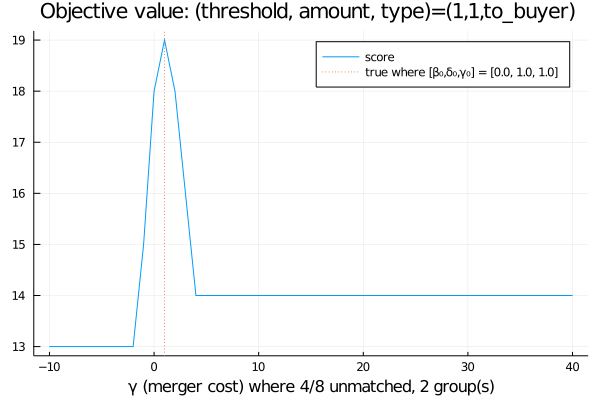}
  \end{center}
 \end{minipage}\\
 \begin{minipage}{0.33\hsize}
  \begin{center}
   \includegraphics[width=50mm]{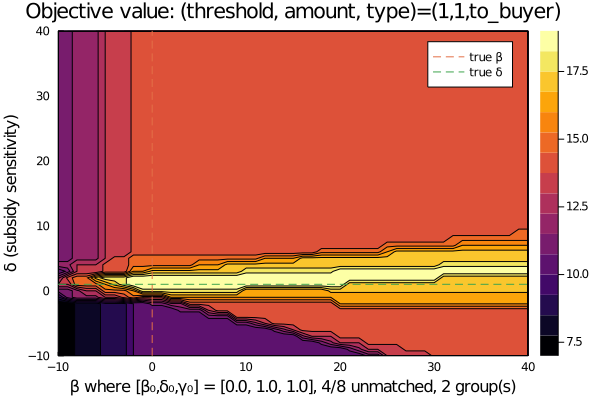}
  \end{center}
 \end{minipage}
 \begin{minipage}{0.33\hsize}
  \begin{center}
   \includegraphics[width=50mm]{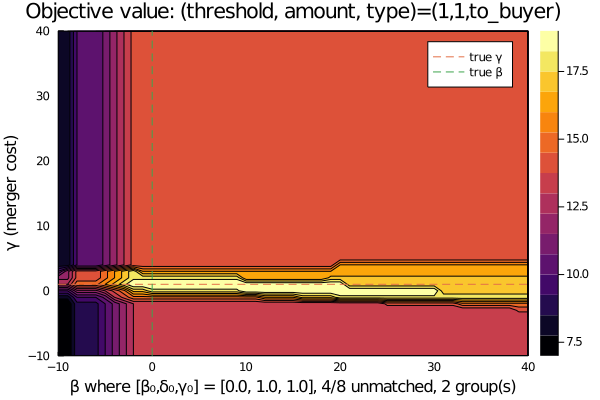}
  \end{center}
 \end{minipage}
 \begin{minipage}{0.33\hsize}
  \begin{center}
   \includegraphics[width=50mm]{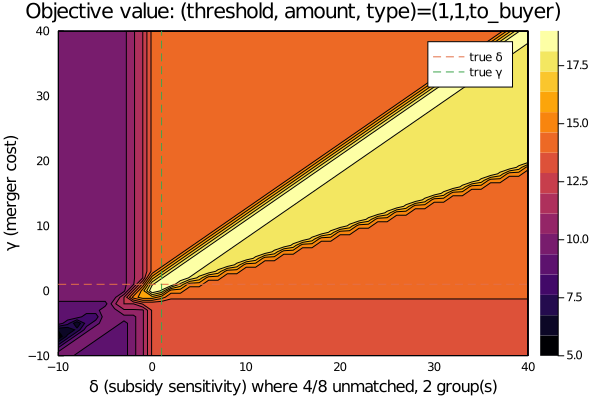}
  \end{center}
 \end{minipage}
 \caption{The maximum rank objective function across different values of a parameter fixing the other parameters to true values (one-dimensional and two-dimensional in the upper and lower panels, respectively). The subsidy is specified as $s_{i,J}^{\text{to buyer}}$. The matching outcome consists of four unmatched firms and two one-to-one matchings. The number of inequalities is 19. }
 \label{fg:plot_beta_delta_gamma_to_buyer_large_firms}
\end{figure}

\begin{figure}[htbp]
 \begin{minipage}{0.33\hsize}
  \begin{center}
   \includegraphics[width=50mm]{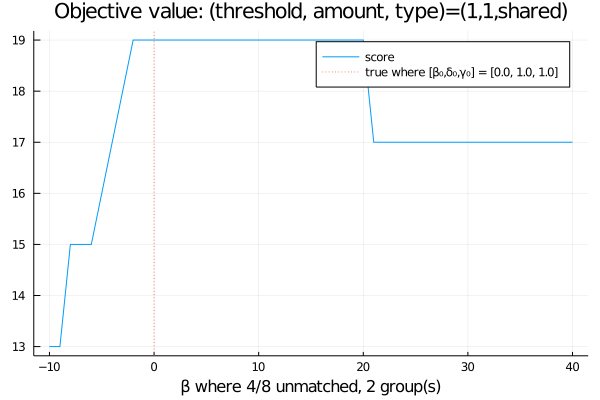}
  \end{center}
 \end{minipage}
 \begin{minipage}{0.33\hsize}
  \begin{center}
   \includegraphics[width=50mm]{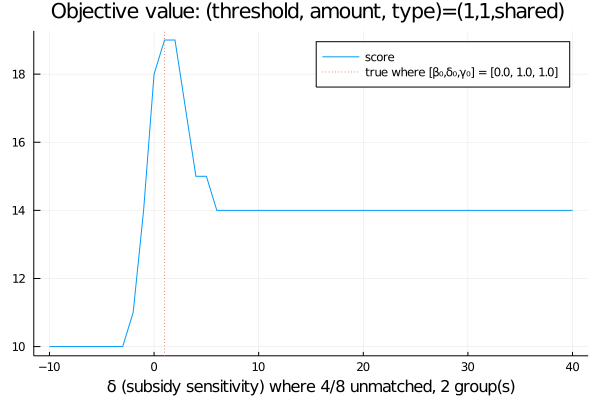}
  \end{center}
 \end{minipage}
 \begin{minipage}{0.33\hsize}
  \begin{center}
   \includegraphics[width=50mm]{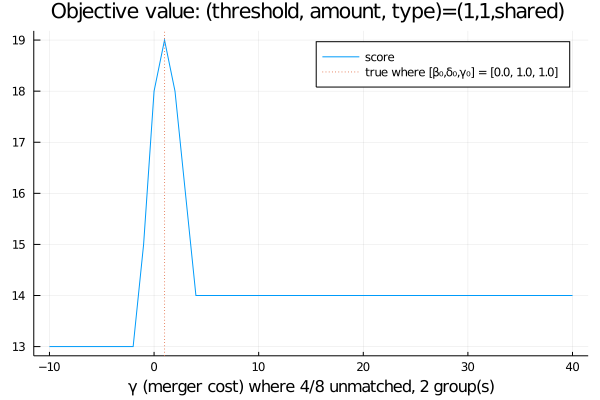}
  \end{center}
 \end{minipage}\\
 \begin{minipage}{0.33\hsize}
  \begin{center}
   \includegraphics[width=50mm]{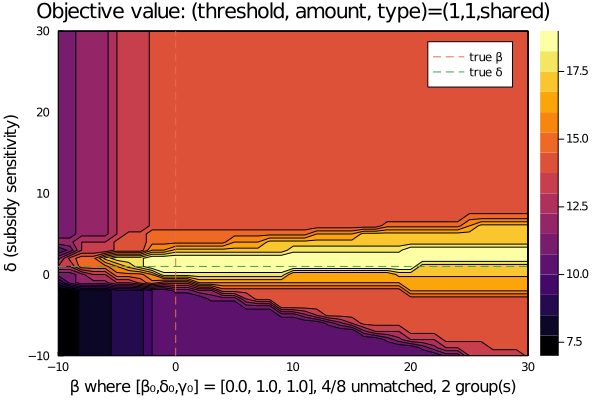}
  \end{center}
 \end{minipage}
 \begin{minipage}{0.33\hsize}
  \begin{center}
   \includegraphics[width=50mm]{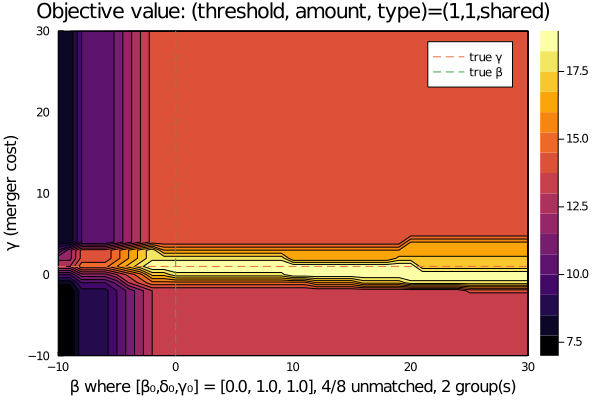}
  \end{center}
 \end{minipage}
 \begin{minipage}{0.33\hsize}
  \begin{center}
   \includegraphics[width=50mm]{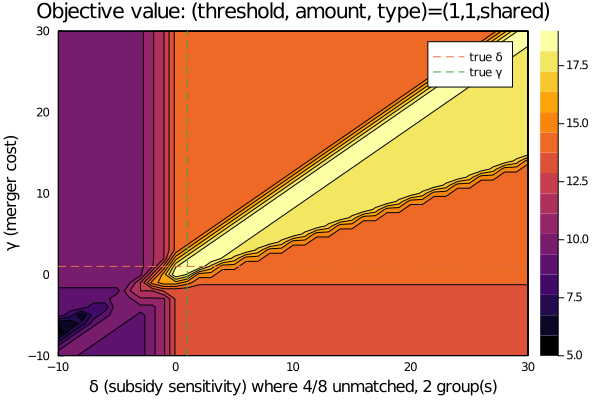}
  \end{center}
 \end{minipage}
 \caption{The maximum rank objective function across different values of a parameter fixing the other parameters to true values (one-dimensional and two-dimensional in the upper and lower panels, respectively). The subsidy is specified as $s_{i,J}^{\text{shared}}$. The matching outcome consists of four unmatched firms and two one-to-one matchings. The number of inequalities is 19. }
 \label{fg:plot_beta_delta_gamma_shared_large_firms}
\end{figure}

\begin{figure}[htbp]
 \begin{minipage}{0.33\hsize}
  \begin{center}
   \includegraphics[width=50mm]{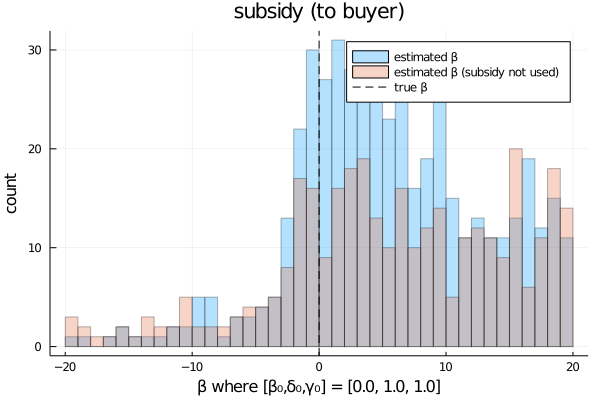}
  \end{center}
 \end{minipage}
 \begin{minipage}{0.33\hsize}
  \begin{center}
   \includegraphics[width=50mm]{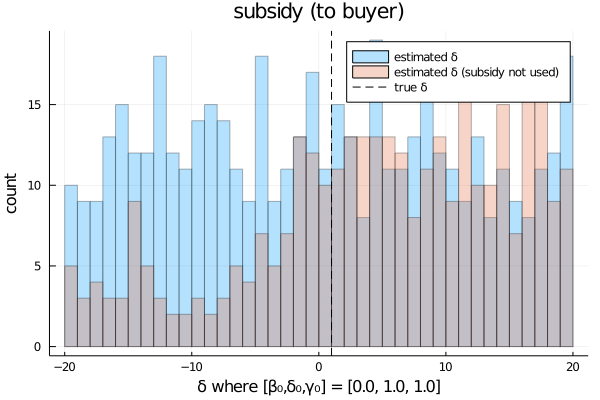}
  \end{center}
 \end{minipage}
 \begin{minipage}{0.33\hsize}
  \begin{center}
   \includegraphics[width=50mm]{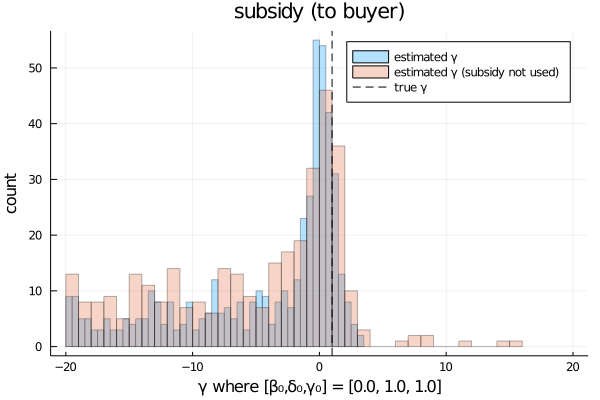}
  \end{center}
 \end{minipage}
 \caption{Distribution of the estimated parameters for 1000 simulated data dropping non-integer matching outcomes. For a DE algorithm with 50 steps, the search domain is restricted to $[-20,20]$. The upper panels use $s_{i,J}^{\text{to buyer}}$.}
 \label{fg:histogram_beta_delta_gamma_to_buyer_shared_large_firms}
\end{figure}

Figures \ref{fg:plot_beta_delta_gamma_to_buyer_large_firms} and \ref{fg:plot_beta_delta_gamma_shared_large_firms} show the same result, which implies that the subsidy specification does not affect the matching outcome in the market where only large firms exist. In addition, the identified regions are wider than the regions in Figures \ref{fg:plot_beta_delta_gamma_to_buyer} and \ref{fg:plot_beta_delta_gamma_shared}, whereas $\delta$ can be identified for some markets. However, Figure \ref{fg:histogram_beta_delta_gamma_to_buyer_shared_large_firms} shows the difficulty in identification of the upper bound of $\delta$. Finally, identification of merger cost $\gamma$ is not affected by the subsidy specification.

In conclusion, under some market situations, where only large firms exist, subsidy sensitivity $\delta$ can be point-identified instead of the loss of estimation accuracy of $\beta$. In the main analysis in Section \ref{sec:estimation}, I take a calibration approach for $\delta$ to avoid the possible nonidentification problem on one side.

\newpage
\section{Comparative statics}\label{sec:comparative_statics}

\subsection{Equilibrium properties in a benchmark market }\label{subsec:comparative_statics_benchmark_markets}

The proposed one-to-many matching model with complementarities is a tractable extension of \cite{azevedo2018existence}. However, the simulated outcome has not been investigated yet in the literature. The purpose of this subsection is not to give definitive answers regarding the implications of mergers and antitrust policies in specific industries, but rather to show that this type of one-sided one-to-many matching model is feasible and useful. First, I examine the equilibrium numbers of groups and unmatched firms by using simulated data, which support the validity of the prediction of matching outcomes at the aggregated market level. Second, I explore what mergers would occur in the simulated data under different parameters to understand merger configurations, which is an open question in the literature. Third, the government expenditure minimization problem is examined to check the optimality of a merger-induced subsidy, which validates the counterfactual simulations important for policymakers.

I compute the equilibrium for a base case parameter vector whose values are summarized in Table \ref{tb:monte_carlo_parameters_list}. Although the choice of parameters and specification is relatively arbitrary, it was made to illustrate three factors different from the computational experiments of \cite{gowrisankaran1999dynamic} in mind. First, the matching model does not incorporate the dynamic effects and welfare measures such as consumer welfare. Instead, the matching model focuses only on merger incentives in a flexible way. Second, I model matching production function with complementarities, which may induce multiple one-to-many merger matchings, whereas \cite{gowrisankaran1999dynamic} assumed a constant-returns-to-scale production technology and fixes the number of firms to a maximum of six to illustrate an industry without many mergers. Third, I introduce an additive merger fixed cost depending on the number of coalition members, whereas \cite{gowrisankaran1999dynamic} modeled fixed costs such as entry costs in a much more complex way with a sequential multi-stage dynamic optimization problem. 

\subsubsection{The equilibrium numbers of groups and unmatched firms}

\begin{figure}[htbp]
 \begin{minipage}{0.33\hsize}
  \begin{center}
   \includegraphics[width=50mm]{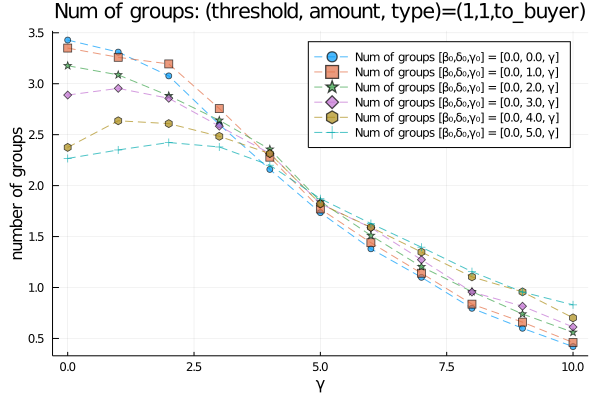}
  \end{center}
 \end{minipage}
 \begin{minipage}{0.33\hsize}
  \begin{center}
   \includegraphics[width=50mm]{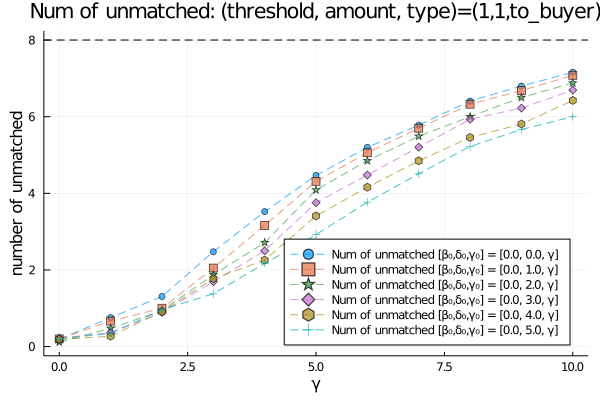}
  \end{center}
 \end{minipage}
 \begin{minipage}{0.33\hsize}
  \begin{center}
   \includegraphics[width=50mm]{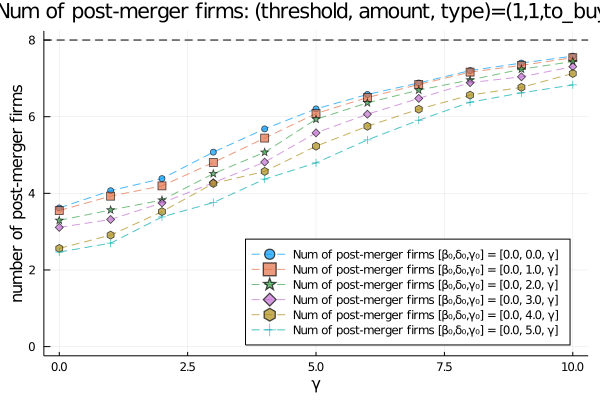}
  \end{center}
 \end{minipage}\\
 \begin{minipage}{0.33\hsize}
  \begin{center}
   \includegraphics[width=50mm]{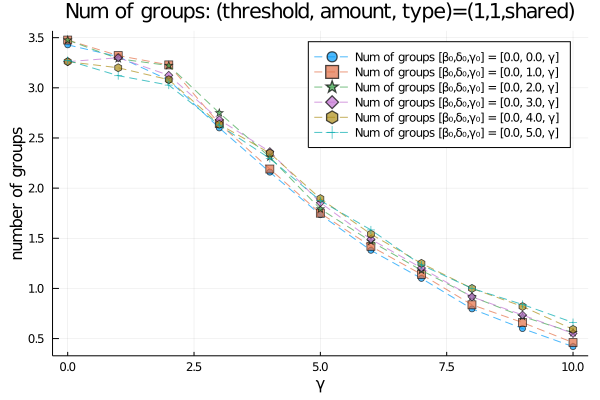}
  \end{center}
 \end{minipage}
 \begin{minipage}{0.33\hsize}
  \begin{center}
   \includegraphics[width=50mm]{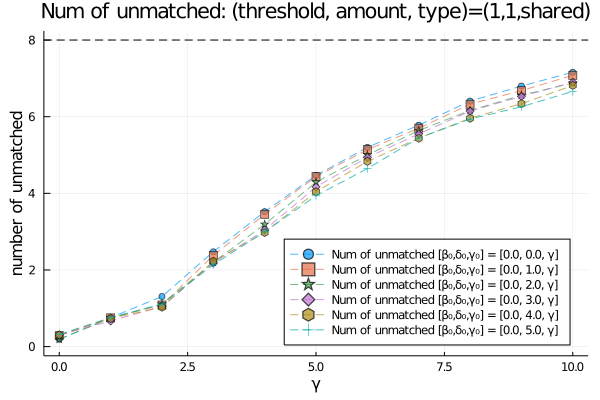}
  \end{center}
 \end{minipage}
 \begin{minipage}{0.33\hsize}
  \begin{center}
   \includegraphics[width=50mm]{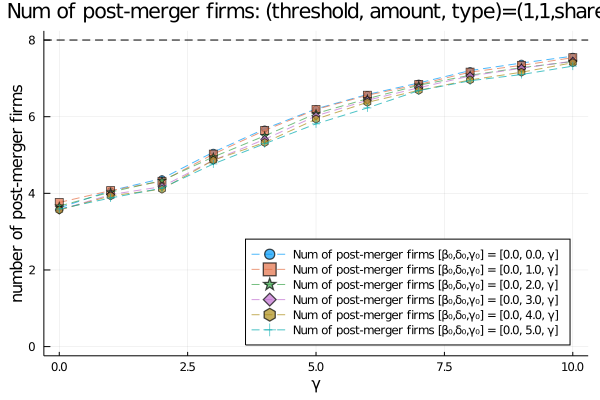}
  \end{center}
 \end{minipage}
 \caption{The equilibrium numbers of groups and unmatched firms. For each parameter pair, I generate 50 simulated data and calculate the average numbers of groups and unmatched firms by using integer matching outcomes. The number of post-merger firms in the right panels includes coalitional groups and unmatched individual firms.}
 \label{fg:plot_num_of_post_merger_firms}
\end{figure}

First, the equilibrium numbers of groups and unmatched firms are examined by using simulated data generated in Section \ref{sec:monte_carlo}. For tractability, I follow the benchmark setup in Table \ref{tb:monte_carlo_parameters_list} and fix $N=8$ with $\beta_1=1$ and $\beta_2=0$, and then change $\delta$, $\gamma$, and $s_{i,J}$. Also, the subsidy amount $M$ is fixed as 1 and subsidy threshold as 1 (i.e., 1 million tons). For each parameter pair, I generate 50 simulated data via equation \eqref{eq:LP} and compute the numbers of groups and unmatched firms on average using integer equilibrium matching data. Table \ref{fg:plot_num_of_post_merger_firms} reports the results under two subsidy specifications. In the left and middle panels, as merger cost $\gamma$ increases, the number of groups decreases but that of unmatched firms increases. These numbers are shifted downward by larger subsidy sensitivity $\delta$, because forming larger groups is easier. Second, the effect of subsidy $s_{i,J}^{\text{shared}}$ on matching outcomes is milder than that of subsidy $s_{i,J}^{\text{to buyer}}$. Finally, the right panels report the equilibrium number of post-merger firms. Without subsidy, two to four groups would appear in the market. On the other hand, if the merger cost $\gamma$ is larger than eight, no endogenous mergers would be observed, which captures the market before introducing the subsidy scheme.

Table \ref{fg:plot_num_of_post_merger_firms} provides a different perspective of blocking mergers at a certain threshold number of firms by antitrust authorities. As \cite{igami2019mergers} mentioned, the Federal Trade Commission (FTC) reports that in merger enforcement concerning high-tech markets between 1996 and 2011, no merger was blocked until the number of ``significant competitors" reached three. This implies that endogenous mergers would induce monopolization if the FTC had not blocked any mergers. On the other hand, as in Section \ref{sec:data_background}, the Japanese government expected that the subsidy scheme would result in the endogenous matching outcome consisting of at least three and a maximum of six groups. Remarkably, the scenario in mind is consistent in the right panels under approximately $\gamma \in [1,6]$ where half of the pre-merger firms are consolidated or merged. This is interpreted as that the government would implicitly consider merger fixed cost to plan and implement the subsidy laws. In Section \ref{sec:counter_factual}, the proposed method is applied to this study's data to quantitatively investigate the anecdotal evidence.

As another useful device for antitrust policy, the proposed method can detect the bound of merger cost $\gamma$ if the present market does not observe any horizontal mergers, that is, it observes only unmatched firms. Given the unique feature of the proposed method using unmatched firms, users can extract useful information from such a market situation. For example, as shown in the upper right panel in Figure \ref{fg:plot_num_of_post_merger_firms}, if the market did not observe any merger out of eight firms, the merger cost would lie in $[8,10]$. Considering that a merger is a rare event, the usefulness of the proposed method is substantial.

\subsubsection{What kind of merger would occur}

Second, this study explores what types of firms would be buyers, sellers, and unmatched and interpret why the merger would or would not occur under certain parameters through numerical comparative statistics. The purpose of this subsection is not to give aggregated implications from thousands of simulated markets but to show the applicability to a specific single industry and market of policy interest. The unique feature of this coalitional model is that it allows one-sidedness, one-to-many features, and complementarities without any restriction on the orders of merger proposal steps. In addition, the coalitional model in this study enables us to illustrate not only aggregate market-level features such as the equilibrium number of groups but also group-level features such as the number and composition of group members.

A few studies have explored theoretical properties and equilibrium merger configurations about why and how firms merge. \cite{qiu2007merger} studied an equilibrium merger configuration under a deterministic dynamic merger game setting under more restricted assumptions on the number of firms and competitions. \cite{gowrisankaran2004mergers} studied long-run equilibrium dynamics and explored to what extent an industry in which mergers are feasible tend toward monopoly. Recently, \cite{hollenbeck2020horizontal} summarized equilibrium properties with his richer model with computational experiments. The following results of the coalitional merger model complement their findings.

\begin{figure}[htbp]
 \begin{minipage}{0.33\hsize}
  \begin{center}
   \includegraphics[width=50mm]{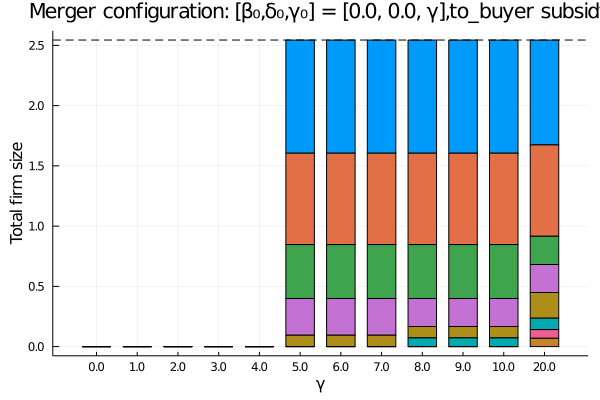}
  \end{center}
 \end{minipage}
 \begin{minipage}{0.33\hsize}
  \begin{center}
   \includegraphics[width=50mm]{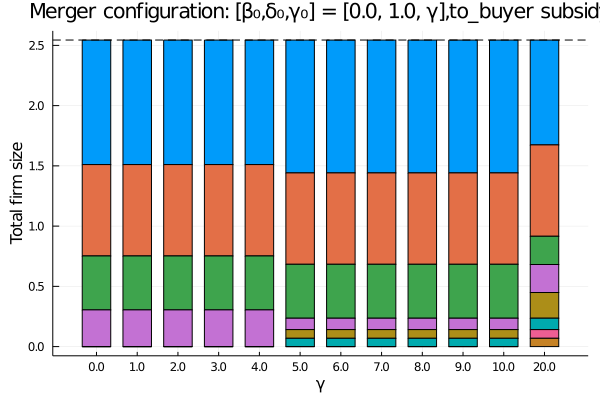}
  \end{center}
 \end{minipage}
 \begin{minipage}{0.33\hsize}
  \begin{center}
   \includegraphics[width=50mm]{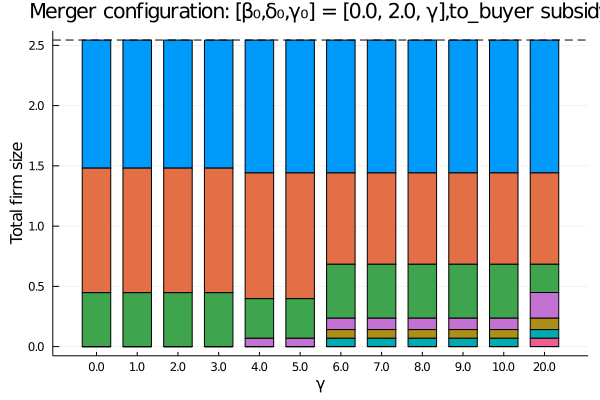}
  \end{center}
 \end{minipage}\\
 \begin{minipage}{0.33\hsize}
  \begin{center}
   \includegraphics[width=50mm]{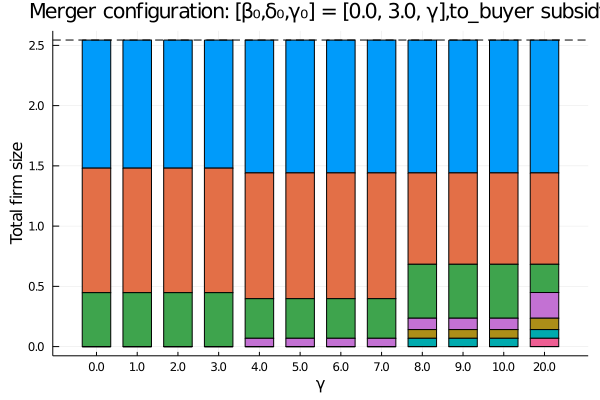}
  \end{center}
 \end{minipage}
 \begin{minipage}{0.33\hsize}
  \begin{center}
   \includegraphics[width=50mm]{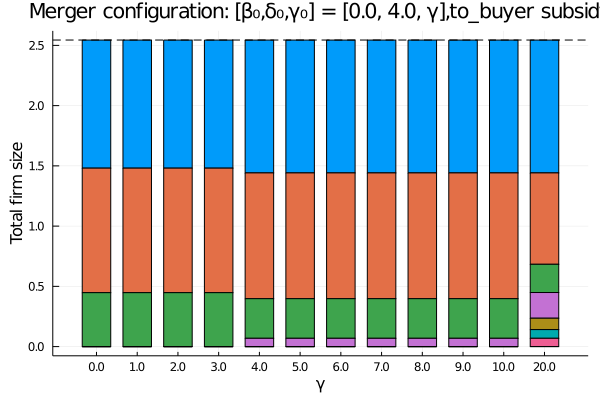}
  \end{center}
 \end{minipage}
 \begin{minipage}{0.33\hsize}
  \begin{center}
   \includegraphics[width=50mm]{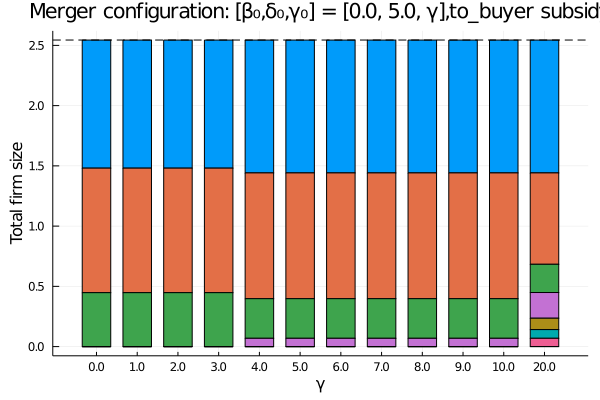}
  \end{center}
 \end{minipage}
 \caption{The merger configuration under different parameters. For each parameter pair, I generate 50 simulated data and pick up one representative data that generate an integer equilibrium for almost all parameters. This study uses subsidy specification $s_{i,J}^{\text{to buyer}}$. Some stacked bars in the upper left panel are missing because the matching outcomes under the parameters are non-integers. The cases should be treated as probabilistic allocations.}
 \label{fg:stacked_plot_merger_configuration}
\end{figure}

I pick up the representative one of the merger configurations from 50 simulated data because it is difficult to generate an integer matching outcome for all parameters given fixed data. The chosen simulated data have approximately 2.5 million tons totally in the market and eight firms are classified into two large firms, three mid-sized firms, and three small-sized firms. Figure \ref{fg:stacked_plot_merger_configuration} depicts how and what types of mergers would occur under different parameters. Note that colored firm identities do not give any sense and the color captures only the configuration process with firm size in the model. 

I focus on the features of merger configuration because the previous subsection already investigated the number of post-merger firms at the market level in the previous subsection. Notice that there are missing values in the upper right panel because the equilibria under the parameters are non-integers, given the chosen data. I find that as merger cost $\gamma$ decreases, large firms acquire small firms, but small firms do not acquire small firms because they would not reach the subsidy threshold. In addition, as the subsidy sensitivity $\delta$ increases, large firms are more likely to form larger coalitional groups. These findings are robust across all specifications. Although the features in Figure \ref{fg:stacked_plot_merger_configuration} are market-specific, it is informative to understand the merger matching that market researchers choose in their applications.

\subsubsection{An optimal merger-induced subsidy}

Finally, I examine the government's expenditure minimization to evaluate the optimality of a merger-induced subsidy. To make Appendix \ref{sec:comparative_statics} self-contained, I reintroduce the main text in Section \ref{sec:counter_factual}.
I specify the objective function of the government as follows. Given the data about observed firm's characteristics $X$, parameters $\theta$, and thresholds $\underline{\mathcal{J}},\overline{\mathcal{J}} \in \mathbb{N}$ such that $\underline{\mathcal{J}}\le\overline{\mathcal{J}}$, the government solves the following expenditure minimization problem:
\begin{align*}
    \min_{\{M_J\}_{J \in \mathcal{J}} \in \mathbb{R}_{+}^{|\mathcal{J}|},\kappa\in \mathbb{R}_{+}} &\sum_{J\in \mathcal{J}(M,\kappa,\epsilon) } M_J \\
    \text{s.t. }& \underline{\mathcal{J}} \le |\mathcal{J}(M,\kappa,\epsilon)| \le \overline{\mathcal{J}}\\
    &\mathcal{J}(M,\kappa,\epsilon) = \Gamma(M,\kappa,\epsilon|\theta,X)\\
    &\epsilon=\{\epsilon_{i,J}\}_{i \in \mathcal{N},J \in \mathcal{J}},\quad \epsilon_{i,J}\sim_{iid} F_{\epsilon}
\end{align*}
where $M_J$ is the subsidy amount paid to group $J$ from the government, $\kappa$ is the predetermined subsidy threshold,  $\overline{\mathcal{J}}$ and $\underline{\mathcal{J}}$ are the predetermined maximum and minimum numbers of groups that the government must satisfy by subsidy design, $\Gamma(M,\kappa,\epsilon|\theta,\text{Data})$ is the operator that determines the matching allocation based on $M$ and $\kappa$ by Equation \eqref{eq:LP} given parameters $\theta$, data, and random term $\epsilon$ drawn from the distribution $F_{\epsilon}$. The analytical or approximated solution is impossible to obtain because $\Gamma$ involves random term $\epsilon$ in equilibrium computation and the domain of control variables, $\mathbb{R}_{+}^{|\mathcal{J}|}$ varies via computation of $\gamma$. Instead, this study focuses on how $|\mathcal{J}(M,\kappa,\epsilon)|$ would change if the government changed $M=M_J$ for all $J\in \mathcal{J}$ and $\kappa$ given drawn $\epsilon$.

\begin{figure}[htbp]
 \begin{minipage}{0.33\hsize}
  \begin{center}
   \includegraphics[width=50mm]{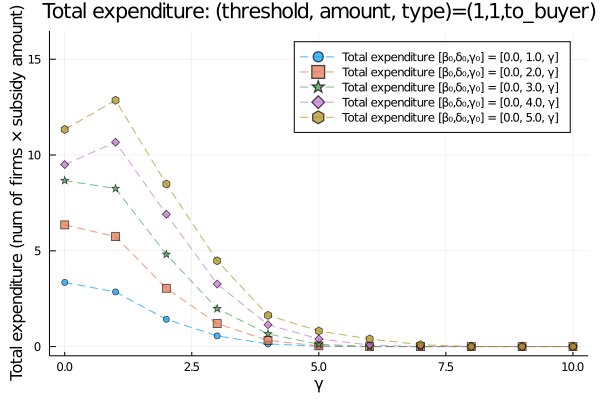}
  \end{center}
 \end{minipage}
 \begin{minipage}{0.33\hsize}
  \begin{center}
   \includegraphics[width=50mm]{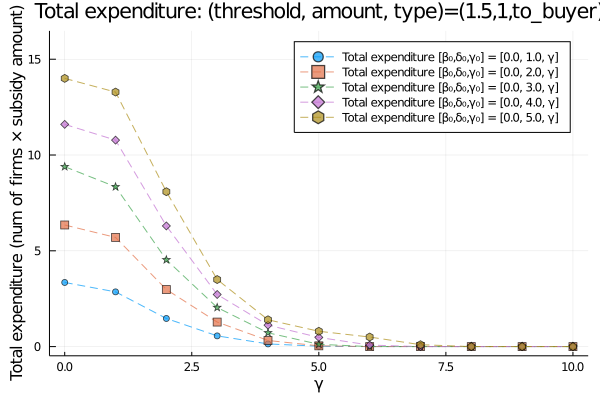}
  \end{center}
 \end{minipage}
 \begin{minipage}{0.33\hsize}
  \begin{center}
   \includegraphics[width=50mm]{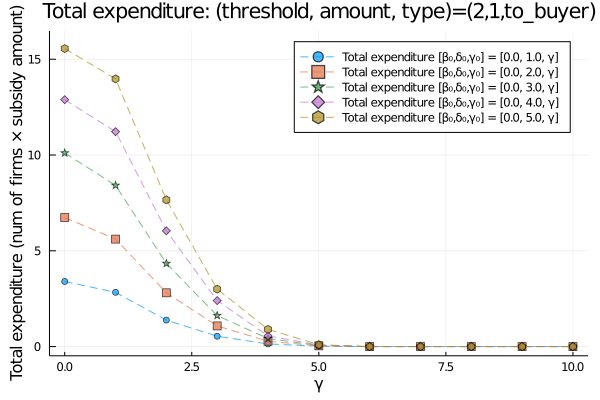}
  \end{center}
 \end{minipage}
 \caption{Total expenditure for subsidy provision under different parameters and subsidy thresholds. Total expenditure $E$ is calculated by $E=\sum_{J\in \mathcal{J}} M_J=|\mathcal{J}|\cdot \delta$ under the specification of $M_J=1$ for all $J\in\mathcal{J}$. I generate 50 simulated data and calculate the total expenditure on average by using integer matching data. This study uses subsidy specification $s_{i,J}^{\text{to buyer}}$.}
 \label{fg:plot_total_expenditure_comparative_statics_to_buyer_subsidy}
\end{figure}

Using the same 50 simulated matching market data in the previous subsections, Figure \ref{fg:plot_total_expenditure_comparative_statics_to_buyer_subsidy} shows the government's total expenditure $E=\sum_{J\in \mathcal{J}} M_J$ across different policy mixes of subsidy amount $M_J$ and threshold $\kappa$. The change of subsidy amount is controlled by the change in subsidy sensitivity $\delta$ under the present specification. The total expenditure increases as not only subsidy amount but also subsidy threshold increase because subsidy threshold with enough subsidy amount may increase the number of qualified groups. This illustrates how the policy mix strategy is related to the government's total expenditure. In addition, the total expenditures across different subsidy thresholds converge to zero as merger cost $\gamma$ increases because the subsidy would not compensate for high merger costs. This implies that policy concern about the government's total expenditure matters only when merger cost is relatively small.

In conclusion, Appendix \ref{sec:comparative_statics} shows the validity of the proposed one-sided one-to-many matching model with complementarities for counterfactual simulation in simulated pseudo-data. All the numerical matching outcomes are reasonable and interpretable. Although this study focuses on the applications to one-to-many horizontal mergers, the proposed model can be applied to various one-to-many matching environments even in a single-market setting with unmatched agents, and not only for industrial organization literature.

\subsection{Equilibrium properties in an oligopoly market with only large firms}\label{subsec:comparative_statics_large_firms}

So far, I simulate the market where two or three large firms and six or five mid-sized or small-sized firms exist. Section \ref{subsec:comparative_statics_large_firms} considers the market where $N=8$ firms are relatively large firms. The market structure is different from the market in Sections \ref{subsec:estimation_two_variables_main_firms_only} and Section \ref{sec:counter_factual}. I use the same simulated data in Appendix \ref{subsec:monte_carlo_large_firms}.

\begin{figure}[htbp]
 \begin{minipage}{0.33\hsize}
  \begin{center}
   \includegraphics[width=50mm]{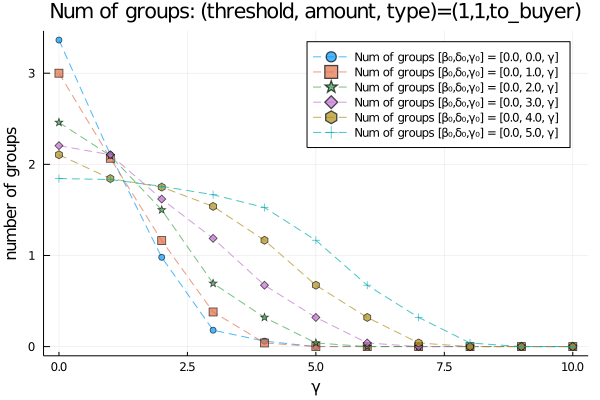}
  \end{center}
 \end{minipage}
 \begin{minipage}{0.33\hsize}
  \begin{center}
   \includegraphics[width=50mm]{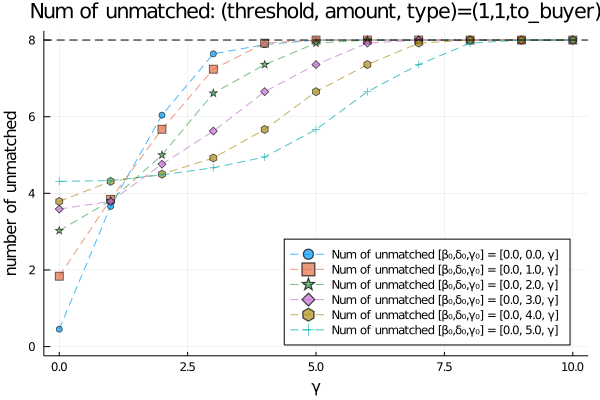}
  \end{center}
 \end{minipage}
 \begin{minipage}{0.33\hsize}
  \begin{center}
   \includegraphics[width=50mm]{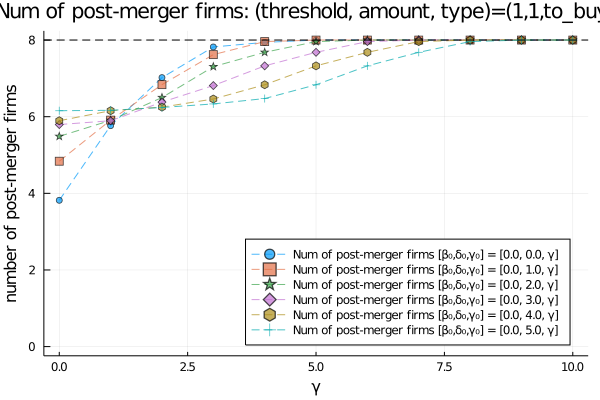}
  \end{center}
 \end{minipage}
 \caption{The equilibrium numbers of groups and unmatched firms. For each parameter pair, I generate 50 simulated data and calculate the average numbers of groups and unmatched firms by using integer matching outcomes. This study uses subsidy specification $s_{i,J}^{\text{to buyer}}$. The number of post-merger firms in the right panels includes coalitional groups and unmatched individual firms.}
 \label{fg:plot_num_of_post_merger_firms_large_firms}
\end{figure}

\begin{figure}[htbp]
 \begin{minipage}{0.33\hsize}
  \begin{center}
   \includegraphics[width=50mm]{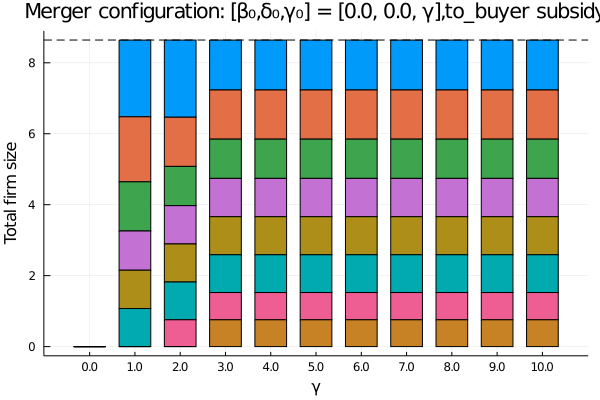}
  \end{center}
 \end{minipage}
 \begin{minipage}{0.33\hsize}
  \begin{center}
   \includegraphics[width=50mm]{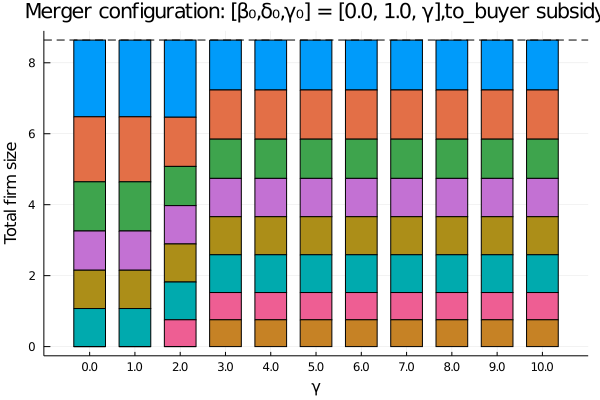}
  \end{center}
 \end{minipage}
 \begin{minipage}{0.33\hsize}
  \begin{center}
   \includegraphics[width=50mm]{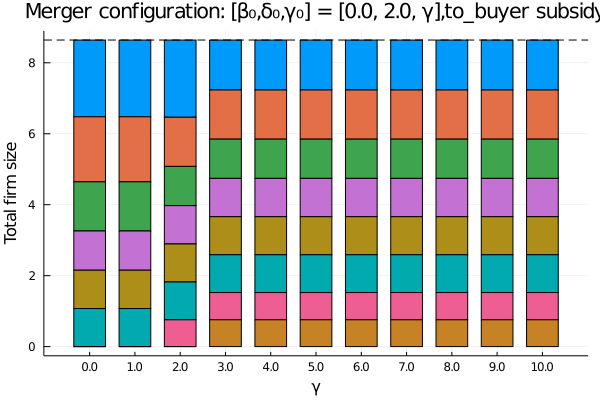}
  \end{center}
 \end{minipage}\\
 \begin{minipage}{0.33\hsize}
  \begin{center}
   \includegraphics[width=50mm]{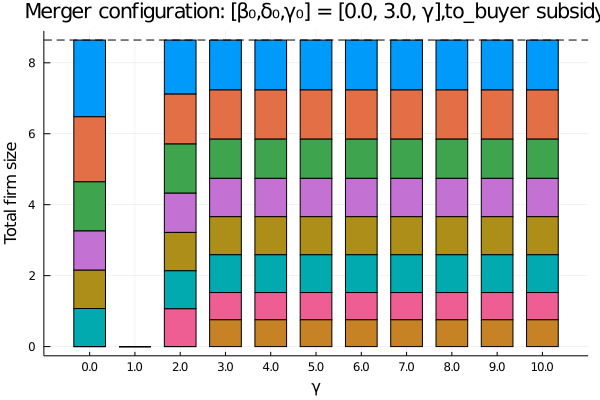}
  \end{center}
 \end{minipage}
 \begin{minipage}{0.33\hsize}
  \begin{center}
   \includegraphics[width=50mm]{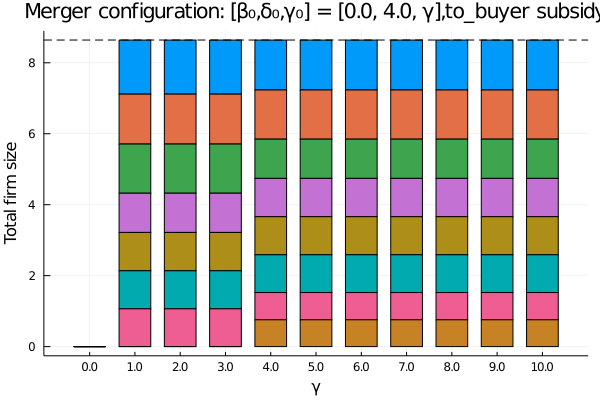}
  \end{center}
 \end{minipage}
 \begin{minipage}{0.33\hsize}
  \begin{center}
   \includegraphics[width=50mm]{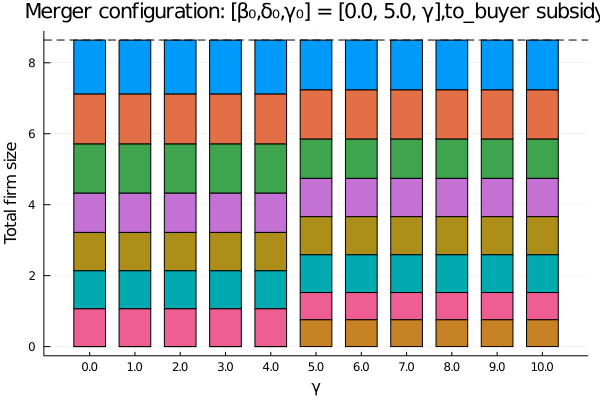}
  \end{center}
 \end{minipage}
 \caption{The merger configuration under different parameters. For each parameter pair, I generate 50 simulated data and pick up one representative data that generate an integer equilibrium for almost all parameters. This study uses subsidy specification $s_{i,J}^{\text{to buyer}}$. Some stacked bars are missing because the matching outcomes under the parameters are non-integers. The cases should be treated as probabilistic allocations, so it is impossible to illustrate configurations.}
 \label{fg:stacked_plot_merger_configuration_large_firms}
\end{figure}

\begin{figure}[htbp]
 \begin{minipage}{0.33\hsize}
  \begin{center}
   \includegraphics[width=50mm]{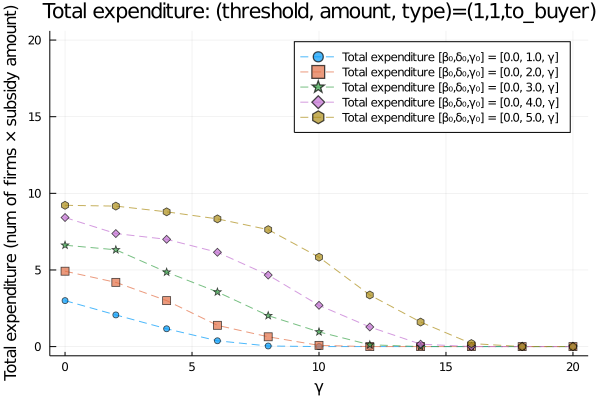}
  \end{center}
 \end{minipage}
 \begin{minipage}{0.33\hsize}
  \begin{center}
   \includegraphics[width=50mm]{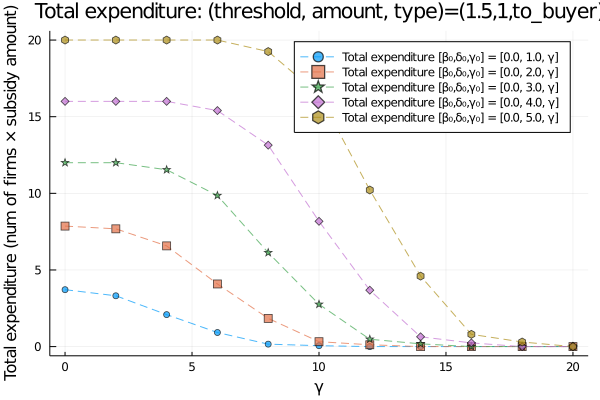}
  \end{center}
 \end{minipage}
 \begin{minipage}{0.33\hsize}
  \begin{center}
   \includegraphics[width=50mm]{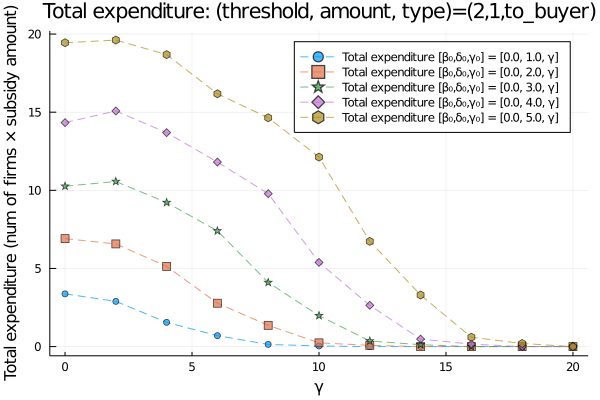}
  \end{center}
 \end{minipage}
 \caption{The total expenditure for subsidy provision under different parameters and subsidy thresholds. Total expenditure $E$ is calculated by $E=\sum_{J\in \mathcal{J}} M_J=|\mathcal{J}|\cdot \delta$ under the specification of $M_J=1$ for all $J\in\mathcal{J}$. I generate 50 simulated data and calculate the total expenditure on average by using integer matching outcomes data. This study uses subsidy specification $s_{i,J}^{\text{to buyer}}$.}
 \label{fg:plot_total_expenditure_comparative_statics_to_buyer_subsidy_large_firms}
\end{figure}

For brevity, I focus only on remarkable differences from the results in the previous sections. Figure \ref{fg:plot_num_of_post_merger_firms_large_firms} shows the equilibrium number of groups. In the left panel, the number of groups is stable across different merger cost $\gamma$ as subsidy sensitivity $\delta$ increases. This implies that the effect of the merger-induced subsidy is sensitive to merger cost only when $\delta$ is low, that is, firms are insensitive to the subsidy.

Figure \ref{fg:stacked_plot_merger_configuration_large_firms} shows the equilibrium merger configuration of illustrative one of the 50 simulated data. The remarkable difference from Figure \ref{fg:plot_num_of_post_merger_firms_large_firms} is that mergers occur only if merger cost $\gamma$ is low because the payoff of staying unmatched is large for all firms. In addition, Figure \ref{fg:plot_total_expenditure_comparative_statics_to_buyer_subsidy_large_firms} shows the stability of total expenditure for subsidy provision across different merger costs as well. The remarkable difference is that the increment of the expenditure of shifting up a subsidy threshold is ignorable only when subsidy sensitivity $\delta$ is low, and is costly proportional to the subsidy amount otherwise. Thus, the subsidy threshold manipulation in the market with only large firms can affect the matching outcome when subsidy sensitivity $\delta$ is high.

\newpage
\section{Supplemental estimation results}\label{sec:supplemental_estimation}

\subsection{Estimation results based on assortativeness of HHI levels}\label{subsec:estimation_HHI_only}

Table \ref{tb:score_results_HHI_only_shared_subsidy} reports the estimation results with only HHI of each buyer firm and coalition, defined in Section \ref{subsec:preliminary_regressions}. The fact that $\beta_{HHI}$ is significantly negative implies that firms aim for technological diversification totally. In addition, as in Section \ref{sec:estimation}, $\gamma$ is significantly large, which shows the existence of merger costs. These results are consistent with Table \ref{tb:score_results_two_variables}.

{
\begin{table}[!htbp] \centering 
  \caption{Results of matching maximum rank estimator based on assortativeness of HHI levels.} 
  \label{tb:score_results_HHI_only_shared_subsidy} 
  \begin{tabular}{@{\extracolsep{5pt}}lcc}
\toprule 
 &  &  \\
 &  & Point Est \\
 &  & [95\% CI] \\
\midrule 
 &  &  \\
total$_{b}$ $\times$ total$_{t}$ & $\beta_0$ & +1 \\
 &  & (S) \\
HHI$_{b}$ $\times$ HHI$_{t}$ & $\beta_{HHI}$ & -47.74 \\
 &  & [-287.9, -28.4] \\
merger cost & -$\gamma$ & -2.3 \\
 &  & [-3.8, -0.7] \\
subsidy sensitivity & $\delta$ & 20 \\
 &  &  \\
\hline 
$\sharp$ Inequalities &  & 17864 \\
\% Inequalities &  & 0.8971 \\
\bottomrule 
\end{tabular}

  \footnotesize
   \begin{tablenotes}
\item[a]\textit{Note:} The objective function was numerically maximized using differential evolution (DE) algorithm in \texttt{BlackBoxOptim.jl} package. For DE algorithm, I require setting the domain of parameters and the number of population seeds so that I fix the former to $[-20, 10]$ for $\gamma$ and $[-300,300]$ for other parameters, and the latter to 100. For point estimation, 100 runs were performed for all specifications. The reported point estimates are the best-found maxima. The parentheses are 95 \% confidence intervals which are computed by taking the 2.5th percentile and the 97.5th percentile of the empirical sampling distribution of estimated parameters. Bootstrap uses 200 replications with 200 population seeds and 118 firms (all 12 main firms and 106 non-main firms sampled with replacement out of 106 non-main firms) per replication. Confidence intervals are not necessarily symmetric around the point estimate. Parameters that can take on only a finite number of values (here +1) converge at an arbitrarily fast rate, then they are superconsistent (denoted by (S)). The unit of measure of assortativeness of scale is a million D/W tonnage. The sub-indices $b$ and $t$ mean the buyer's and target coalition's covariates. 
   \end{tablenotes}
\end{table} 
}

\subsection{Estimation results on multivariate observed characteristics}\label{subsec:estimation_multivariate}

Table \ref{tb:score_results_multivariate} reports the estimation results with multivariate observed characteristics. In this study, the existence of a set of global maximizers cannot be verified so that the point estimation may not be reliable. The numbers in parentheses are 95 \% confidence intervals via bootstrap. Estimation details are summarized in the footnotes of Tables. In this study, I normalized $\beta_0$ to $+1$ for identification. The other parameters are estimated separately under the normalization, and the vector of the parameters with the highest number of satisfied inequalities is chosen. The number of inequalities for the full sample is $17,864$. 

{
\begin{table}[!htbp] \centering 
  \caption{Results of matching maximum rank estimator.} 
  \label{tb:score_results_multivariate} 
  \scriptsize{}
  \begin{tabular}{@{\extracolsep{5pt}}lcccccc}
\toprule 
 &  &  & Value Function &  &  &  \\
 &  & (1) & (2) & (3) &  &  \\
 &  & Point Estimate & Point Estimate & Point Estimate & Use Column 3 &  \\
 &  & [95\% CI] & [95\% CI] & [95\% CI] & $\bar{x}$  & $\beta(\bar{x}\times\bar{x})/\gamma$ \\
\midrule 
Scale variables &  &  &  &  &  &  \\
 &  &  &  &  &  \\
total$_{b}$ $\times$ total$_{t}$ & $\beta_0$ & +1 & +1 & +1 & 0.11 & 0.001 \\
 &  & (S) & (S) & (S) &  &  \\
liner$_{b}$ $\times$ liner$_{t}$ & $\beta_1$ & 297.64 &  & 67.74 & 0.031 & 0.005 \\
 &  & [8.76, 296.16] &  & [-285.41, 256.78] &  &  \\
tramper$_{b}$ $\times$ tramper$_{t}$ & $\beta_2$ & 86.47 &  & 266.17 & 0.015 & 0.005 \\
 &  & [-140.09, 288.7] &  & [-282.79, 295.86] &  &  \\
special$_{b}$ $\times$ special$_{t}$ & $\beta_3$ & 289.17 &  & -121.43 & 0.03 & -0.009 \\
 &  & [23.37, 296.29] &  & [-278.51, 295.32] &  &  \\
tanker$_{b}$ $\times$ tanker$_{t}$ & $\beta_4$ & 13.64 &  & 152.96 & 0.035 & 0.015 \\
 &  & [-180.16, 260.22] &  & [-258.33, 269.98] &  &  \\
Share variables &  &  &  &  &  &  \\
 &  &  &  &  &  &  \\
liner$_{b}$ $\times$ liner$_{t}$ & $\beta_5$ &  & -23.8 & -67.06 & 0.104 & -0.056 \\
 &  &  & [-292.34, -12.62] & [-289.45, -23.64] &  &  \\
tramper$_{b}$ $\times$ tramper$_{t}$ & $\beta_6$ &  & -79.47 & -36.06 & 0.117 & -0.038 \\
 &  &  & [-292.64, -3.44] & [-282.32, -3.97] &  &  \\
special$_{b}$ $\times$ special$_{t}$ & $\beta_7$ &  & -28.07 & -194.54 & 0.192 & -0.558 \\
 &  &  & [-292.48, -11.48] & [-294.97, -9.23] &  &  \\
tanker$_{b}$ $\times$ tanker$_{t}$ & $\beta_8$ &  & -57.63 & -76.95 & 0.587 & -2.063 \\
 &  &  & [-290.01, -3.59] & [-293.72, -9.35] &  &  \\
 &  &  &  &  &  &  \\
 &  &  &  &  &  &  \\
merger cost & -$\gamma$ & -11.2 & -6.12 & -12.85 &  &  \\
 &  & [-19.33, 2.05] & [-18.94, -0.43] & [-18.76, 0.81] &  &  \\
subsidy sensitivity & $\delta$ & 20 & 20 & 20 &  &  \\
 &  &  &  &  &  &  \\
\hline 
$\sharp$ Inequalities &  & 17864 & 17864 & 17864 &  &  \\
\% Inequalities &  & 0.815 & 0.902 & 0.904 &  &  \\
\bottomrule 
\end{tabular}

\begin{tablenotes}
\item[a]\textit{Note:} The objective function was numerically maximized using differential evolution (DE) algorithm in \texttt{BlackBoxOptim.jl} package. For DE algorithm, I am required to set the domain of parameters and the number of population seeds so that I fix the former to $[-20, 10]$ for $\gamma$ and $[-300,300]$ for other parameters, and the latter to 100. For point estimation, 100 runs were performed for all specifications. The reported point estimates are the best-found maxima. The parentheses are 95 \% confidence intervals which are computed by taking the 2.5th percentile and the 97.5th percentile of the empirical sampling distribution of estimated parameters. Bootstrap uses 200 replications and 118 firms (all 12 main firms and sampled 106 non-main firms out of 106) per replication (sampled without replacement). The confidence regions are not necessarily symmetric around the point estimate. Parameters that can take on only a finite number of values (here +1) converge at an arbitrarily fast rate, then they are superconsistent (denoted by (S)). The unit of measure of economies of scale is a million D/W tonnage. The sub-indices $b$ and $t$ mean the buyer's and target coalition's covariates. The subsidy term is specified as the form of $s_{i,J}^{\text{to buyer}}$.
   \end{tablenotes}
\end{table} 
}

Three specifications are considered. The robust finding in Table \ref{tb:score_results_two_variables} is that merger cost $\gamma$ is significantly positive and assortativeness of specialization ($\beta_5,\beta_6,\beta_7$ and $\beta_8$) is negative for all specifications, but the signs of parameters regarding assortativeness of size are ambiguous. This is because many possible combinations of the observed characteristics that are potentially correlated can achieve the maximum score of the objective function. In addition, if the specification of the joint production function includes not only positive assortative features but also negative elements such as merger costs, the estimation will be harder because the positive and negative terms are balanced at many points. Based on these findings, as in \cite{bajari2008evaluating}, I proceed in Section \ref{sec:estimation} with a small set of observed characteristics separately while confirming the shape of the objective function. 

Instead of loss of robustness of signs of share's coefficients, Column 3 in Table \ref{tb:score_results_multivariate} gives comparable results of each variable in a single specification. Column 5 shows the total value of the joint production function when a hypothetical average firm merges another hypothetical average firm based on estimates in Column 3. First, this illustrates that the size of liner and tramper has the same importance for merger incentives. Also, the share variables are 10 times more important than the scale variables. Second, the importance of the share of special and tanker shipping is fairly large negatively. This captures that firms consider the diversification on tanker and special shipping as the first criteria for merger decisions, which seems consistent with the post-merger compositions of five groups, except Showa Line Group in Figure \ref{fg:carrier_dist_eachgroup}.

\section{Robustness checks}\label{sec:robustness_check}

\subsection{Different calibrated $\delta$}\label{subsec:robustness_check_delta}

This subsection shows corresponding results with Table \ref{tb:score_results_two_variables_main_firms_only} under different calibrated parameters $\delta\in\{10,100,200,300,400,500,1000\}$. Tables \ref{tb:score_results_two_variables_main_firms_only_different_delta} and \ref{tb:score_results_two_variables_main_firms_only_different_delta_2} show that the number of correctly predicted inequalities increases as $\delta$ increases, and the improvement is weakly monotonic in $\delta$. In Table \ref{tb:score_results_two_variables_main_firms_only}, I use $\delta=400$ as an approximate lower bound of $\delta$ achieving the maximum matching score uniquely.\footnote{Table \ref{tb:score_results_two_variables_main_firms_only_different_delta} shows that $\delta=300$ achieve approximately the same maximum values of the objective function. If the value is taken as a calibration parameter, the rescaled subsidy amount may give the different conclusion that subsidy provision is perfectly optimal. However, allowing smaller $\delta$ with corresponding $\beta$ and $\gamma$ generate matching outcomes inconsistent with the actual post merger data. In other words, the counterfactual simulation under smaller $\delta$ does not replicate the actual matching outcome under actual subsidy design. Thus, $\delta=400$ is the best from the perspective of prediction accuracy of inequalities and point-identification.} The full-sample estimation in Tables \ref{tb:score_results_two_variables}, \ref{tb:score_results_HHI_only_shared_subsidy}, and \ref{tb:score_results_multivariate} show the similar monotonic property, so I take $\delta=20$ in the main analysis.

{
\begin{table}[!htbp] \centering 
  \caption{\textbf{Results of matching maximum rank estimator with two observed variables for 12 main firms.} The numbers in parentheses are lower and upper bounds of a set of maximizers of the maximum rank estimator. $\delta=10,100,200,300$.} 
  \label{tb:score_results_two_variables_main_firms_only_different_delta} 
  {\fontsize{6.0}{8.0}\selectfont
  \begin{tabular}{@{\extracolsep{5pt}}lccccccccc}
\toprule 
 &  &  &  &  &  &  &  &  &  \\
 &  & (1) & (2) & (3) & (4) & (5) & (6) & (7) & (8) \\
 &  & [LB, UB] & [LB, UB] & [LB, UB] & [LB, UB] & [LB, UB] & [LB, UB] & [LB, UB] & [LB, UB] \\
\midrule 
Scale variables &  &  &  &  &  &  &  \\
 &  &  &  &  &  &  &  &  \\
total$_{b}$ $\times$ total$_{t}$ & $\beta_0$ & +1 & +1 & +1 & +1 & +1 & +1 & +1 & +1 \\
 &  & (S) & (S) & (S) & (S) & (S) & (S) & (S) & (S) \\
liner$_{b}$ $\times$ liner$_{t}$ & $\beta_1$ & [124.0, 124.0] &  &  &  &  &  &  &  \\
tramper$_{b}$ $\times$ tramper$_{t}$ & $\beta_2$ &  & [54.1, 293.4] &  &  &  &  &  &  \\
special$_{b}$ $\times$ special$_{t}$ & $\beta_3$ &  &  & [-151.8, -42.6] &  &  &  &  &  \\
tanker$_{b}$ $\times$ tanker$_{t}$ & $\beta_4$ &  &  &  & [-220.8, -220.8] &  &  &  &  \\
Share variables &  &  &  &  &  &  &  &  &  \\
 &  &  &  &  &  &  &  &  &  \\
liner$_{b}$ $\times$ liner$_{t}$ & $\beta_5$ &  &  &  &  & [33.9, 37.4] &  &  &  \\
tramper$_{b}$ $\times$ tramper$_{t}$ & $\beta_6$ &  &  &  &  &  & [241.9, 279.9] &  &  \\
special$_{b}$ $\times$ special$_{t}$ & $\beta_7$ &  &  &  &  &  &  & [278.9, 292.6] &  \\
tanker$_{b}$ $\times$ tanker$_{t}$ & $\beta_8$ &  &  &  &  &  &  &  & [286.4, 286.4] \\
 &  &  &  &  &  &  &  &  &  \\
 &  &  &  &  &  &  &  &  &  \\
merger cost & -$\gamma$ & [-15.4, -15.4] & [-10.3, -5.2] & [-9.1, -1.8] & [-8.4, -8.4] & [-13.6, -3.5] & [-9.1, -3.8] & [-7.8, -5.2] & [-8.9, -8.9] \\
subsidy sensitivity & $\delta$ & 10 & 10 & 10 & 10 & 10 & 10 & 10 & 10 \\
 &  &  &  &  &  &  &  &  &  \\
\hline 
$\sharp$ Inequalities (Point) &  & 330 & 330 & 330 & 330 & 330 & 330 & 330 & 330 \\
\% Inequalities &  & 0.9636 & 0.9271 & 0.9394 & 0.9152 & 0.9756 & 0.9514 & 0.9635 & 0.9515 \\
\bottomrule 
\end{tabular}

  \begin{tabular}{@{\extracolsep{5pt}}lccccccccc}
\toprule 
 &  &  &  &  &  &  &  &  &  \\
 &  & (1) & (2) & (3) & (4) & (5) & (6) & (7) & (8) \\
 &  & [LB, UB] & [LB, UB] & [LB, UB] & [LB, UB] & [LB, UB] & [LB, UB] & [LB, UB] & [LB, UB] \\
\midrule 
Scale variables &  &  &  &  &  &  &  \\
 &  &  &  &  &  &  &  &  \\
total$_{b}$ $\times$ total$_{t}$ & $\beta_0$ & +1 & +1 & +1 & +1 & +1 & +1 & +1 & +1 \\
 &  & (S) & (S) & (S) & (S) & (S) & (S) & (S) & (S) \\
liner$_{b}$ $\times$ liner$_{t}$ & $\beta_1$ & [124.0, 124.0] &  &  &  &  &  &  &  \\
tramper$_{b}$ $\times$ tramper$_{t}$ & $\beta_2$ &  & [54.1, 293.4] &  &  &  &  &  &  \\
special$_{b}$ $\times$ special$_{t}$ & $\beta_3$ &  &  & [-151.8, -42.6] &  &  &  &  &  \\
tanker$_{b}$ $\times$ tanker$_{t}$ & $\beta_4$ &  &  &  & [-220.8, -220.8] &  &  &  &  \\
Share variables &  &  &  &  &  &  &  &  &  \\
 &  &  &  &  &  &  &  &  &  \\
liner$_{b}$ $\times$ liner$_{t}$ & $\beta_5$ &  &  &  &  & [33.9, 37.4] &  &  &  \\
tramper$_{b}$ $\times$ tramper$_{t}$ & $\beta_6$ &  &  &  &  &  & [241.9, 279.9] &  &  \\
special$_{b}$ $\times$ special$_{t}$ & $\beta_7$ &  &  &  &  &  &  & [278.9, 292.6] &  \\
tanker$_{b}$ $\times$ tanker$_{t}$ & $\beta_8$ &  &  &  &  &  &  &  & [286.4, 286.4] \\
 &  &  &  &  &  &  &  &  &  \\
 &  &  &  &  &  &  &  &  &  \\
merger cost & -$\gamma$ & [-15.4, -15.4] & [-10.3, -5.2] & [-9.1, -1.8] & [-8.4, -8.4] & [-13.6, -3.5] & [-9.1, -3.8] & [-7.8, -5.2] & [-8.9, -8.9] \\
subsidy sensitivity & $\delta$ & 100 & 100 & 100 & 100 & 100 & 100 & 100 & 100 \\
 &  &  &  &  &  &  &  &  &  \\
\hline 
$\sharp$ Inequalities (Point) &  & 330 & 330 & 330 & 330 & 330 & 330 & 330 & 330 \\
\% Inequalities &  & 0.9636 & 0.9271 & 0.9394 & 0.9152 & 0.9756 & 0.9514 & 0.9635 & 0.9515 \\
\bottomrule 
\end{tabular}

  \begin{tabular}{@{\extracolsep{5pt}}lccccccccc}
\toprule 
 &  &  &  &  &  &  &  &  &  \\
 &  & (1) & (2) & (3) & (4) & (5) & (6) & (7) & (8) \\
 &  & [LB, UB] & [LB, UB] & [LB, UB] & [LB, UB] & [LB, UB] & [LB, UB] & [LB, UB] & [LB, UB] \\
\midrule 
Scale variables &  &  &  &  &  &  &  \\
 &  &  &  &  &  &  &  &  \\
total$_{b}$ $\times$ total$_{t}$ & $\beta_0$ & +1 & +1 & +1 & +1 & +1 & +1 & +1 & +1 \\
 &  & (S) & (S) & (S) & (S) & (S) & (S) & (S) & (S) \\
liner$_{b}$ $\times$ liner$_{t}$ & $\beta_1$ & [124.0, 124.0] &  &  &  &  &  &  &  \\
tramper$_{b}$ $\times$ tramper$_{t}$ & $\beta_2$ &  & [54.1, 293.4] &  &  &  &  &  &  \\
special$_{b}$ $\times$ special$_{t}$ & $\beta_3$ &  &  & [-151.8, -42.6] &  &  &  &  &  \\
tanker$_{b}$ $\times$ tanker$_{t}$ & $\beta_4$ &  &  &  & [-220.8, -220.8] &  &  &  &  \\
Share variables &  &  &  &  &  &  &  &  &  \\
 &  &  &  &  &  &  &  &  &  \\
liner$_{b}$ $\times$ liner$_{t}$ & $\beta_5$ &  &  &  &  & [33.9, 37.4] &  &  &  \\
tramper$_{b}$ $\times$ tramper$_{t}$ & $\beta_6$ &  &  &  &  &  & [241.9, 279.9] &  &  \\
special$_{b}$ $\times$ special$_{t}$ & $\beta_7$ &  &  &  &  &  &  & [278.9, 292.6] &  \\
tanker$_{b}$ $\times$ tanker$_{t}$ & $\beta_8$ &  &  &  &  &  &  &  & [286.4, 286.4] \\
 &  &  &  &  &  &  &  &  &  \\
 &  &  &  &  &  &  &  &  &  \\
merger cost & -$\gamma$ & [-15.4, -15.4] & [-10.3, -5.2] & [-9.1, -1.8] & [-8.4, -8.4] & [-13.6, -3.5] & [-9.1, -3.8] & [-7.8, -5.2] & [-8.9, -8.9] \\
subsidy sensitivity & $\delta$ & 200 & 200 & 200 & 200 & 200 & 200 & 200 & 200 \\
 &  &  &  &  &  &  &  &  &  \\
\hline 
$\sharp$ Inequalities (Point) &  & 330 & 330 & 330 & 330 & 330 & 330 & 330 & 330 \\
\% Inequalities &  & 0.9636 & 0.9271 & 0.9394 & 0.9152 & 0.9756 & 0.9514 & 0.9635 & 0.9515 \\
\bottomrule 
\end{tabular}

  \begin{tabular}{@{\extracolsep{5pt}}lccccccccc}
\toprule 
 &  &  &  &  &  &  &  &  &  \\
 &  & (1) & (2) & (3) & (4) & (5) & (6) & (7) & (8) \\
 &  & [LB, UB] & [LB, UB] & [LB, UB] & [LB, UB] & [LB, UB] & [LB, UB] & [LB, UB] & [LB, UB] \\
\midrule 
Scale variables &  &  &  &  &  &  &  \\
 &  &  &  &  &  &  &  &  \\
total$_{b}$ $\times$ total$_{t}$ & $\beta_0$ & +1 & +1 & +1 & +1 & +1 & +1 & +1 & +1 \\
 &  & (S) & (S) & (S) & (S) & (S) & (S) & (S) & (S) \\
liner$_{b}$ $\times$ liner$_{t}$ & $\beta_1$ & [124.0, 124.0] &  &  &  &  &  &  &  \\
tramper$_{b}$ $\times$ tramper$_{t}$ & $\beta_2$ &  & [54.1, 293.4] &  &  &  &  &  &  \\
special$_{b}$ $\times$ special$_{t}$ & $\beta_3$ &  &  & [-151.8, -42.6] &  &  &  &  &  \\
tanker$_{b}$ $\times$ tanker$_{t}$ & $\beta_4$ &  &  &  & [-220.8, -220.8] &  &  &  &  \\
Share variables &  &  &  &  &  &  &  &  &  \\
 &  &  &  &  &  &  &  &  &  \\
liner$_{b}$ $\times$ liner$_{t}$ & $\beta_5$ &  &  &  &  & [33.9, 37.4] &  &  &  \\
tramper$_{b}$ $\times$ tramper$_{t}$ & $\beta_6$ &  &  &  &  &  & [241.9, 279.9] &  &  \\
special$_{b}$ $\times$ special$_{t}$ & $\beta_7$ &  &  &  &  &  &  & [278.9, 292.6] &  \\
tanker$_{b}$ $\times$ tanker$_{t}$ & $\beta_8$ &  &  &  &  &  &  &  & [286.4, 286.4] \\
 &  &  &  &  &  &  &  &  &  \\
 &  &  &  &  &  &  &  &  &  \\
merger cost & -$\gamma$ & [-15.4, -15.4] & [-10.3, -5.2] & [-9.1, -1.8] & [-8.4, -8.4] & [-13.6, -3.5] & [-9.1, -3.8] & [-7.8, -5.2] & [-8.9, -8.9] \\
subsidy sensitivity & $\delta$ & 300 & 300 & 300 & 300 & 300 & 300 & 300 & 300 \\
 &  &  &  &  &  &  &  &  &  \\
\hline 
$\sharp$ Inequalities (Point) &  & 330 & 330 & 330 & 330 & 330 & 330 & 330 & 330 \\
\% Inequalities &  & 0.9636 & 0.9271 & 0.9394 & 0.9152 & 0.9756 & 0.9514 & 0.9635 & 0.9515 \\
\bottomrule 
\end{tabular}

  }
\end{table} 
}

{
\begin{table}[!htbp] \centering 
  \caption{\textbf{Results of matching maximum rank estimator with two observed variables for 12 main firms.} The numbers in parentheses are lower and upper bounds of a set of maximizers of the maximum rank estimator. $\delta=400,500,1000$.} 
  \label{tb:score_results_two_variables_main_firms_only_different_delta_2}
  {\fontsize{7.0}{10.0}\selectfont
  \begin{tabular}{lccccccccc}
\toprule 
 &  & (1) & (2) & (3) & (4) & (5) & (6) & (7) & (8) \\
 &  & [LB, UB] & [LB, UB] & [LB, UB] & [LB, UB] & [LB, UB] & [LB, UB] & [LB, UB] & [LB, UB] \\
\midrule 
Scale variables &  &  &  &  &  &  &  \\
total$_{b}$ $\times$ total$_{t}$ & $\beta_0$ & +1 & +1 & +1 & +1 & +1 & +1 & +1 & +1 \\
 &  & (S) & (S) & (S) & (S) & (S) & (S) & (S) & (S) \\
liner$_{b}$ $\times$ liner$_{t}$ & $\beta_1$ & [124.0, 124.0] &  &  &  &  &  &  &  \\
tramper$_{b}$ $\times$ tramper$_{t}$ & $\beta_2$ &  & [54.1, 293.4] &  &  &  &  &  &  \\
special$_{b}$ $\times$ special$_{t}$ & $\beta_3$ &  &  & [-151.8, -42.6] &  &  &  &  &  \\
tanker$_{b}$ $\times$ tanker$_{t}$ & $\beta_4$ &  &  &  & [-220.8, -220.8] &  &  &  &  \\
Share variables &  &  &  &  &  &  &  &  &  \\
liner$_{b}$ $\times$ liner$_{t}$ & $\beta_5$ &  &  &  &  & [33.9, 37.4] &  &  &  \\
tramper$_{b}$ $\times$ tramper$_{t}$ & $\beta_6$ &  &  &  &  &  & [241.9, 279.9] &  &  \\
special$_{b}$ $\times$ special$_{t}$ & $\beta_7$ &  &  &  &  &  &  & [278.9, 292.6] &  \\
tanker$_{b}$ $\times$ tanker$_{t}$ & $\beta_8$ &  &  &  &  &  &  &  & [286.4, 286.4] \\
 &  &  &  &  &  &  &  &  &  \\
 &  &  &  &  &  &  &  &  &  \\
merger cost & -$\gamma$ & [-15.4, -15.4] & [-10.3, -5.2] & [-9.1, -1.8] & [-8.4, -8.4] & [-13.6, -3.5] & [-9.1, -3.8] & [-7.8, -5.2] & [-8.9, -8.9] \\
subsidy sensitivity & $\delta$ & 400 & 400 & 400 & 400 & 400 & 400 & 400 & 400 \\
 &  &  &  &  &  &  &  &  &  \\
\hline 
$\sharp$ Inequalities (Point) &  & 330 & 330 & 330 & 330 & 330 & 330 & 330 & 330 \\
\% Inequalities &  & 0.9636 & 0.9271 & 0.9394 & 0.9152 & 0.9756 & 0.9514 & 0.9635 & 0.9515 \\
\bottomrule 
\end{tabular}

  \begin{tabular}{@{\extracolsep{5pt}}lccccccccc}
\toprule 
 &  &  &  &  &  &  &  &  &  \\
 &  & (1) & (2) & (3) & (4) & (5) & (6) & (7) & (8) \\
 &  & [LB, UB] & [LB, UB] & [LB, UB] & [LB, UB] & [LB, UB] & [LB, UB] & [LB, UB] & [LB, UB] \\
\midrule 
Scale variables &  &  &  &  &  &  &  \\
 &  &  &  &  &  &  &  &  \\
total$_{b}$ $\times$ total$_{t}$ & $\beta_0$ & +1 & +1 & +1 & +1 & +1 & +1 & +1 & +1 \\
 &  & (S) & (S) & (S) & (S) & (S) & (S) & (S) & (S) \\
liner$_{b}$ $\times$ liner$_{t}$ & $\beta_1$ & [124.0, 124.0] &  &  &  &  &  &  &  \\
tramper$_{b}$ $\times$ tramper$_{t}$ & $\beta_2$ &  & [54.1, 293.4] &  &  &  &  &  &  \\
special$_{b}$ $\times$ special$_{t}$ & $\beta_3$ &  &  & [-151.8, -42.6] &  &  &  &  &  \\
tanker$_{b}$ $\times$ tanker$_{t}$ & $\beta_4$ &  &  &  & [-220.8, -220.8] &  &  &  &  \\
Share variables &  &  &  &  &  &  &  &  &  \\
 &  &  &  &  &  &  &  &  &  \\
liner$_{b}$ $\times$ liner$_{t}$ & $\beta_5$ &  &  &  &  & [33.9, 37.4] &  &  &  \\
tramper$_{b}$ $\times$ tramper$_{t}$ & $\beta_6$ &  &  &  &  &  & [241.9, 279.9] &  &  \\
special$_{b}$ $\times$ special$_{t}$ & $\beta_7$ &  &  &  &  &  &  & [278.9, 292.6] &  \\
tanker$_{b}$ $\times$ tanker$_{t}$ & $\beta_8$ &  &  &  &  &  &  &  & [286.4, 286.4] \\
 &  &  &  &  &  &  &  &  &  \\
 &  &  &  &  &  &  &  &  &  \\
merger cost & -$\gamma$ & [-15.4, -15.4] & [-10.3, -5.2] & [-9.1, -1.8] & [-8.4, -8.4] & [-13.6, -3.5] & [-9.1, -3.8] & [-7.8, -5.2] & [-8.9, -8.9] \\
subsidy sensitivity & $\delta$ & 500 & 500 & 500 & 500 & 500 & 500 & 500 & 500 \\
 &  &  &  &  &  &  &  &  &  \\
\hline 
$\sharp$ Inequalities (Point) &  & 330 & 330 & 330 & 330 & 330 & 330 & 330 & 330 \\
\% Inequalities &  & 0.9636 & 0.9271 & 0.9394 & 0.9152 & 0.9756 & 0.9514 & 0.9635 & 0.9515 \\
\bottomrule 
\end{tabular}

  \begin{tabular}{@{\extracolsep{5pt}}lccccccccc}
\toprule 
 &  &  &  &  &  &  &  &  &  \\
 &  & (1) & (2) & (3) & (4) & (5) & (6) & (7) & (8) \\
 &  & [LB, UB] & [LB, UB] & [LB, UB] & [LB, UB] & [LB, UB] & [LB, UB] & [LB, UB] & [LB, UB] \\
\midrule 
Scale variables &  &  &  &  &  &  &  \\
 &  &  &  &  &  &  &  &  \\
total$_{b}$ $\times$ total$_{t}$ & $\beta_0$ & +1 & +1 & +1 & +1 & +1 & +1 & +1 & +1 \\
 &  & (S) & (S) & (S) & (S) & (S) & (S) & (S) & (S) \\
liner$_{b}$ $\times$ liner$_{t}$ & $\beta_1$ & [124.0, 124.0] &  &  &  &  &  &  &  \\
tramper$_{b}$ $\times$ tramper$_{t}$ & $\beta_2$ &  & [54.1, 293.4] &  &  &  &  &  &  \\
special$_{b}$ $\times$ special$_{t}$ & $\beta_3$ &  &  & [-151.8, -42.6] &  &  &  &  &  \\
tanker$_{b}$ $\times$ tanker$_{t}$ & $\beta_4$ &  &  &  & [-220.8, -220.8] &  &  &  &  \\
Share variables &  &  &  &  &  &  &  &  &  \\
 &  &  &  &  &  &  &  &  &  \\
liner$_{b}$ $\times$ liner$_{t}$ & $\beta_5$ &  &  &  &  & [33.9, 37.4] &  &  &  \\
tramper$_{b}$ $\times$ tramper$_{t}$ & $\beta_6$ &  &  &  &  &  & [241.9, 279.9] &  &  \\
special$_{b}$ $\times$ special$_{t}$ & $\beta_7$ &  &  &  &  &  &  & [278.9, 292.6] &  \\
tanker$_{b}$ $\times$ tanker$_{t}$ & $\beta_8$ &  &  &  &  &  &  &  & [286.4, 286.4] \\
 &  &  &  &  &  &  &  &  &  \\
 &  &  &  &  &  &  &  &  &  \\
merger cost & -$\gamma$ & [-15.4, -15.4] & [-10.3, -5.2] & [-9.1, -1.8] & [-8.4, -8.4] & [-13.6, -3.5] & [-9.1, -3.8] & [-7.8, -5.2] & [-8.9, -8.9] \\
subsidy sensitivity & $\delta$ & 1000 & 1000 & 1000 & 1000 & 1000 & 1000 & 1000 & 1000 \\
 &  &  &  &  &  &  &  &  &  \\
\hline 
$\sharp$ Inequalities (Point) &  & 330 & 330 & 330 & 330 & 330 & 330 & 330 & 330 \\
\% Inequalities &  & 0.9636 & 0.9271 & 0.9394 & 0.9152 & 0.9756 & 0.9514 & 0.9635 & 0.9515 \\
\bottomrule 
\end{tabular}

  }
\end{table} 
}

\newpage

\bibliographystyle{aer}
\bibliography{00ship_merger}

\end{document}